\documentclass[longauth]{aa}  

\usepackage{graphicx}
\usepackage{txfonts}
\usepackage{xcolor}
\usepackage{multirow}
\usepackage{hyperref}
\graphicspath{ {Figures/} }
\begin{document}

   \title{The inflated, eccentric warm Jupiter TOI-4914 b orbiting a metal-poor star, and the hot Jupiters TOI-2714 b and TOI-2981 b   }

   \subtitle{}

   \author{G. Mantovan
          \inst{\ref{inst1},\ref{inst2}}$^{\orcid{0000-0002-6871-6131}}$ \and
          T. G. Wilson\inst{\ref{inst3}}$^{\orcid{0000-0001-8749-1962}}$ \and
          L. Borsato\inst{\ref{inst2}}$^{\orcid{0000-0003-0066-9268}}$ \and
          T. Zingales\inst{\ref{inst1},\ref{inst2}}$^{\orcid{0000-0001-6880-5356}}$ \and
          K. Biazzo\inst{\ref{inst4}}$^{\orcid{0000-0002-1892-2180}}$ \and
          D. Nardiello\inst{\ref{inst1},\ref{inst2}}$^{\orcid{0000-0003-1149-3659}}$ \and
          L. Malavolta\inst{\ref{inst1},\ref{inst2}}$^{\orcid{0000-0002-6492-2085}}$ \and
          S. Desidera\inst{\ref{inst2}}$^{\orcid{0000-0001-8613-2589}}$ \and
          F. Marzari\inst{\ref{inst1}} \and
          A. Collier Cameron\inst{\ref{inst5}}$^{\orcid{0000-0002-8863-7828}}$ \and
          V. Nascimbeni\inst{\ref{inst2}}$^{\orcid{0000-0001-9770-1214}}$ \and
          F. Z. Majidi\inst{\ref{inst6}}$^{\orcid{0000-0002-8407-5282}}$ \and
          M. Montalto\inst{\ref{inst7}}$^{\orcid{0000-0002-7618-8308}}$ \and
          G. Piotto\inst{\ref{inst1}}$^{\orcid{0000-0002-9937-6387}}$ \and
          K. G. Stassun\inst{\ref{inst8}}$^{\orcid{0000-0002-3481-9052}}$ \and
          J. N. Winn\inst{\ref{inst9}}$^{\orcid{0000-0002-4265-047X}}$ \and
          J. M. Jenkins\inst{\ref{inst10}}$^{\orcid{0000-0002-4715-9460}}$ \and
          L. Mignon\inst{\ref{inst11}, \ref{inst11b}}$^{\orcid{0000-0002-5407-3905}}$ \and
          A. Bieryla\inst{\ref{inst12}}$^{\orcid{0000-0001-6637-5401}}$ \and
          D.~W.~Latham\inst{\ref{inst12}}$^{\orcid{0000-0001-9911-7388}}$ \and 
          K. Barkaoui\inst{\ref{inst13},\ref{inst14},\ref{inst15}}$^{\orcid{0000-0003-1464-9276}}$ \and
          K. A. Collins\inst{\ref{inst12}}$^{\orcid{0000-0001-6588-9574}}$ \and
          P. Evans\inst{\ref{inst16}}$^{\orcid{0000-0002-5674-2404}}$ \and
          M.~M.~Fausnaugh\inst{\ref{inst17}}$^{\orcid{0000-0002-9113-7162}}$ \and
          V. Granata\inst{\ref{inst18},\ref{inst19},\ref{inst2}}$^{\orcid{0000-0002-1425-4541}}$ \and
          V. Kostov\inst{\ref{inst20},\ref{inst21}}$^{\orcid{0000-0001-9786-1031}}$ \and
          A. W. Mann\inst{\ref{inst22}}$^{\orcid{0000-0003-3654-1602}}$ \and
          F. J. Pozuelos\inst{\ref{inst23}}$^{\orcid{0000-0003-1572-7707}}$ \and
          D. J. Radford\inst{\ref{inst24}} \and
          H. M. Relles\inst{\ref{inst12}}$^{\orcid{0009-0009-5132-9520}}$ \and
          P. Rowden\inst{\ref{inst25}}$^{\orcid{0000-0002-4829-7101}}$ \and
          S. Seager\inst{\ref{inst26},\ref{inst14},\ref{inst27}}$^{\orcid{0000-0002-6892-6948}}$ \and
          T. -G. Tan\inst{\ref{inst28}}$^{\orcid{0000-0001-5603-6895}}$ \and
          M. Timmermans\inst{\ref{inst13}}$^{\orcid{0009-0008-2214-5039}}$ \and
          C. N. Watkins\inst{\ref{inst12}}$^{\orcid{0000-0001-8621-6731}}$ 
          }

   \institute{Dipartimento di Fisica e Astronomia ``Galileo Galilei'', Università di Padova, Vicolo dell'Osservatorio 3, IT-35122, Padova, Italy\\
              \email{giacomo.mantovan@unipd.it}\label{inst1}
         \and
             Istituto Nazionale di Astrofisica - Osservatorio Astronomico di Padova, Vicolo dell'Osservatorio 5, IT-35122, Padova, Italy\label{inst2}
             \and
             Department of Physics, University of Warwick, Gibbet Hill Road, Coventry CV4 7AL, United Kingdom\label{inst3}
             \and 
             INAF – Osservatorio Astronomico di Roma, Via Frascati 33, 00078, Monte Porzio Catone (Roma), Italy\label{inst4}
             \and 
             Centre for Exoplanet Science, SUPA School of Physics \& Astronomy, University of St Andrews, North Haugh, St Andrews KY169SS, UK\label{inst5}
             \and
             Blue Skies Space SRL., Milan, Italy\label{inst6}
             \and 
             INAF - Osservatorio Astrofisico di Catania, Via S. Sofia 78, 95123, Catania, Italy\label{inst7} 
             \and
             Department of Physics and Astronomy, Vanderbilt University, Nashville, TN 37235, USA\label{inst8} 
             \and
             Department of Astrophysical Sciences, Princeton University, Princeton, NJ 08544, USA\label{inst9} 
             \and 
             NASA Ames Research Center, Moffett Field, CA  94035, USA\label{inst10} 
             \and
             Departement d'astronomie, Université de Genève, Chemin Pegasi, 51, CH-1290 Versoix, Switzerland\label{inst11}
             \and
             Univ. Grenoble Alpes, CNRS, IPAG, 38000 Grenoble, France\label{inst11b}
             \and 
             Center for Astrophysics \textbar Harvard \& Smithsonian, 60 Garden Street, Cambridge, MA, 02138, USA\label{inst12}
             \and 
             Astrobiology Research Unit, Universit\'e de Li\`ege, 19C All\'ee du 6 Ao\^ut, 4000 Li\`ege, Belgium\label{inst13}
             \and 
             Department of Earth, Atmospheric and Planetary Science, Massachusetts Institute of Technology, 77 Massachusetts Avenue, Cambridge, MA 02139, USA\label{inst14}
             \and 
             Instituto de Astrof\'{i}sica de Canarias (IAC), 38205 La Laguna, Tenerife, Spain\label{inst15}
             \and 
             Phil Evans, El Sauce Observatory, Coquimbo Province, Chile\label{inst16}
             \and
             Department of Physics \& Astronomy, Texas Tech University, Lubbock TX, 79409-1051, USA\label{inst17}
             \and
             Centro di Ateneo di Studi e Attività Spaziali "Giuseppe Colombo" (CISAS)\label{inst18}
             \and
             Università degli Studi di Padova, Via Venezia 15, 35131, Padova, Italy\label{inst19}
             \and
             NASA Goddard Space Flight Center, 8800 Greenbelt Road, Greenbelt, MD 20771, USA\label{inst20}
             \and
             SETI Institute, 189 Bernardo Ave, Suite 200, Mountain View, CA 94043, USA\label{inst21}
             \and 
             Department of Physics and Astronomy, The University of North Carolina at Chapel Hill, Chapel Hill, NC 27599, USA\label{inst22}
             \and
             Instituto de Astrof\'isica de Andaluc\'ia (IAA-CSIC), Glorieta de la Astronom\'ia s/n, 18008 Granada, Spain \label{inst23}
             \and
             American Association of Variable Star Observers, 49 Bay State Road, Cambridge, MA 02138, USA\label{inst24}
             \and 
             Royal Astronomical Society, Burlington House, Piccadilly, London W1J 0BQ, UK\label{inst25}
             \and
             Department of Physics and Kavli Institute for Astrophysics and Space Research, Massachusetts Institute of Technology, Cambridge, MA 02139, USA\label{inst26}
             \and
             Department of Aeronautics and Astronautics, MIT, 77 Massachusetts Avenue, Cambridge, MA 02139, USA\label{inst27}
             \and
             Perth Exoplanet Survey Telescope, Perth, Western Australia, Australia\label{inst28}
\\
             }

   \date{Compiled: \today}

 
  \abstract
 { Recent observations of giant planets have revealed unexpected bulk densities. Hot Jupiters, in particular, appear larger than expected for their masses compared to planetary evolution models, while warm Jupiters seem denser than expected. These differences are often attributed to the influence of the stellar incident flux, but could they also result from different planet formation processes? Is there a trend linking the planetary density to the chemical composition of the host star?
 
 In this work we present the confirmation of three giant planets in orbit around solar analogue stars. TOI-2714 b ($P \simeq 2.5$ d, $R_{\rm p} \simeq 1.22~R_{\rm J}$, $M_{\rm p} = 0.72~M_{\rm J}$) and TOI-2981 b ($P \simeq 3.6$ d, $R_{\rm p} \simeq 1.2~R_{\rm J}$, $M_{\rm p} = 2~M_{\rm J}$) are hot Jupiters on nearly circular orbits, while TOI-4914 b ($P \simeq 10.6$ d, $R_{\rm p} \simeq 1.15~R_{\rm J}$, $M_{\rm p} = 0.72~M_{\rm J}$) is a warm Jupiter with a significant eccentricity ($e = 0.41 \pm 0.02$) that orbits a star more metal-poor ([Fe/H]$~= -0.13$) than most of the stars known to host giant planets. Similarly, TOI-2981 b orbits a metal-poor star ([Fe/H]$~= -0.11$), while TOI-2714 b orbits a metal-rich star ([Fe/H]$~= 0.30$). Our radial velocity (RV) follow-up with the HARPS spectrograph allows us to detect their Keplerian signals at high significance (7, 30, and 23$\sigma$, respectively) and to place a strong constraint on the eccentricity of TOI-4914 b (18$\sigma$). TOI-4914 b, with its large radius ($R_{\rm p} \simeq 1.15~R_{\rm J}$) and low insolation flux ($F_\star < 2 \times 10^8~{\rm erg~s^{-1}~cm^{-2}}$), appears to be more inflated than what is supported by current theoretical models for giant planets. Moreover, it does not conform to the previously noted trend that warm giant planets orbiting metal-poor stars have low eccentricities. This study thus provides insights into the diverse orbital characteristics and formation processes of giant exoplanets, in particular the role of stellar metallicity in the evolution of planetary systems.   }

   \keywords{ Planets and satellites: fundamental parameters -- Stars: fundamental parameters -- Techniques: photometric -- Techniques: radial velocities -- Planets and satellites: gaseous planets 
                               }

    \titlerunning{Confirmation of TOI-2714 b, TOI-2981 b, and TOI-4914 b}

   \maketitle
%

\section{Introduction}

The population of giant planets in close orbit has long been studied since the discovery of the first planet orbiting a main sequence star in 1995, the hot Jupiter (HJ) 51 Pebasi b \citep{1995Natur.378..355M}. Despite the unstopping growth of discoveries and many published papers, there are still some key questions about HJs ($P_{\rm orb} \lesssim 10$ d, $R_{\rm p} > 8~R_\oplus$, \citealt{2015ApJ...799..229W}) and their slightly longer-period counterparts, the warm Jupiters (WJs, $10 < P_{\rm orb} < 100$ d, $T_{\rm eq} < 1000$ K) planets: e.g., \textit{what is the formation channel that contributes predominantly to the WJ population? Do their eccentricities follow the predictions of this formation scenario? Do the observed differences (for example in their radii) with the HJs come from a different process of planet formation? Is there a trend linking the planetary density to the chemical composition of the host star? And do the giant planets follow a mass-metallicity relationship?} These questions require a full characterisation of the planets and their host stars, focusing on WJs, which are presumably not affected by the HJ radius inflation mechanism. WJs – scarce among the confirmed planets\footnote{Only 40 WJs have precise bulk densities (density determination better than 20 per cent) in the NASA Exoplanet Archive.} – are crucial to further understand the planetary bulk composition and look for trends with planetary and stellar properties.

The \textit{Transiting Exoplanet Survey Satellite} (\textit{TESS}, \citealt{2015JATIS...1a4003R}) searches for transiting exoplanets orbiting bright, nearby stars and yields a huge archive of light curves. These light curves are analysed with several transit-search pipelines: the Quick-Look Pipeline (QLP, \citealt{2020RNAAS...4..204H,2020RNAAS...4..206H}) and the QLP-based FAINT search pipeline at MIT \citep{2021RNAAS...5..234K}, and the Science Processing Operations Center \citep[SPOC,][]{2016SPIE.9913E..3EJ} at NASA Ames Research Center. The most prominent targets showing credible transit-like signals are classified as \textit{TESS} Objects of Interest (TOIs, \citealt{2021ApJS..254...39G, 2020RNAAS...4..204H}). High-precision radial velocities (RVs) are needed to derive masses (and eccentricities), which, combined with radius from transit, allow measurement of precise inner bulk densities and explore the differences in planetary structure and evolution, from inflated HJ exoplanets to ``over-dense'' WJs \citep{2021JGRE..12606629F}. While the prediction of hotter interiors and larger radii for HJs at old ages \citep{1996ApJ...459L..35G}, compared to Jupiter itself, has proven true, it is challenging to understand the magnitude of their extreme radii, termed ``radius anomaly'' \citep{2018AJ....155..214T}, and the mechanism(s)/source(s) of internal heat that seem necessary to keep them large for billions of years. On the other hand, for the same incident stellar flux, giant planets denser than Jupiter are thought to be simply more enriched in heavy elements \citep{2007ApJ...659.1661F}. For non-inflated giant planets ($F_\star < 2 \times 10^8~{\rm erg~s^{-1}~cm^{-2}}$, or $T_{\rm eq}$ < 1000 K), \cite{2016ApJ...831...64T} found a relation between planet mass and bulk metallicity, confirming a key prediction of the core-accretion planet formation model \citep{2014A&A...566A.141M} and reproducing the Solar system trend in which more massive giant planets are less metal-rich. However, this relation also shows a surprising amount of metals within planets heavier than Jupiter. A recent study of warm giants \citep{2023FrASS..1079000M} presents the current knowledge of mass-metallicity trends for this class of planets and raises doubts about its extent and existence. \cite{2023FrASS..1079000M} link this ambiguity to theoretical uncertainties associated with the assumed models and the need for accurate atmospheric measurements. Obtaining atmospheric measurements and information on metal enrichment is crucial to breaking the degeneracy in determining the planetary bulk composition. This is essential to validate/disprove formation and evolution theories. \cite{2023FrASS..1079000M} show that atmospheric measurements by JWST, which is particularly promising for warm giant planets \citep{2023FrASS..1079000M}, can significantly reduce this degeneracy. The atmospheric characterisation of exoplanets requires accurate mass determination (e.g., \citealt{2023A&A...669A.150D}). Moreover, an accurate determination of the planetary surface gravity can lead to a better description of the atmospheric properties and a reliable interpretation of the radius anomaly.

In this paper we characterise two HJs and one WJ orbiting solar analogue stars using a combination of \textit{TESS} and ground-based photometry and RVs collected with the High Accuracy Radial velocity Planet Searcher (HARPS, \citealt{2003Msngr.114...20M}) spectrograph at the ESO La Silla 3.6m telescope (Sect. \ref{sec:obs}). In Sect. \ref{sec:sel}, we describe the target selection and statistical validation of the three planets characterised in the present work, each of which was first identified by the \textit{TESS} pipelines. Section \ref{sec:stellar} reports the stellar properties determined by two independent methods. In Sect. \ref{sec:analysis} we show how we identified and confirmed the three planets by outlining the joint modelling of photometry and spectroscopy. Section \ref{sec:discussion} discusses our results and presents the features and peculiarities of each new planet, highlighting the unusual properties of TOI-4914 b -- an eccentric, inflated WJ hosted by a star more metal-poor than many gas giants hosting stars. Concluding remarks are given in Sect. \ref{sec:conclusions}. 

\section{Target selection}
\label{sec:sel}
From the list of TOIs, we selected solar analogues by generating a colour-magnitude diagram in the \textit{Gaia} bands and adopting the spectral classes reported in \citep{2013ApJS..208....9P}, as done in \cite{2022MNRAS.516.4432M}. We chose only solar analogues in order to consider exoplanets with similar incident flux from the star, and to compare host star properties and their effect on exoplanets. For this paper, we selected stars observable from La Silla that host gas giant planets favourable for atmospheric characterisation with JWST. Our selection yielded two stars hosting HJs (TOI-2714, TOI-2981) and one hosting a WJ (TOI-4914), with the peculiarity of being a star that is more metal-poor than most of the WJ hosting stars (further explained in Sect. \ref{sec:discussion}). None of our candidate planets had previously been analysed using high-precision RVs, and their masses were still unknown; therefore, we confirmed the planetary nature of these three candidates through HARPS observations. We emphasise that our selection focused exclusively on candidate planets that were favourable for atmospheric characterisation by JWST -- crucial for assessing their bulk metallicity and validating or disproving formation theories -- with a Transmission Spectroscopy Metric (TSM, predicted value\footnote{Value predicted by the \textit{TESS} atmospheric characterisation working group (ACWG) from a deterministic mass-radius relation \citep{2017ApJ...834...17C}.} in Table \ref{table:tsm}) greater than 90 \citep{2018PASP..130k4401K}. The TSM parameter is proportional to the expected signal-to-noise ratio in transmission spectroscopy for a given planet, and according to \cite{2018PASP..130k4401K}, giant planets are considered suitable for JWST atmospheric observations if their TSM value is above 90.

\section{Observations and data reduction}
\label{sec:obs}

\subsection{TESS photometry}

Each planet described in this paper was identified as a transiting planet candidate in \textit{TESS} photometry. In particular, \textit{TESS} observed TOI-2714 (TIC 332534326) at 30~min cadence in Sector 5 and at 10~min cadence in Sector 31, TOI-2981 (TIC 287145649) at 30~min cadence in Sectors 9 and at 10~min cadence in Sector 36, at 2~min cadence in Sector 63, and TOI-4914 (TIC 49254857) at 10~min cadence in Sector 37 and 200 sec cadence in Sector 64. Table \ref{table:tess-obs} summarises the \textit{TESS} observations for each target.

\begin{table}
\caption{Observations from TESS.}             
\label{table:tess-obs}      
\centering                          
\begin{tabular}{c | c c c}        
\hline\hline                 
Target & Sector(s) & Source & Cadence \rule{0pt}{2.5ex} \rule[-1ex]{0pt}{0pt} \\    
\hline                        
\multirow{2}{*}{TOI-2714} &  5 & SPOC & 1800~s  \rule{0pt}{2.5ex} \rule[-1ex]{0pt}{0pt}\\ 
&  31 & SPOC & 600~s  \rule{0pt}{2.5ex} \rule[-1ex]{0pt}{0pt}\\
   \hline
\multirow{3}{*}{TOI-2981} &  9 & QLP & 1800~s  \rule{0pt}{2.5ex} \rule[-1ex]{0pt}{0pt}\\ 
 &  36 & QLP & 600~s  \rule{0pt}{2.5ex} \rule[-1ex]{0pt}{0pt}\\ 
&  63 & SPOC & 120~s  \rule{0pt}{2.5ex} \rule[-1ex]{0pt}{0pt}\\ 
\hline
\multirow{2}{*}{TOI-4914} &  37 & SPOC & 600~s  \rule{0pt}{2.5ex} \rule[-1ex]{0pt}{0pt}\\ 
&  64 & SPOC & 200~s  \rule{0pt}{2.5ex} \rule[-1ex]{0pt}{0pt}\\ 
\hline                                   
\end{tabular}
\end{table}

The long cadence light curves were extracted using the QLP, while the 2-minute cadence light curves were reduced by the \textit{TESS} SPOC pipeline. Starting from sector 36, some targets observed at long cadence were analysed with transit search pipelines developed by the SPOC \citep{2020RNAAS...4..201C}, and we used these light curves when available. We used Presearch Data Conditioning Simple Aperture Photometry (PDCSAP; \citealt{2012PASP..124.1000S,2012PASP..124..985S,2014PASP..126..100S}) light curves from the SPOC pipeline, that are corrected for systematic effects. When SPOC light curves were not available, we exploit those produced by the QLP.

Both the PDCSAP and QLP light curves were extracted while taking into account contamination from stars within the same aperture. Diagnostic tests were also performed to assess the planetary nature of the three signals by QLP. Each transit signal passed all \textit{TESS} data validation tests and the \textit{TESS} Science Office issued alerts for TOI-2714.01, TOI-2981.01 and TOI-4914.01 on 2021 June 03, 2021 June 04, and 2021 December 21 respectively. The SPOC subsequently detected the transit signatures for TOI-2981 and TOI-4914 in the TESS-SPOC light curves. Additionally, the difference image centroiding test performed by the SPOC Data Validation module \citep{2018PASP..130f4502T} constrained the location of the host stars to within $1.5 \pm 2.5$ arcsec of the transit source for TOI-2981 in sector 36, and to within $0.9 \pm 2.5$ arcsec of the transit source for TOI-4914 in sector 37. Figures \ref{fig:2714lc}, \ref{fig:2981lc}, and \ref{fig:4914lc} show the \textit{TESS} photometric time series.

\subsection{Probabilistic validation}
\begin{figure}
   \centering
   \includegraphics[width=\hsize,trim=20 150 5 120]%
   {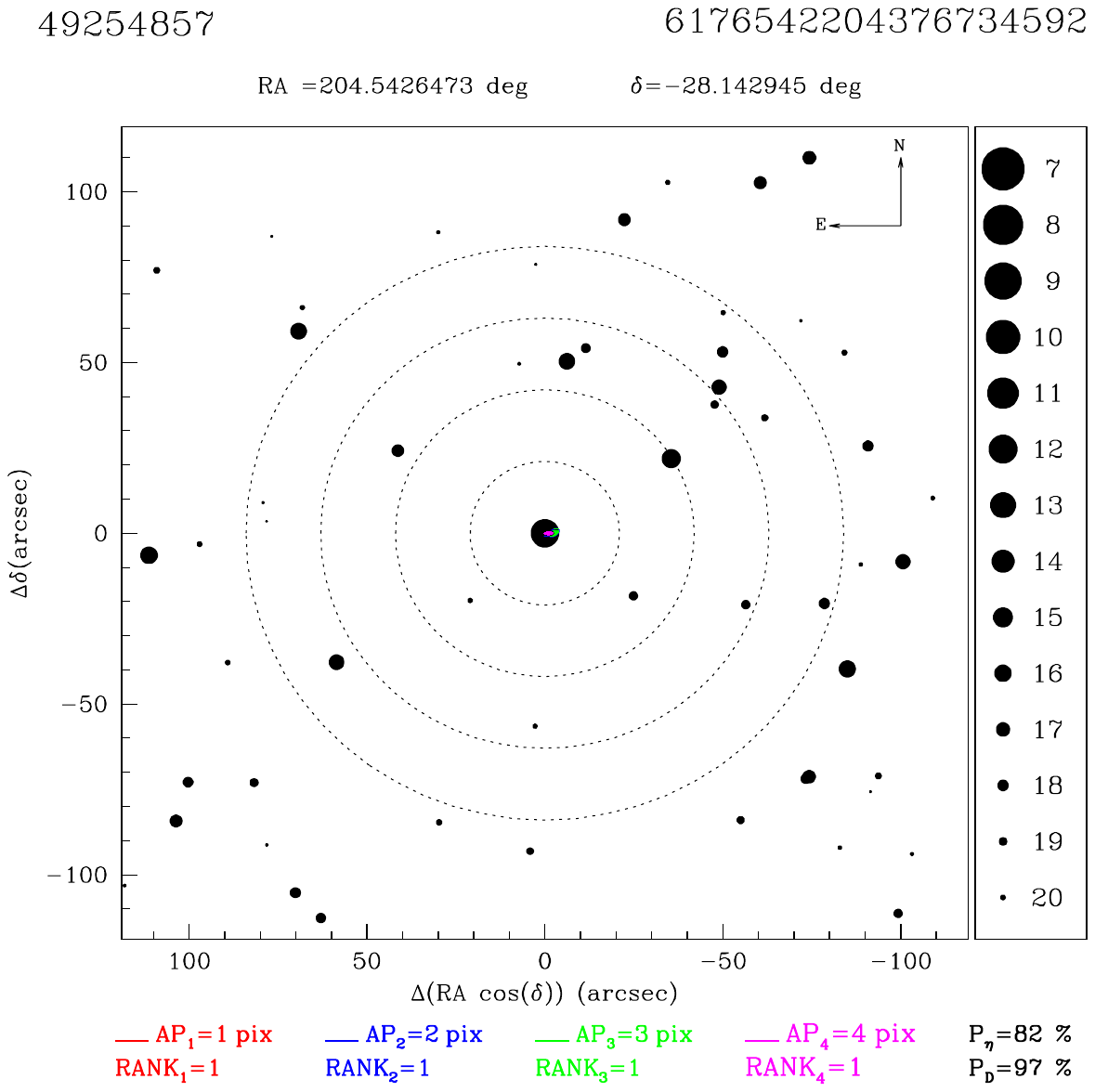}
   \caption{Calculation of the in- and out-of-transit centroid test for TOI-4914 (TIC 49254857). It represents a passed test. The colour-coded ellipses represent the position and dispersion of the centroid metric measurements relative to this target for four concentric apertures, as discussed in \cite{2020MNRAS.498.1726M}. The probability of source association for this target is 97 per cent.}
   \label{fig:4914_centroid}
\end{figure}

To optimise follow-up observations, we performed the probabilistic validation procedure, which aids in distinguishing between a planet and a false positive (FP) from a particular transiting candidate \citep[e.g.,][]{2011ApJ...727...24T,2012ApJ...761....6M,2014MNRAS.441..983D}. First, we exploited the \textit{Gaia} DR3 photometry to check the stellar neighbourhood and exclude each \textit{Gaia} source as a possible blended eclipsing binary. As additional evidence for the origin of the planetary transit sources, we also performed centroid motion tests \citep{2020MNRAS.498.1726M,2020MNRAS.495.4924N}. We show an example in Fig. \ref{fig:4914_centroid}. To ensure that each planet is not a FP, we used the VESPA\footnote{\url{https://github.com/timothydmorton/VESPA}} software \citep{2012ApJ...761....6M} and followed the procedure adopted in \cite{2022MNRAS.516.4432M}, which takes into account the main issues reported in \cite{2023RNAAS...7..107M} and allows us to obtain reliable results while using VESPA. Our selected targets had low False Positive Probability (FPP, the two HJs had $< 1$ in $ 10^6 $ while the WJ had less than 3\%), enough to claim a statistical vetting for both TOI-2714.01 and TOI-2981.01, while leaving the WJ candidate TOI-4914.01 with some risk of being a FP. This statistical validation led to the follow-up observations detailed in Sect. \ref{phot}, \ref{sec:harps} and \ref{sec:modelling}, which ultimately confirmed the planetary nature of each candidate. We will now refer to them as TOI-2714 b, TOI-2981 b, and TOI-4914 b.

\subsection{Photometric follow-up}
\label{phot}

The \textit{TESS} pixel scale is $\sim$ 21 arcsec / pixel and photometric apertures typically extend to about 1 arcmin, which generally results in multiple stars blending into the \textit{TESS} aperture. To determine the true source of the transit signals in the \textit{TESS} data, to improve the transit ephemerides, to monitor transit timing variations, and to check the transit depth after accounting for crowding, we conducted ground-based light-curve follow-up observations (see also Sect. \ref{sec:modelling} and ﬁgures therein) of the ﬁeld around TOI-2714, TOI-2981, and TOI-4914.

As part of the \textit{TESS} Follow-up Observing Program Sub Group 1 (TFOP; \citealt{2019AAS...23314005C}), we conducted the ground-based light-curve follow-up. We used the \textit{TESS} Transit Finder, a customised version of the Tapir software package \citep{2013ascl.soft06007J}, to schedule our transit observations. All image data were extracted using {\tt AstroImageJ} \citep{2017AJ....153...77C} with the exception of PEST which used a custom pipeline based on {\tt C-Munipack}\footnote{\url{http://c-munipack.sourceforge.net}}. We used circular photometric apertures centred on TOI-2714, TOI-2981, and TOI-4914. Figures \ref{fig:2714lc_tfop}, \ref{fig:2981lc_tfop}, and \ref{fig:4914lc_tfop} show the ground-based photometric time series.

\subsubsection{LCOGT}

For both TOI-2714 b and TOI-2981 b, we observed two transit windows, in Sloan $i'$ and $g'$ bands using the Las Cumbres Observatory Global Telescope (LCOGT; \citealt{2013PASP..125.1031B}) 1.0 m network nodes at Teide Observatory (TEID) on the island of Tenerife, and McDonald Observatory (MCD) near Fort Davis, Texas, United States. We observed full transits of TOI-2714 b on UTC 2021 October 11 and October 16 from MCD in two filters. We observed full transits of TOI-2981 b on UTC 2023 April 20 from two TEID telescopes. The 1 m telescopes are equipped with 4096 $\times$ 4096 Sinistro cameras with an image scale of 0.389 arcsec / pixel, resulting in a 26 arcmin $\times$ 26 arcmin ﬁeld of view. The images were calibrated using the standard LCOGT BANZAI pipeline \citep{2018SPIE10707E..0KM}. We used circular photometric apertures with radii in the range of 7.4 to 8.0 arcsec. 

\subsubsection{TRAPPIST}

The TRAnsiting Planets and PlanetesImals Small Telescope-South (TRAPPIST-S, \citealt{2011EPJWC..1106002G,2011Msngr.145....2J}) is a 0.6 m robotic telescope based at ESO's La Silla Observatory. TRAPPIST-S has a 2K $\times$ 2K detector with a 0.6 arcsec pixel scale and a field of view of 22 $\times$ 22 arcminutes. We observed two transit windows of TOI-2981 b on UT 2022 April 3 and  UT 2023 March 11 in the Sloan-$z'$ and B filters, respectively. While a full transit of TOI-4914 b was observed on UT 2022 March 6 in the $I+z'$ filter. The data reduction and photometric extraction were performed using the {\tt AstroImageJ} software.

\subsubsection{El Sauce}

We observed, using a Cousins R filter ($R_{\rm c}$), one full transit of TOI-2981 b on UT 2022 February 27 using the Evans CDK 0.51 m telescope at El Sauce Observatory, Chile. The telescope was equipped with an SBIG STT 1603 CCD camera with 1536 $\times$ 1024 pixels giving an image scale of 1.08 arcsec pixel$^{-1}$ when binned 2x2 in camera. On UT 2023 June 07 we used the Evans 0.51 m telescope and $R_{\rm c}$ filter, but now equipped with a Moravian C3-26000 CMOS camera, to observe one full transit of TOI-4914 b. The CMOS camera has 6252 $\times$ 4176 pixels, binned 2x2 in camera, for a pixel scale of 0.449 arcsec pixel$^{-1}$. The exposure times were 120 seconds for TOI-2981 b and 90 seconds for TOI-4914 b. 

\subsubsection{PEST}

The Perth Exoplanet Survey Telescope (PEST) is located near Perth, Australia. The 0.3 m telescope is equipped with a $5544\times3694$ QHY183M camera. Images are binned 2x2 in software giving an image scale of 0.7 arcsec pixel$^{-1}$ resulting in a 32 arcmin $\times$ 21 arcmin field of view. A custom pipeline based on {\tt C-Munipack} was used to calibrate the images and extract the differential photometry. PEST observed a full transit of TOI-4914 b on 2023 May 26. The observation was conducted with a Sloan $g'$ filter and 120 s integration times.

\subsubsection{Brierfield}

The Brierfield Observatory, located in Bowral, New South Wales, Australia, houses the 14 inches (0.36 m) Planewave Corrected Dall-Kirkham Astrograph telescope mounted with the instrument Moravian G4-16000 KAF-16803. Brierfield observed TOI-4914 b on 2023 May 26 UT. The observation was conducted with a Bessel $B$ filter and 240 s integration times.

\subsection{HARPS spectroscopic follow-up}
\label{sec:harps}

We collected observations of TOI-2714, TOI-2981, and TOI-4914 with HARPS at ESO's 3.6 m telescope \citep{2003Msngr.114...20M} between April and September 2023 (proposal 111.254A, PI: G. Mantovan), obtaining between 8 and 17 spectra for each target, with exposure times of 1800 s. These spectra cover the wavelength range 378-691 nm with a resolving power of R $\sim$ 115~000. We report details of the observations and typical S/N in Table \ref{table:obsHARPS}. 
\begin{table}[t]
\caption{Observations from HARPS summarised.}             
\label{table:obsHARPS}      
\centering                          
\begin{tabular}{c | c c c}        
\hline\hline                 
 HARPS & TOI-2714 &  TOI-2981 & TOI-4914\rule{0pt}{2.5ex} \rule[-1ex]{0pt}{0pt} \\
\hline                        

     N$^{\circ}$ spectra & 8 & 15 & 17 \rule{0pt}{2.5ex} \rule[-1ex]{0pt}{0pt} \\
     Time-span (days)   & 32 & 59 & 88 \rule{0pt}{2.5ex} \rule[-1ex]{0pt}{0pt} \\
     $\langle \rm RV_{\rm err} \rangle$ (m s$^{-1}$) & 20.7 & 22.6 & 7.3     \rule{0pt}{2.5ex} \rule[-1ex]{0pt}{0pt} \\
     $\langle \rm S/N \rangle_{5460\AA}$  & 7.7 & 9.4 & 19     \rule{0pt}{2.5ex} \rule[-1.2ex]{0pt}{0pt} \\ 
\hline                                   
\end{tabular}
\end{table}
For our observations we used the standard high accuracy mode with a 1 arcsec science fibre on the target and fibre B on the sky (see \citealt{2003Msngr.114...20M}). As a precaution, we avoided observing when the fraction of lunar illumination was higher than 0.9 and when the absolute difference between the systematic RV of the target and the Barycentric Earth RV correction was lower than 15 km s$^{-1}$ \citep[e.g.,][]{2017AJ....153..224M}.

\begin{figure}[!t]
   \centering
   \includegraphics[width=\hsize]%
    {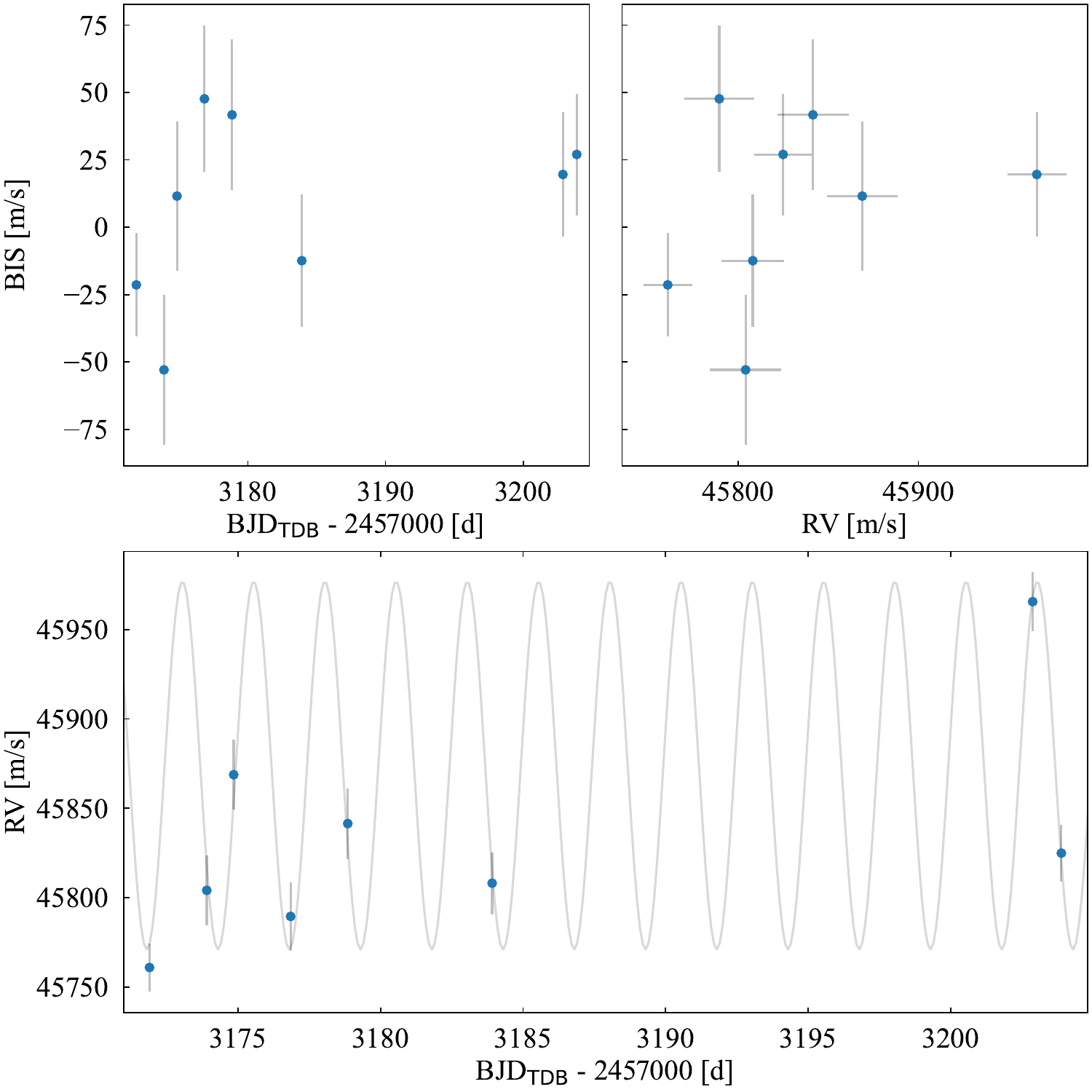}
   \caption{HARPS-N spectroscopic time series of TOI-2714. \textit{Bottom:} RV time series with Keplerian fit superimposed. \textit{Top Left:} BIS time series. \textit{Top Right:} RV versus BIS time series. There is no clear correlation between BIS and RV time series. }
   \label{fig:2714_gls}
\end{figure}

\begin{figure}
   \centering
   \includegraphics[width=\hsize]%
   {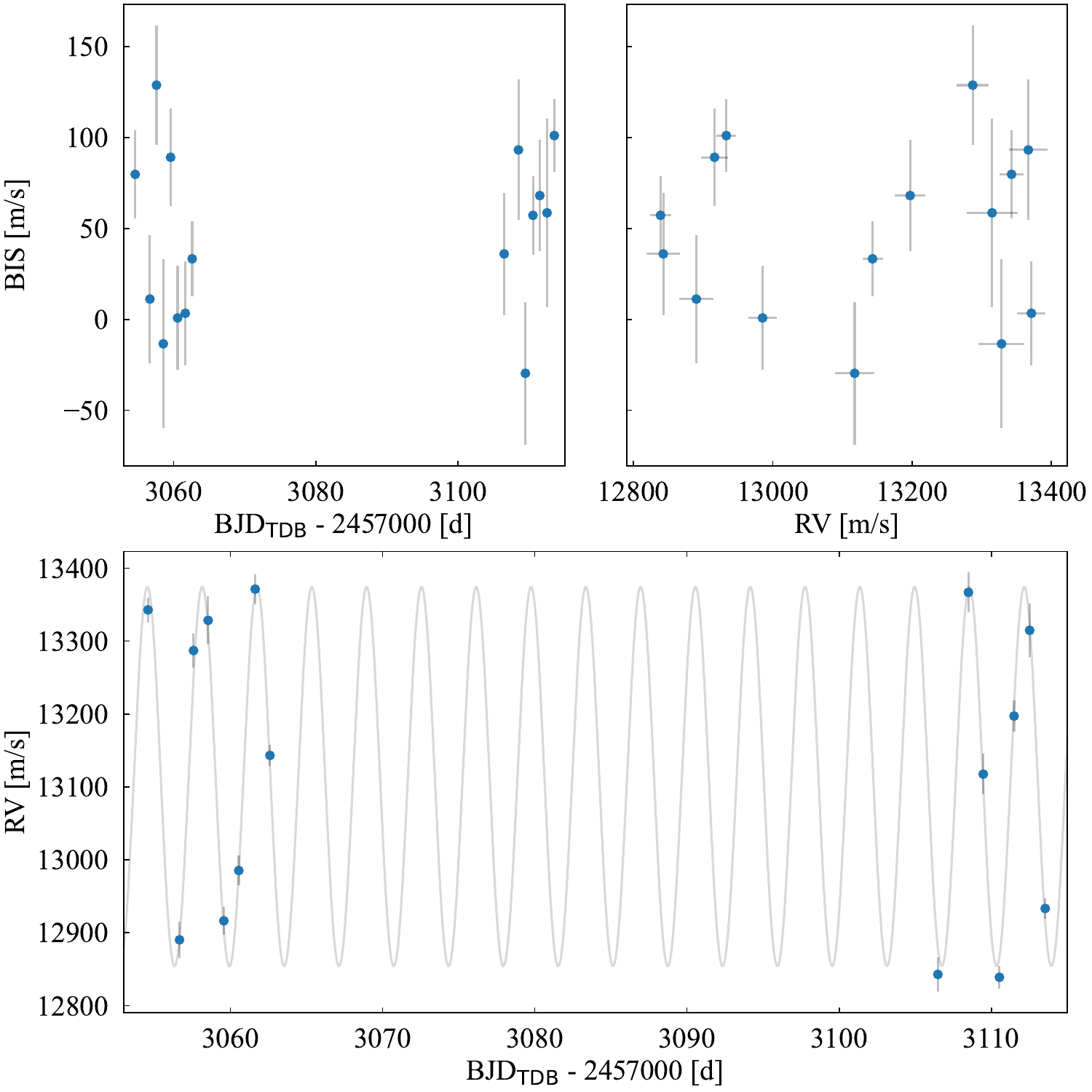}
   \caption{As in Fig. \ref{fig:2714_gls}, but for TOI-2981.}
   \label{fig:2981_gls}
\end{figure}

\begin{figure}
   \centering
   \includegraphics[width=\hsize]%
   {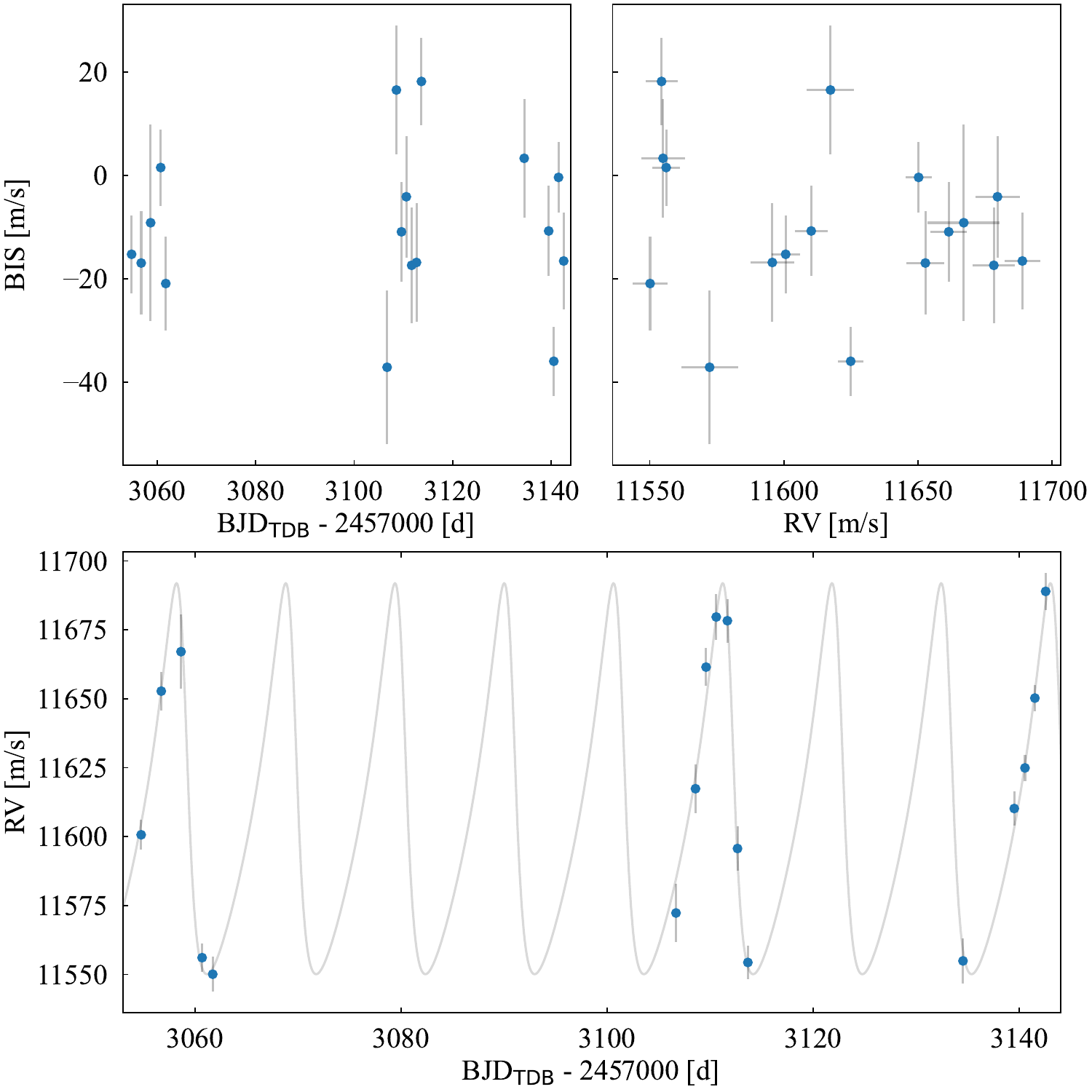}
   \caption{As in Fig. \ref{fig:2714_gls}, but for TOI-4914.}
   \label{fig:4914_gls}
\end{figure}

The data were reduced using the HARPS pipeline and radial velocities (RVs) computed using the cross-correlation function (CCF) method (\citealt{2002Msngr.110....9P}, and references therein). In this method, the scientific spectra are cross-correlated with a binary mask that describes the typical features of a star of a chosen spectral type. We used a G2 mask for each target. The pipeline also produces values for the CCF bisector span (BIS), the full width at half-maximum (FWHM) depth of the CCF, and its equivalent width ($W_{\rm CCF}$, see \citealt{2019MNRAS.487.1082C} for more details). We extracted the $H_\alpha$ and $\log R'_{HK}$ indices using the ACTIN 2 code \citep{2018JOSS....3..667G, 2021A&A...646A..77G}. Figures \ref{fig:2714_gls}, \ref{fig:2981_gls}, and \ref{fig:4914_gls} show the spectroscopic time series. There is no clear correlation between BIS and RV time series.

\subsection{High-angular-resolution data}
\label{sec:imaging}

\begin{figure*}
   \centering
   \includegraphics[width=\hsize]%
   {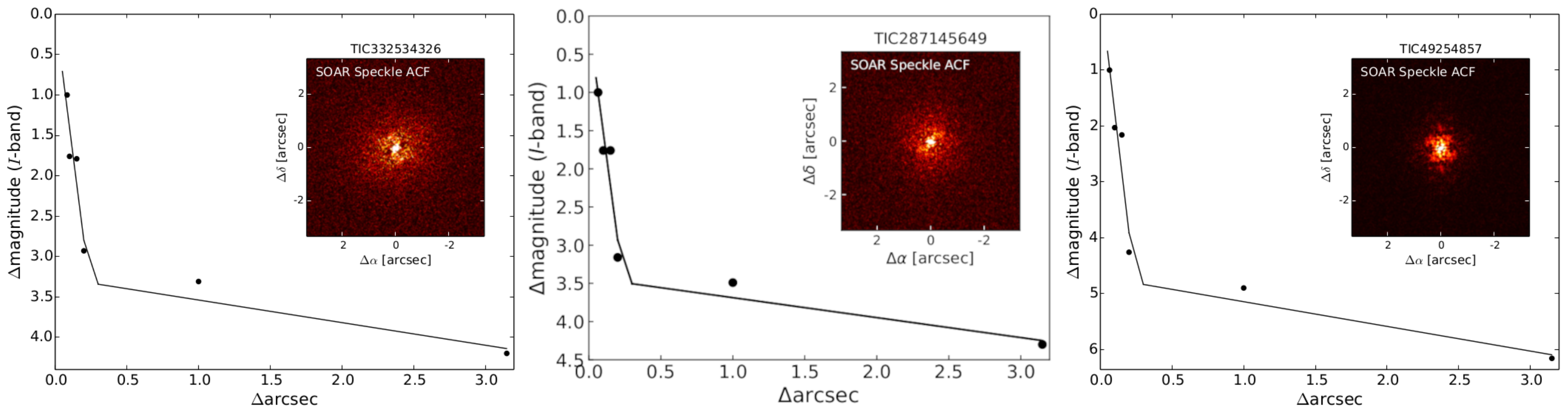}
   \caption{From \textit{left} to \textit{right}: High resolution imaging data for TOI-2714, TOI-2981 and TOI-4914. Each image shows a SOAR HRCam speckle sensitivity curve (solid line) and an auto-correlation function (ACF, inset). }
   \label{fig:imaging}
\end{figure*}

Within the framework of the follow-up observations organised by the TFOP high-resolution imaging Sub-Group 3 (SG3), we acquired high angular resolution imaging of all the targets mentioned in this paper. Specifically, the observations were conducted with the High-Resolution Camera (HRCam; \citealt{2008PASP..120..170T}) speckle imaging instrument on the 4.1 m Southern Astrophysical Research (SOAR) telescope. The observing strategy and data reduction procedures for the SOAR observations are detailed in \cite{2018PASP..130c5002T,2020AJ....159...19Z}, and \cite{2021AJ....162..192Z}. These imaging observations are summarised in Table \ref{table:imaging} and Fig. \ref{fig:imaging}. No companions are detected in the high angular resolution imaging down to the detection limits.

\begin{table}
\caption{High-resolution imaging observations using HRCam on the SOAR telescope.}             
\label{table:imaging}      
\centering                          
\begin{tabular}{c | c c c}        
\hline\hline                 
Target & Date & Filter & Contrast \rule{0pt}{2.5ex} \rule[-1ex]{0pt}{0pt} \\    
\hline                        
TOI-2714 &  2021-10-01 & $I_c$ & $\Delta 4.2~{\rm mag}$ at 1 arcsec  \rule{0pt}{2.5ex} \rule[-1ex]{0pt}{0pt}\\ 
TOI-2981 &  2022-06-10 & $I_c$ & $\Delta 4.3~{\rm mag}$ at 1 arcsec  \rule{0pt}{2.5ex} \rule[-1ex]{0pt}{0pt}\\ 
TOI-4914 &  2022-03-20 & $I_c$ & $\Delta 6.2~{\rm mag}$ at 1 arcsec  \rule{0pt}{2.5ex} \rule[-1ex]{0pt}{0pt}\\ 
\hline                                   
\end{tabular}
\end{table}

\section{Stellar parameters}
\label{sec:stellar}

Here we outline the methods we used to determine stellar parameters by combining photometric, spectroscopic, astrometric, and additional ancillary data. In particular, we produced a co-added spectrum using all the spectra, giving an average signal-to-noise ratio (S/N) of $\sim$ 30-40 per extracted pixel at around 6000 Å for TOI-2714, 45-55 for TOI-2981, and 85-100 for TOI-4914. 

We analysed the combined HARPS-N spectrum using the same methodology as in \cite{2022A&A...664A.161B} and we derived the effective temperature $T_{\rm eff}$, the surface gravity $\log{g}$, the microturbolence velocity $\xi$, the iron abundance [Fe/H] and the projected rotational velocity $v \sin{i}_{\star}$ of TOI-2714, TOI-2981, and TOI-4914.

\subsection{Atmospheric parameters and metallicity}
\label{sec:atm_param}

Specifically, for $T_{\rm eff}$, $\log{g}$, $\xi$, and [Fe/H] we applied a method based on the equivalent widths (EWs) of \ion{Fe}{i} and \ion{Fe}{ii} lines from \cite{2015A&A...583A.135B} and \cite{2022A&A...664A.161B}. The final parameters were derived using the version 2019 of MOOG code \citep{1973ApJ...184..839S}, adopting the ATLAS9 grids of model atmospheres with solar-scaled chemical composition and new opacities \citep{2003IAUS..210P.A20C}. The parameters $T_{\rm eff}$ and $\xi$ were derived by imposing that the abundance of the \ion{Fe}{i} lines does not depend on the line excitation potentials and the reduced equivalent widths (i.e. EW/$\lambda$), respectively, while $\log{g}$ was obtained by imposing the ionisation equilibrium condition between the abundances of the \ion{Fe}{i} and \ion{Fe}{ii} lines.

As an example, for TOI-2714 our spectroscopic analysis yields final atmospheric parameters of $T_{\rm eff}$ = 5665 $\pm$ 115 K, $\log{g}$ = 4.25 $\pm$ 0.27 dex, and $\xi$ = 0.58 $\pm$ 0.29 km s$^{-1}$. The derived iron abundance (obtained with respect to the solar Fe abundance as in \citealt{2022A&A...664A.161B}) is [Fe/H]= 0.30 $\pm$ 0.11, where the error includes the scatter due to the EW measurements and the uncertainties in the stellar parameters. The same stellar parameters for TOI-2981 and TOI-4914 are provided in Table \ref{tab:star_param}. It is worth noting that we quadratically added a systematic error of 60 K \citep{2011A&A...526A..99S} to the effective temperature precision error of TOI-4914, which is intrinsic to our spectroscopic method \citep[e.g.,][]{2018MNRAS.481.1839M,2022ApJ...927...31T}. We only did this for TOI-4914 because its intrinsic error was much smaller than the value commonly accepted as the temperature uncertainty (i.e., the systematic error). The $v \sin{i}_{\star}$ of the three stars are discussed in Sect. \ref{sec:vsini}.

\subsection{Spectral energy distribution}
\label{sec:sed}

\begin{figure}[h!]
\centering
\includegraphics[width=\linewidth,trim=80 70 50 50,clip]{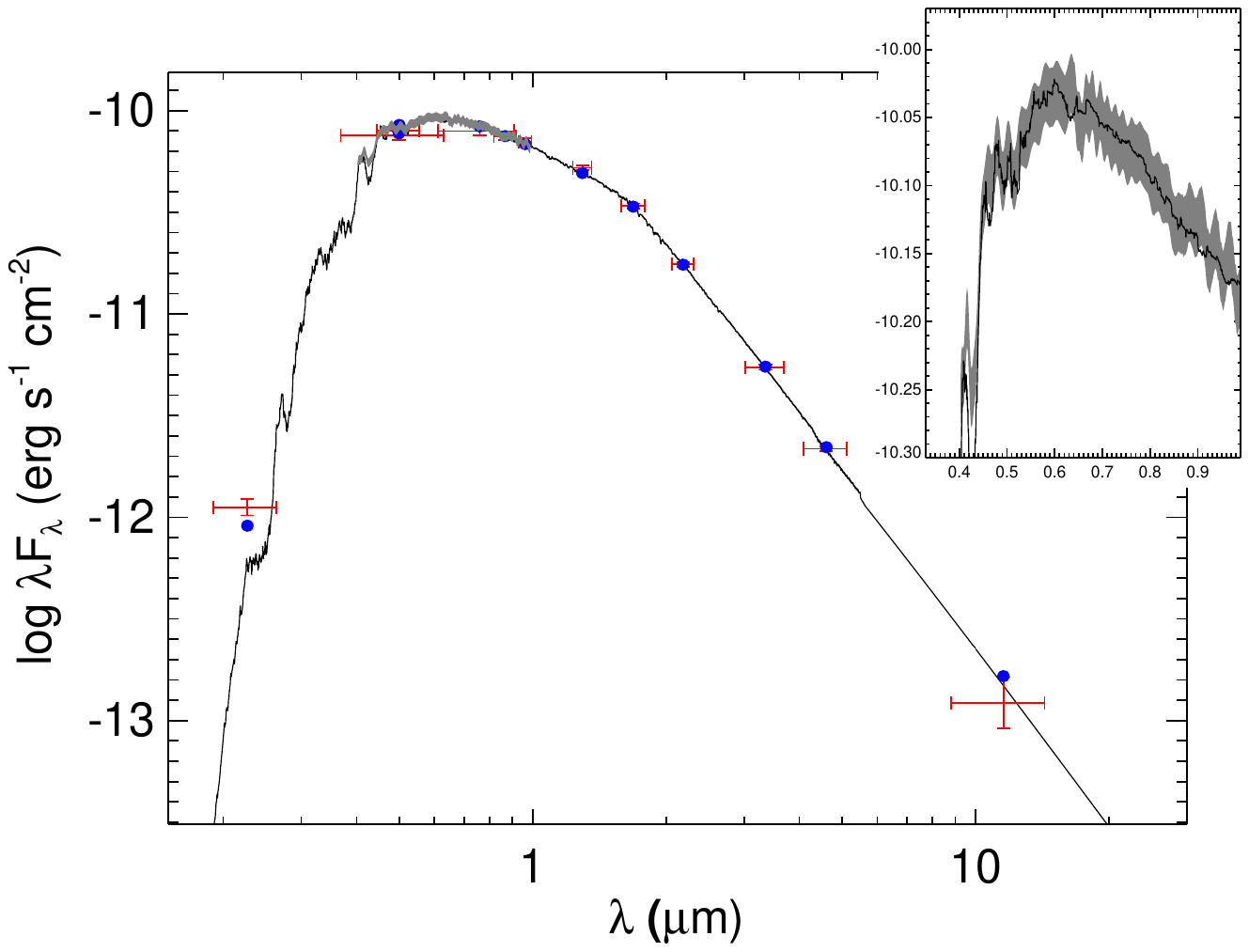}
\includegraphics[width=\linewidth,trim=80 70 50 50,clip]{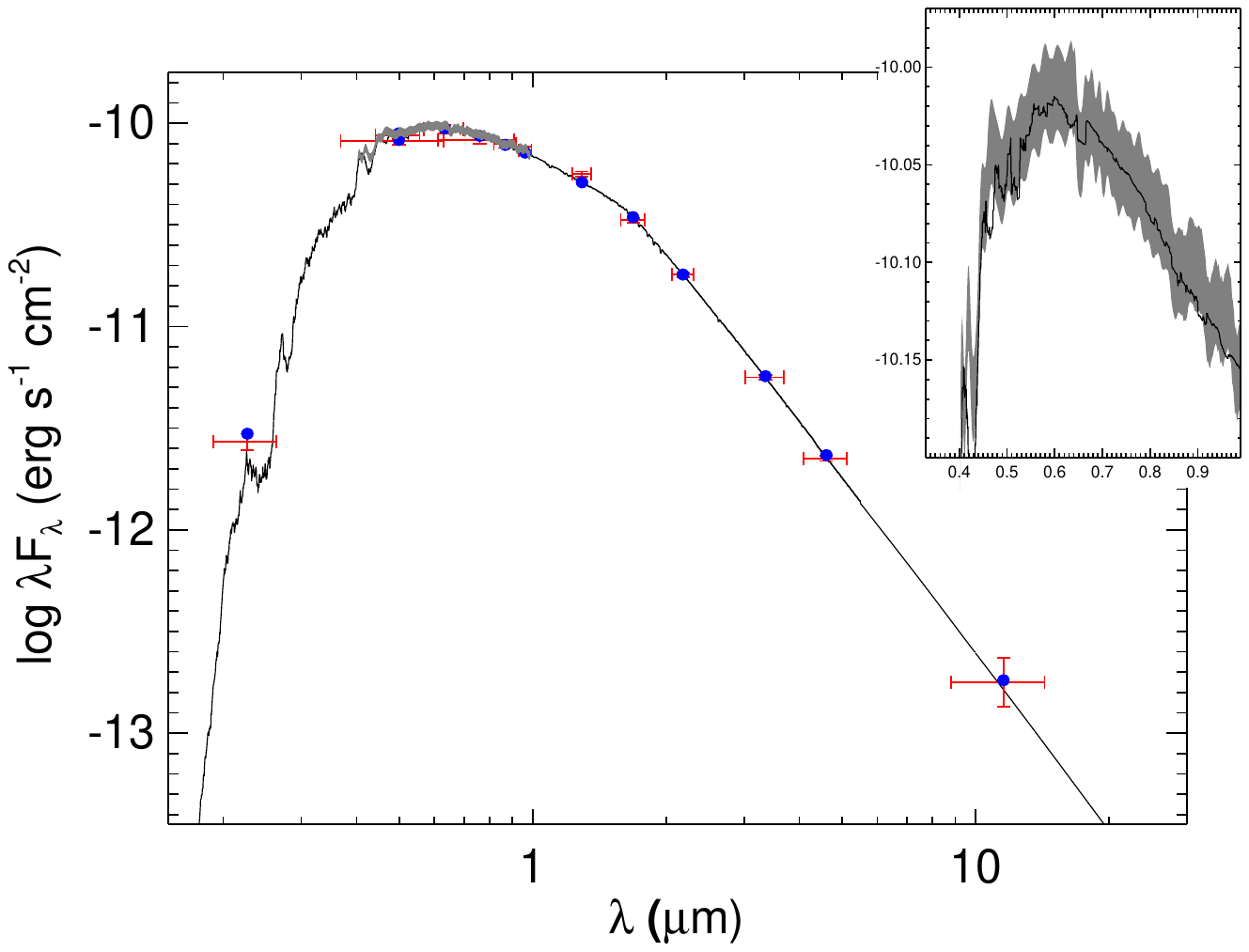}
\includegraphics[width=\linewidth,trim=80 70 50 50,clip]{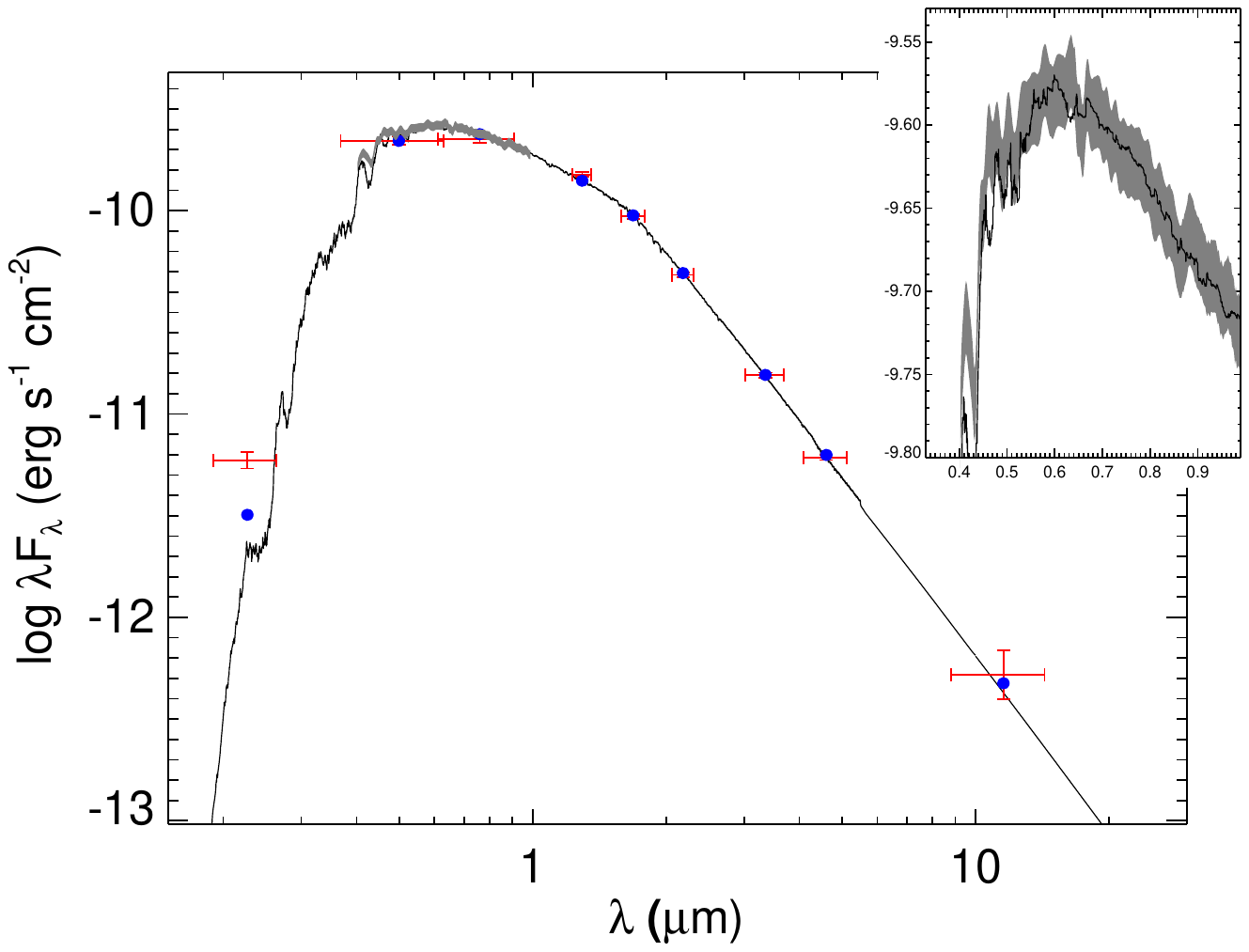}
\caption{Spectral energy distributions for TOI-2714 (top), TOI-2981 (middle), and TOI-4914 (bottom). Red symbols represent the observed photometric measurements, where the horizontal bars represent the effective width of the passband. Blue symbols are the model fluxes from the best-fit PHOENIX atmosphere model (black). The absolute flux-calibrated {\it Gaia\/} spectrum is shown as a grey swathe in the inset figure. \label{fig:sed}}
\end{figure}

We performed an analysis of the broadband spectral energy distribution (SED) of each star together with the {\it Gaia\/} DR3 parallax \citep[without adjustment; see][]{StassunTorres:2021}, in order to determine an empirical measurement of the stellar radius \citep{Stassun:2016,Stassun:2017,Stassun:2018}. We pulled the $JHK_S$ magnitudes from {\it 2MASS}, the W1--W4 magnitudes from {\it WISE}, the $G_{\rm BP}\,G_{\rm RP}$ magnitudes from {\it Gaia}, and when available the FUV and/or NUV fluxes from {\it GALEX} \citep{2005ApJ...619L...1M}. We also utilised the absolute flux-calibrated {\it Gaia\/} spectrum. Together, the available photometry spans the full stellar SED over the wavelength range 0.4--10~$\mu$m in all cases, and as much as 0.2--20~$\mu$m in some cases (see Fig.~\ref{fig:sed}). 

We performed a fit using PHOENIX stellar atmosphere models \citep{2013A&A...553A...6H}, adopting from the spectroscopic analysis the effective temperature ($T_{\rm eff}$), metallicity ([Fe/H]), and surface gravity ($\log g$). We fitted for the extinction $A_V$, limited to the maximum line-of-sight value from the Galactic dust maps of \citet{Schlegel:1998}. The resulting fits are shown in Fig.~\ref{fig:sed}.  Integrating the (unreddened) model SED gives the bolometric flux at Earth, $F_{\rm bol}$. Taking the $F_{\rm bol}$ together with the {\it Gaia\/} parallax directly gives the bolometric luminosity, $L_{\rm bol}$. The Stefan-Boltzmann relation then gives the stellar radius, $R_\star$. In addition, we estimated the stellar mass, $M_\star$, using the empirical relations of \citet{Torres:2010}. 
All resulting values are summarised in Table~\ref{tab:star_param}.

\subsection{Lithium abundance}
\label{sec:li}

From the same co-added spectra we also derived the abundance of the lithium line at $\sim$6707.8 Å ($\log n({\rm Li})_{\rm NLTE}$), after applying the non-LTE calculations of \cite{2009A&A...503..541L} and considering the stellar parameters derived in Sect. \ref{sec:atm_param}.

For TOI-2714, we obtained a lithium abundance of $\log n({\rm Li})_{\rm NLTE} =$ 2.13 $\pm$ 0.13 dex. With this value, together with the effective temperature of the target, the position of TOI-2714 in the $\log n({\rm Li}) - T_{\rm eff} $ diagram seems to be between that of the NGC752 cluster ($\sim$ 2 Gyr) and that of the M67 cluster ($\sim$ 4.5 Gyr). 

As for TOI-2981, we found $\log n({\rm Li})_{\rm NLTE} =$ 2.59 $\pm$ 0.06 dex. The position of TOI-2981 in the $\log n({\rm Li}) - T_{\rm eff} $ diagram appears to be intermediate between the Hyades (625 Myr) and NGC 752 ($\sim 2$ Gyr) open clusters, compatible with both if observational errors and intrinsic scatter of individual members are taken into account \citep{Boesgaard2020,Boesgaard2022}.  

TOI-4914 has a lithium abundance of $\log n({\rm Li})_{\rm NLTE} =$ 2.14 $\pm$ 0.06 dex, while its position in the $\log n({\rm Li}) - T_{\rm eff} $ diagram appears to lie between the NGC 752 cluster ($\sim$ 2 Gyr) and M67 ($\sim$ 4.5 Gyr).

\subsection{Chromospheric activity}
\label{sec:chrom}

We calculated the mean S-index values calibrated in the Mt. Wilson scale \citep{1995ApJ...438..269B} and found 0.15, 0.14, and 0.13 for TOI-2714, TOI-2981, and TOI-4914, respectively. These three values correspond to $\log R'_{HK}$ values of $-4.93\,\pm\,0.26$, $-5.09\,\pm\,0.37$, and $-5.32\,\pm\,0.37$ (arithmetic mean and standard deviation). According to the relations reported in \cite{2018A&A...616A.108B}, the $\log R'_{HK}$ values we obtained suggest that these stars are inactive.

\subsection{Rotation period}
\label{sec:rot_per}

\begin{figure}
   \centering
   \includegraphics[width=\hsize, bb=7 510 567 693]{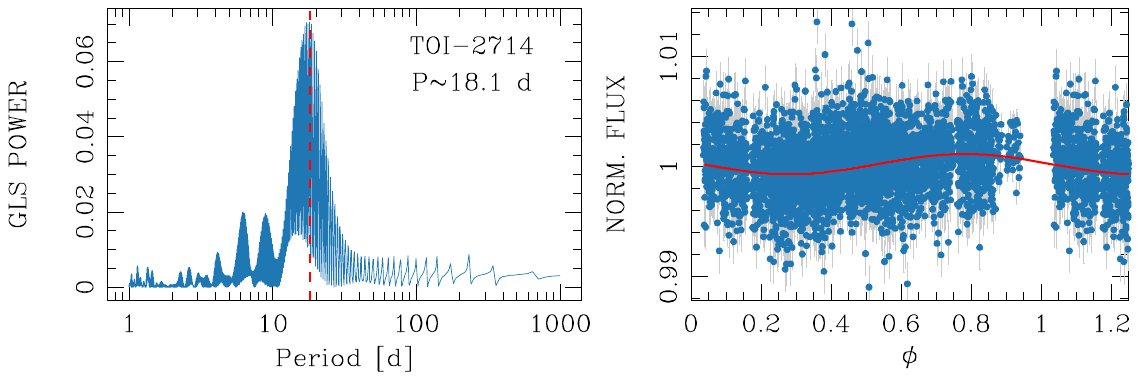} \\   
   \includegraphics[width=\hsize, bb=7 510 567 693]{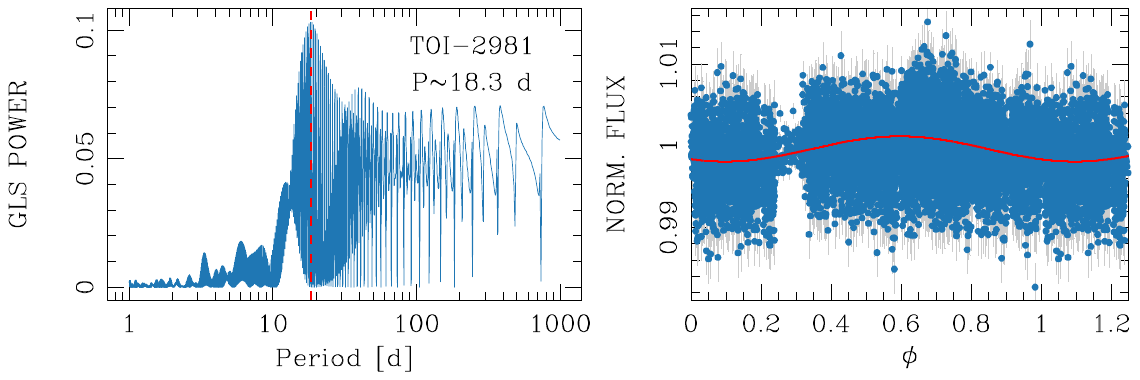} \\   
   \includegraphics[width=\hsize, bb=7 510 567 693]{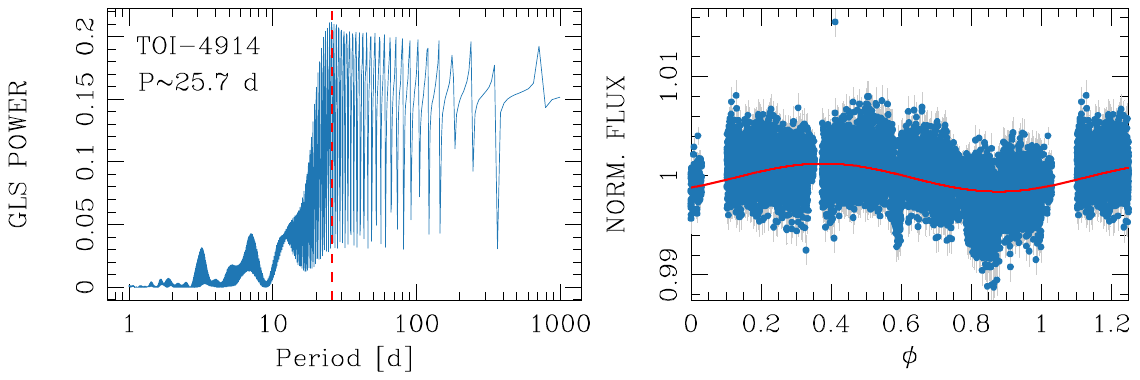}   
   \caption{Left panels show the GLS periodograms extracted from the photometric time series of TOI-2714 (top panel), TOI-2981 (middle panel), and TOI-4914 (bottom panel). Right panels show the phased light curves adopting the period corresponding to the peak of the periodogram. }
   \label{fig:lc_gls}
\end{figure}

We attempted to estimate the stellar rotation periods of the three targets, a key parameter for age estimation. In particular, we extracted the Generalised Lomb-Scargle (GLS) periodograms \citep{2009A&A...496..577Z} of the \textit{TESS} light curves obtained with the PATHOS pipeline \citep{2020MNRAS.495.4924N}.
We sampled periods between 1~d and 1000~d. The results are shown in Fig.~\ref{fig:lc_gls}. For TOI-2714 we found a peak at P$\sim 18.1$~d, for TOI-2981, after removing a systematic trend due to the X/Y position of the star on the CCD in sectors 9 and 36, we found a peak in the periodogram at P$\sim 18.3$~d, and for TOI-4914 we found a peak at P$\sim 25.7$~d. However, the presence of multiple peaks in the periodograms accompanied by low power, and a visual inspection of the light curves, indicate that these periods may represent a lower limit of the true rotational periods. In fact, the search for periods longer than $\sim 14$~days is complicated by the difficulty of stitching light curves of different TESS orbits. Furthermore, residual systematic errors in the light curves can generate peaks in the periodogram of low amplitude variable light curves. We can assert that no variability is detectable for periods shorter than $\sim 14$~days, which increases the probability that the targets are not young.

\subsection{Projected rotational velocity}
\label{sec:vsini}

The projected rotational velocities of the stars studied in this work were derived using the calibration of the FWHM of the CCF into $v\sin{i_{\star}}$ developed by \citet{Rainer2023}. All stars are moderately slow rotators, in general agreement with the tentative rotation periods derived in Sect. \ref{sec:rot_per}.

\subsection{Kinematics and multiplicity}
\label{sec:kinematics}

The kinematic space velocities $U$, $V$, $W$ were computed following the formalism of \citet{Johnson1987}.
The three targets are found to be part of the thin disk but outside the typical kinematic space of young stars \citep[e.g.,][]{Montes2001}.

The search for additional companions besides the transiting ones does not yield any convincing candidates, considering the imaging observations described in Sect. \ref{sec:imaging} and the sources in \textit{Gaia} DR3. Furthermore, there is no evidence for close stellar companions around TOI-2714 and TOI-4914 from the intrinsic astrometric scatter in \textit{Gaia} DR3 and no significant differences in the absolute RV of \textit{Gaia} DR3 and our HARPS spectra, with a time baseline of about 7.5 years. For TOI-4914, the additional measurement by \citet{GALAH_DR3} is also compatible within errors. 

For TOI-2981 the situation is more ambiguous, with a RV difference of 3.8 $\pm$ 2.7 km s$^{-1}$ and a Renormalised Unit Weight Error (RUWE) of 1.22, which is slightly below the threshold (RUWE = 1.4) for high confidence detection of astrometric companions \citep{Lindegren2021}. However, on the time base of our observations (59 days), there is no indication of long-term RV trends (see Fig. \ref{fig:2981} and further analysis in Sect. \ref{sec:additionalplanets}). Finally, we note that for our targets there is still a significant incompleteness in the detectability of stellar companions due to their large distances from the Sun.

\subsection{Stellar age}
\label{sec:age}

We performed the isochrone fitting using the web interface PARAM\footnote{\url{http://stev.oapd.inaf.it/cgi-bin/param_1.3}} \citep{param}, which interpolates the stellar models from \citet{Bressan2012} and infers the most probable solution in a Bayesian framework, taking into account the observational errors and the lifetimes of the various evolutionary phases. We used as inputs the spectroscopic effective temperature and metallicity, the $V$ magnitude corrected for interstellar absorption, and the \textit{Gaia} DR3 trigonometric parallax. 

Both the radius measurement and the isochrone fit seem to indicate that TOI-2714 is evolved outside the main sequence. The isochrone age is 5.5$\pm$3.0 Gyr. The gyrochronology is marginally discrepant, suggesting an age of about 2 Gyr. However, given the uncertainties in determining the rotation period, we do not consider this to be significant. The lithium content is slightly above, but fully compatible within the error bars with the locus of the members of the open cluster M67. We then adopt the isochrone age as the other methods are either inconclusive or compatible with it. 

TOI-4914 also appears to be old and slightly evolved, with an estimated age of 5.3$\pm$3.4 Gyr, according to the isochrone fitting. The tentative rotation period, lithium content and kinematics are consistent with the found isochrone age, but may indicate that the star is on the younger side of the inferred age range. 

TOI-2981 has an isochrone age of 4.5$\pm$2.9 Gyr. However, an age on the younger side of this range, and possibly even younger, is supported by the moderately large Li, as discussed in Sect. \ref{sec:li}. The tentative rotation period would suggest an age of the order of 3 Gyr, while the stellar activity is low and the kinematics are quite far from the locus of young stars, ruling out a younger age ($\leq$ 1 Gyr). Considering the possibility of some changes in angular momentum and stellar mixing due to a close and moderately massive planet, we adopt the isochrone age. 

The stellar parameters outlined in this and the previous subsections serve as the reference for this study; they are listed in Table \ref{tab:star_param}.

\begin{table*}[!t]
   \caption[]{Stellar properties of TOI-2714, TOI-2981, and TOI-4914}
     \label{tab:star_param}
     \small
     \centering
       \begin{tabular}{lcccc}
         \hline
         \noalign{\smallskip}
         Parameter   &  \object{TOI-2714} &  \object{TOI-2981} &  \object{TOI-4914} & Reference  \\
         \noalign{\smallskip}
         \hline
         \noalign{\smallskip}
$\alpha$ (J2000)          &   04:05:36.288    & 10:53:56.983 &  13:38:10.257 & {\it Gaia} DR3    \\
$\delta$ (J2000)          &   $-$14:55:36.361  & $-$25:31:28.834 & $-$28:08:34.278 & {\it Gaia} DR3  \\
$\mu_{\alpha}$ (mas yr$^{-1}$)  &    11.27$\pm$0.01 & $-$7.23$\pm$0.02& $-$13.40$\pm$0.02 & {\it Gaia} DR3  \\
$\mu_{\delta}$ (mas yr$^{-1}$)  &    4.92$\pm$0.01 & 17.97$\pm$0.02 & $-$15.23$\pm$0.01 & {\it Gaia} DR3  \\
RV     (km s$^{-1}$)            &    43.99$\pm$2.24  & 9.31$\pm$2.70 & 11.99$\pm$0.70 & {\it Gaia} DR3   \\
$\pi$  (mas)             &    1.62$\pm$0.01 & 1.87$\pm$0.02 & 3.38$\pm$0.02 & {\it Gaia} DR3  \\
$U$   (km s$^{-1}$)             &       $-$50.0$\pm$1.5   & $-$38.8$\pm$0.4 &  $-$2.3$\pm$0.4   & This paper (Sect. \ref{sec:kinematics}) \\
$V$   (km s$^{-1}$)             &      $-$26.8$\pm$0.8    & 5.4$\pm$2.3 &  $-$29.9$\pm$0.4  & This paper (Sect. \ref{sec:kinematics}) \\
$W$   (km s$^{-1}$)             &     $-$3.5$\pm$1.6    & 31.0$\pm$1.4 & $-$7.3$\pm$0.4  & This paper (Sect. \ref{sec:kinematics})  \\
\noalign{\medskip}
V (mag)                  &    13.421$\pm$0.024   & 13.33$\pm$0.07 & 12.24$\pm$0.03 &   APASS DR10 \citep{2018AAS...23222306H}   \\
$B-V$ (mag)                &   0.759$\pm$0.048  & 0.63$\pm$0.08 & 0.646$\pm$0.038 & APASS DR10 \citep{2018AAS...23222306H}  \\
$G$ (mag)                  &    13.2421$\pm$0.0002 & 13.1776$\pm$0.0002 & 12.0930$\pm$0.0003 & {\it Gaia} DR3  \\
$G_{BP}-G_{RP}$ (mag)              &         0.8569      & 0.8157 &  0.8228 & {\it Gaia} DR3  \\
$J$ (mag)    &   12.172$\pm$0.025 & 12.095$\pm$0.024 & 11.025$\pm$0.024 & 2MASS  \\
$H$ (mag)    &   11.84$\pm$0.03 & 11.86$\pm$0.023 & 10.737$\pm$0.023 & 2MASS  \\
$K$ (mag)    &   11.807$\pm$0.025 & 11.78$\pm$0.02& 10.706$\pm$0.023 & 2MASS  \\
\noalign{\medskip}
$T_{\rm eff}$ (K)        &  5665$\pm$115     & 5940$\pm$75& 5805$\pm$62 & This paper (Sect. \ref{sec:atm_param}) \\  
$\log g$                 &  4.25$\pm$0.27   &4.28$\pm$0.21  & 4.30$\pm$0.14 & This paper (Sect. \ref{sec:atm_param}) \\ 
${\rm [Fe/H]}$ (dex)     &  +0.30$\pm$0.11  & $-$0.11$\pm$0.10 & $-$0.13$\pm$0.08  & This paper (Sect. \ref{sec:atm_param}) \\ 
\noalign{\medskip}
$\log R^{'}_{\rm HK}$    &     $-$4.93$\pm$0.26 & $-$5.09$\pm$0.37 & $-$5.32$\pm$0.37 &  This paper (Sect. \ref{sec:chrom}) \\  
$v\sin{i_{\star}}$ (km s$^{-1}$)      &   2.9$\pm$0.5 & 3.4$\pm$0.5 & 2.4$\pm$0.6 & This paper (Sect. \ref{sec:vsini}) \\  
$P_{\rm rot}$ (d)  &   $\gtrsim$18.1  & $\gtrsim$18.3 & $\gtrsim$25.7 & This paper (Sect. \ref{sec:rot_per}) \\
$EW_{\rm Li}$ (m\AA)     &     40$\pm$8 & 71$\pm$5 & 36$\pm$5 &  This paper (Sect. \ref{sec:li})  \\
A(Li)                    &  2.13$\pm$0.13 & 2.59$\pm$0.06 & 2.14$\pm$0.06 &  This paper  (Sect. \ref{sec:li}) \\
\noalign{\medskip}
Extinction ($A_V$ mag)     &    0.02$\pm$0.02 & 0.17$\pm$0.05 & 0.10$\pm$0.08 & This paper (Sect. \ref{sec:sed}) \\
Bolometric Flux ($10^{-10}$ erg~s$^{-1}$~cm$^{-2}$)     &    1.223$\pm$0.028 & 1.469$\pm$0.034 & 3.744$\pm$0.043 & This paper (Sect. \ref{sec:sed}) \\
Luminosity (L$_{\odot}$) &    1.446$\pm$0.036 & 1.306$\pm$0.033 & 1.021$\pm$0.013 & This paper (Sect. \ref{sec:sed}) \\
Radius (R$_{\odot}$)     &    1.250$\pm$0.054 & 1.080$\pm$0.032 & 1.000$\pm$0.023 & This paper (Sect. \ref{sec:sed}) \\
Mass (M$_{\odot}$)       &    1.07$\pm$0.06  & 1.03$\pm$0.06 & 1.03$\pm$0.06 & This paper (Sect. \ref{sec:sed}) \\
Age  (Gyr)               &    5.5$\pm$3.0 & 4.5$\pm$2.9 &  5.3$\pm$3.4 & This paper (Sect. \ref{sec:age}) \\
\noalign{\medskip}
Distance  (pc)               &    593$\pm$5 & 534$\pm$5\tablefootmark{a} &  291.7$\pm$1.6 & {\it Gaia} DR3 \\

         \noalign{\smallskip}
         \hline
      \end{tabular}

\tablefoot{\tablefoottext{a}{{\it Gaia} EDR3.}}
\end{table*}

\subsection{Independent stellar parameters using TRES spectra}
As an independent measurement, TOI-4914 was also observed with the Tillinghast Reflector Echelle Spectrograph \citep[TRES,][]{gaborthesis} on UT 2022-02-11 and 2022-02-16. TRES is a R $=$ 44,000 spectrograph mounted on the 1.5 m Tillinghast Reflector located at the Fred Lawrence Whipple Observatory (FLWO) in Arizona, USA. The spectra were extracted as described in \cite{buchhave2010}. We derived stellar parameters using the Stellar Parameter Classification \citep[SPC,][]{buchhave2012} tool. SPC cross correlates an observed spectrum against a grid of synthetic spectra based on Kurucz atmospheric models \citep{kurucz1992}. The average parameters are $T_{\rm eff} = 5783 \pm 51$ K, $\log g = 4.49 \pm 0.10$, ${\rm [m/H]} = -0.07 \pm 0.08$, $v\sin{i_{\star}} = 4.4 \pm 0.5$ km s$^{-1}$. The stellar parameters align with the estimates given in Sect. \ref{sec:atm_param} and Table \ref{tab:star_param}. The only exception is the $v\sin{i_{\star}}$, but this depends strongly on the choice of the stellar parameters and the values of the microturbolence and macroturbolence velocities. Therefore, we adopted the parameters presented in the previous sections as a reference for this work.

\section{Analysis}
\label{sec:analysis}
\subsection{Planetary systems characterisation}
\label{sec:modelling}

To characterise the properties of TOI-2714 b, TOI-2981 b and TOI-4914 b, we simultaneously studied all ground-based photometry along with the transits observed by \textit{TESS} and the HARPS-N RV time series. This analysis was conducted in a Bayesian framework using \texttt{PyORBIT}\footnote{\url{https://github.com/LucaMalavolta/PyORBIT}} \citep{2016A&A...588A.118M,2018AJ....155..107M}, a Python package for modelling planetary transits and RVs while simultaneously taking into account the effects of stellar activity.

We selected each space-based transit event with a window of three times the transit duration centred on the transit times predicted by a linear ephemeris. 

We simultaneously modelled each transit with the \texttt{BATMAN} code \citep{2015PASP..127.1161K}, fitting the following parameters: the planetary-to-star radius ratio ($R_p /R_{\star}$), the reference transit time ($T_0$), the orbital period ($P$), the impact parameter ($b$), the stellar density ($\rho_\star$, in solar units), the RV semi-amplitude ($K$), the quadratic limb-darkening (LD) coefficients $ u_1$ and $u_2$ adopting the LD parameterisation ($q_1$ and $ q_2$) introduced by \cite{2013MNRAS.435.2152K}, the systemic RV (offset), and a jitter term added in quadrature to the photometric and RV errors to account for any effects not included in our model (e.g. short term stellar activity) or any underestimation of the error bars. Given the well-determined periods provided by photometry and the wide boundaries used, we fitted the periods and semi-amplitudes of the RV signals in linear space. Additionally, we calculated the eccentricity $e$ and periastron argument $\omega$ by fitting $\sqrt{e}\cos{\omega}$ and $\sqrt{e}\sin{\omega}$ \citep{2013PASP..125...83E}. 

We estimated $u_1$ and $u_2$ using \texttt{PyLDTk}\footnote{\url{https://github.com/hpparvi/ldtk}} \citep{2013A&A...553A...6H,2015MNRAS.453.3821P} -- applying the specific filters used during the observations -- and used them as Gaussian priors. We added 0.1 in quadrature to the associated Gaussian error to account for the known underestimation by models. In addition, while performing the joint fit, we detrended each ground-based light curve against the different parameters listed in Table \ref{table:detrending}. We imposed a Gaussian prior on the stellar density and uniform priors on the period, $T_0$, and eccentricity; see list of priors on Table \ref{table:model-lcrv}.

\begin{table*}
\caption{Ground-based photometry observations and detrending parameters.}             
\label{table:detrending}      
\centering                          
\begin{tabular}{c | c c c c c}        
\hline\hline                 
Target & Telescope & Date & Cadence & Parameter & Detrending \rule{0pt}{2.5ex} \rule[-1ex]{0pt}{0pt} \\    
\hline                        
\multirow{2}{*}{TOI-2714} &  McD ($i$) & 2021-10-11 & 300 s& X1 & polynomial  \rule{0pt}{2.5ex} \rule[-1ex]{0pt}{0pt}\\ 
 &  McD ($g$) & 2021-10-16 & 300 s& X1 & polynomial  \rule{0pt}{2.5ex} \rule[-1ex]{0pt}{0pt}\\ 
   \hline
\multirow{8}{*}{TOI-2981} &  El Sauce & 2022-02-27 & 120 s & airmass & exponential  \rule{0pt}{2.5ex} \rule[-1ex]{0pt}{0pt}\\ 

&  TRAPPIST ($z$) & 2022-04-03 & 105 s& airmass & exponential\rule{0pt}{2.5ex} \rule[-1ex]{0pt}{0pt}\\
& ...  & ... & ... & fwhm & polynomial\rule{0pt}{2.5ex} \rule[-1ex]{0pt}{0pt}\\
&  TRAPPIST ($B$) & 2023-03-11 & 100 s& airmass & exponential  \rule{0pt}{2.5ex} \rule[-1ex]{0pt}{0pt}\\
& ... & ... & ... & sky/pixel & polynomial  \rule{0pt}{2.5ex} \rule[-1ex]{0pt}{0pt}\\
&  Teid & 2023-04-20 & 340 s& fbjd & polynomial  \rule{0pt}{2.5ex} \rule[-1ex]{0pt}{0pt}\\
\hline
\multirow{5}{*}{TOI-4914} &  TRAPPIST & 2022-03-06 & 20 s& airmass &  exponential \rule{0pt}{2.5ex} \rule[-1ex]{0pt}{0pt}\\
&  PEST & 2023-05-26 & 120 s& airmass &  exponential \rule{0pt}{2.5ex} \rule[-1ex]{0pt}{0pt}\\
&  Brierfield & 2023-05-26 & 240 s& fwhm &  polynomial \rule{0pt}{2.5ex} \rule[-1ex]{0pt}{0pt}\\
&  El Sauce & 2023-06-07 & 90 s & counts &  polynomial \rule{0pt}{2.5ex} \rule[-1ex]{0pt}{0pt}\\
\hline                                   
\end{tabular}
\end{table*}

\begin{figure}
   \centering
   \includegraphics[width=\hsize]%
   {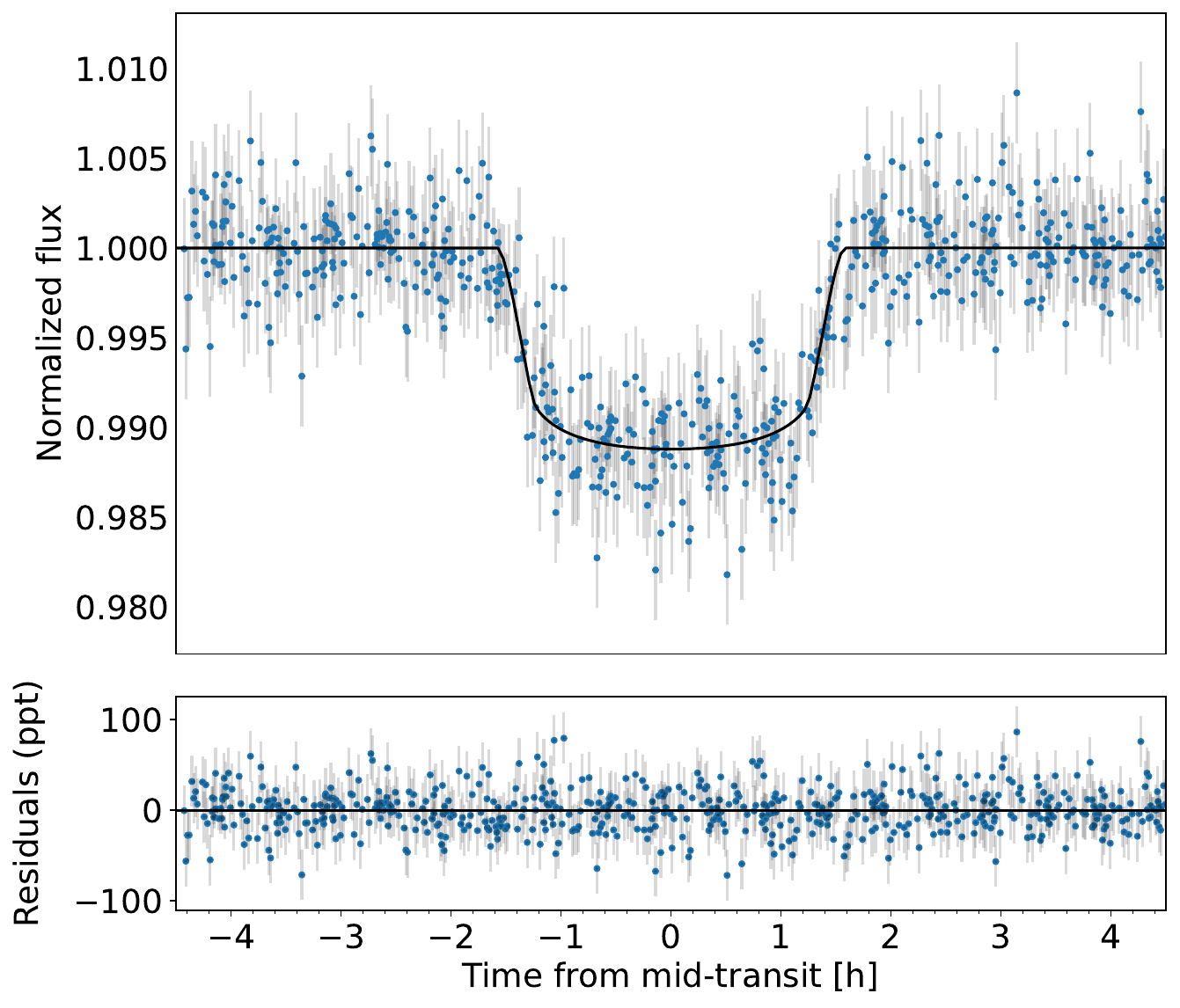}
   \caption{\textit{Top:} \textit{TESS} phase-folded transits of TOI-2714 b after normalisation together with the transit model (black line). \textit{Bottom:} Residuals of the joint fit.}
   \label{fig:2714lc}
\end{figure}

\begin{figure}
   \centering
   \includegraphics[width=\hsize]%
   {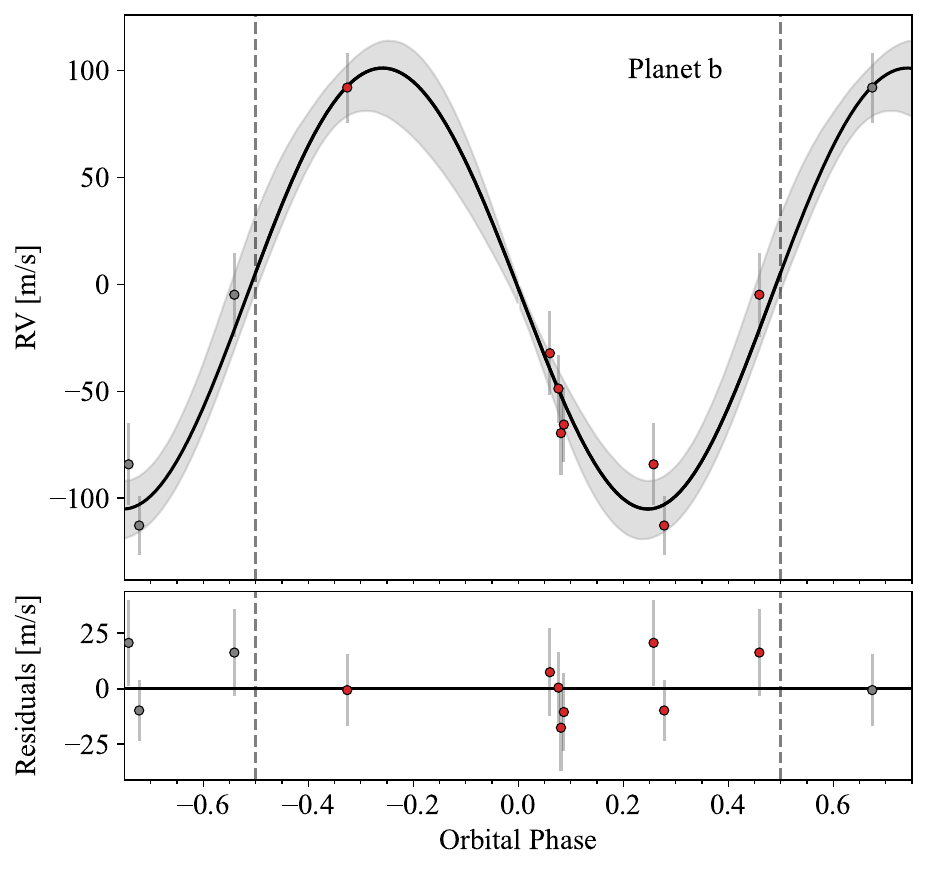}
   \caption{Phase-folded RV fit of TOI-2714 b planetary signal. The shaded area represents the $\pm 1\sigma$ uncertainties of the RV model. 
   The \textit{bottom} panel shows the residuals of the fit.}
   \label{fig:2714}
\end{figure}

\begin{figure}
   \centering
   \includegraphics[width=\hsize]%
   {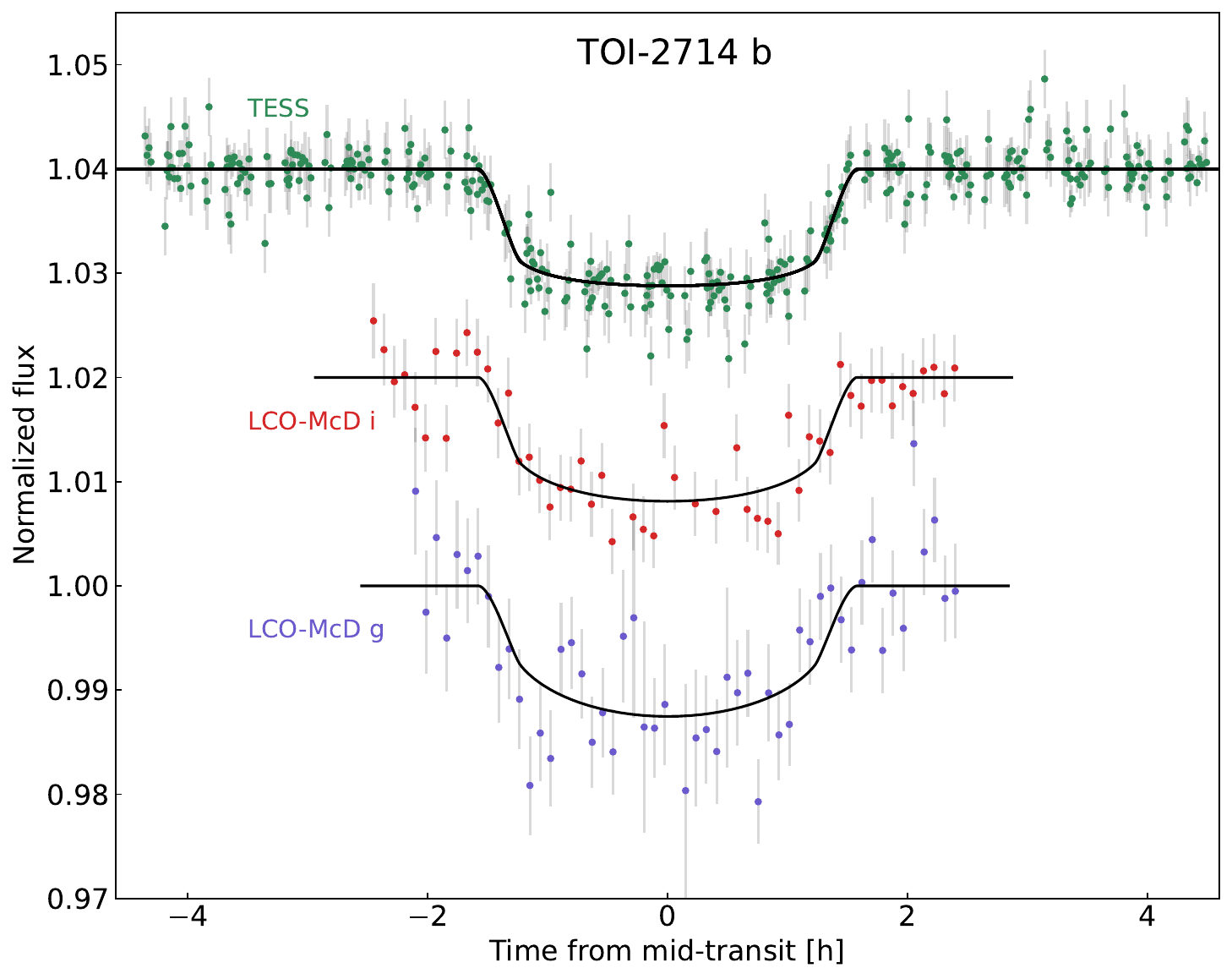}
   \caption{Ground-based photometric data jointly modelled with \textit{TESS} transits of TOI-2714 b. Different datasets are shown in different colours. }
   \label{fig:2714lc_tfop}
\end{figure}

\begin{figure}
   \centering
   \includegraphics[width=\hsize]%
   {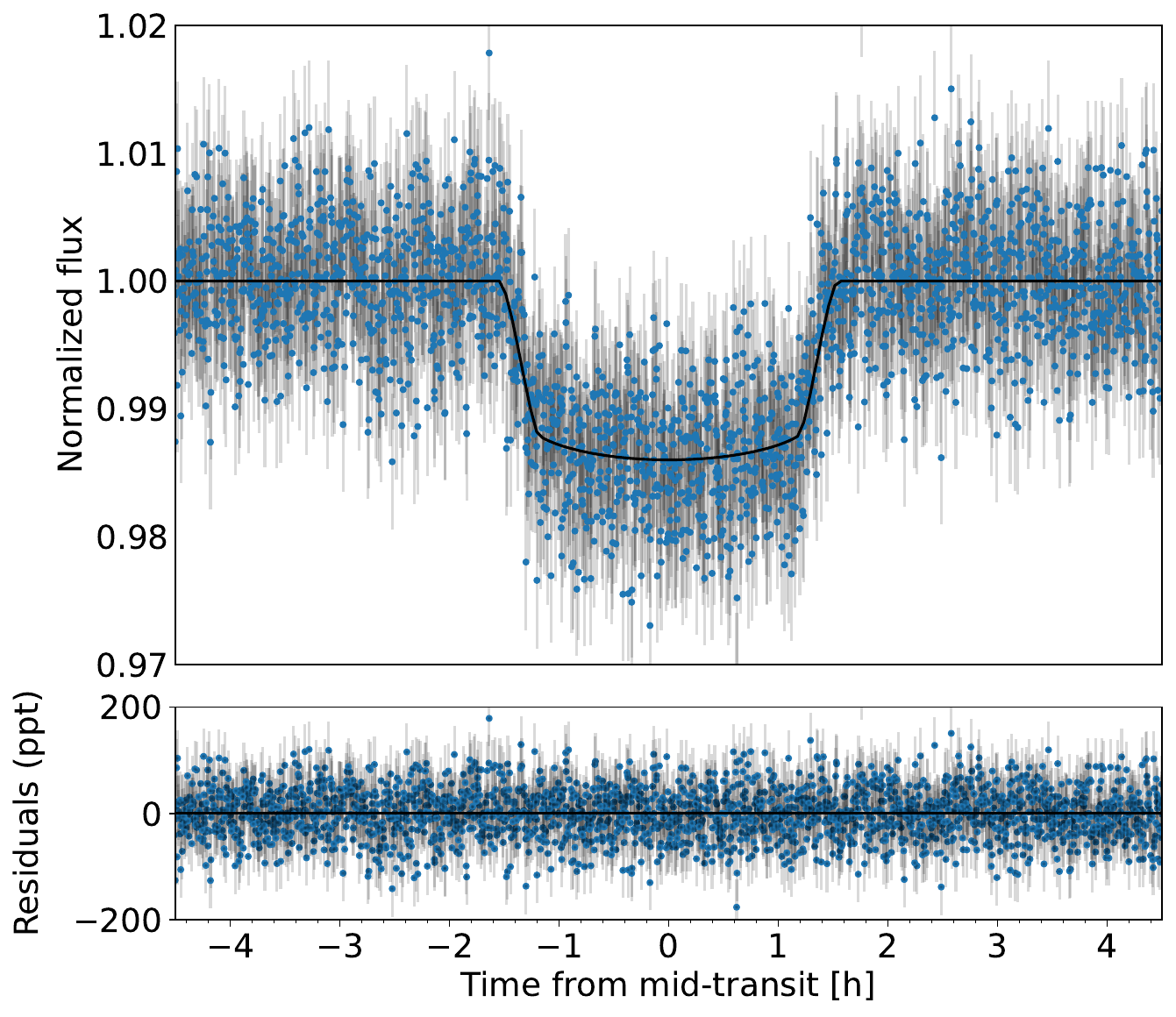}
   \caption{As in Fig. \ref{fig:2714lc}, but for TOI-2981 b.}
   \label{fig:2981lc}
\end{figure}

\begin{figure}
   \centering
   \includegraphics[width=\hsize]%
   {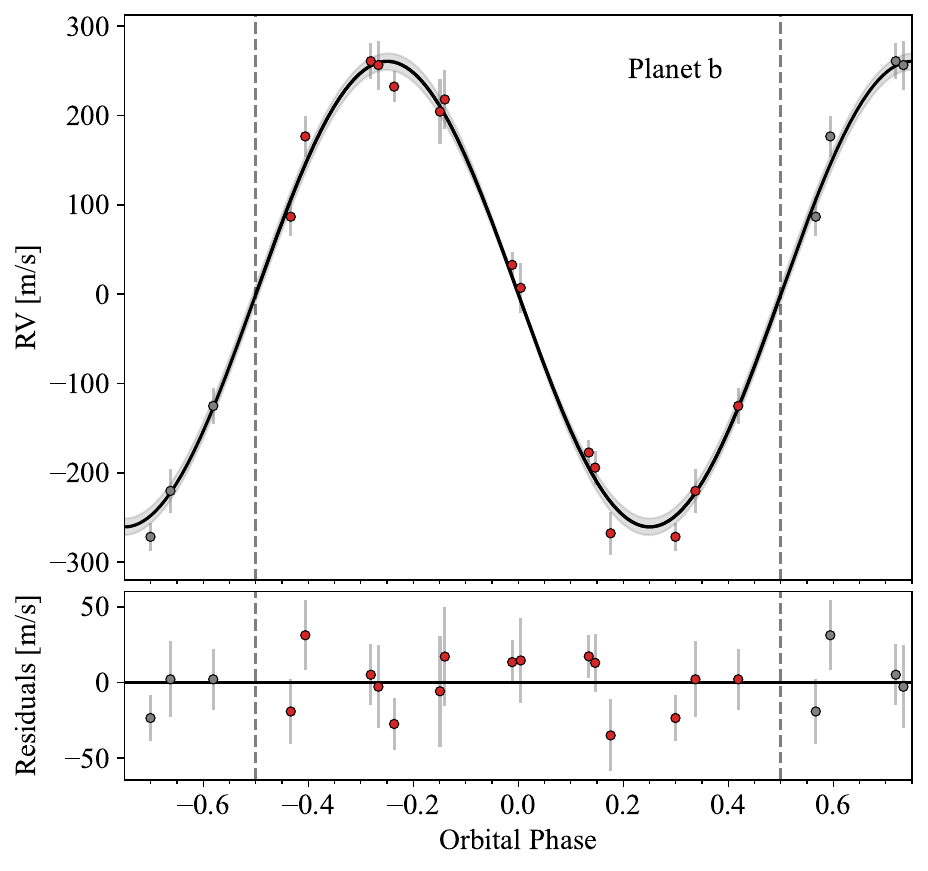}
   \caption{Phase-folded RV fit of TOI-2981 b planetary signal. The shaded area represents the $\pm 1\sigma$ uncertainties of the RV model. The \textit{bottom} panel shows the residuals of the fit.}
   \label{fig:2981}
\end{figure}

\begin{figure}
   \centering
   \includegraphics[width=\hsize]%
   {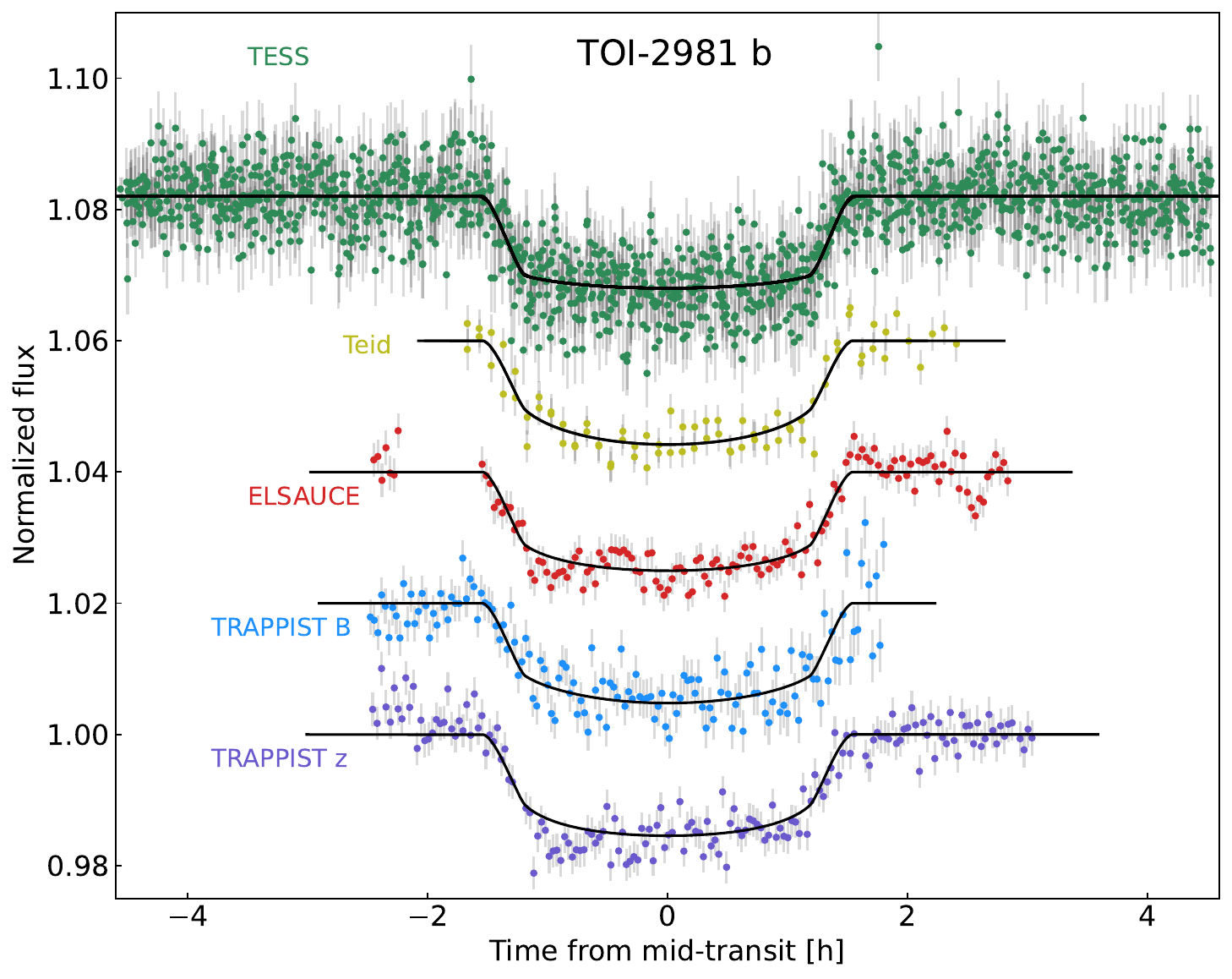}
   \caption{As in Fig. \ref{fig:2714lc_tfop}, but for TOI-2981 b.}
   \label{fig:2981lc_tfop}
\end{figure}

\begin{figure}
   \centering
   \includegraphics[width=\hsize]%
   {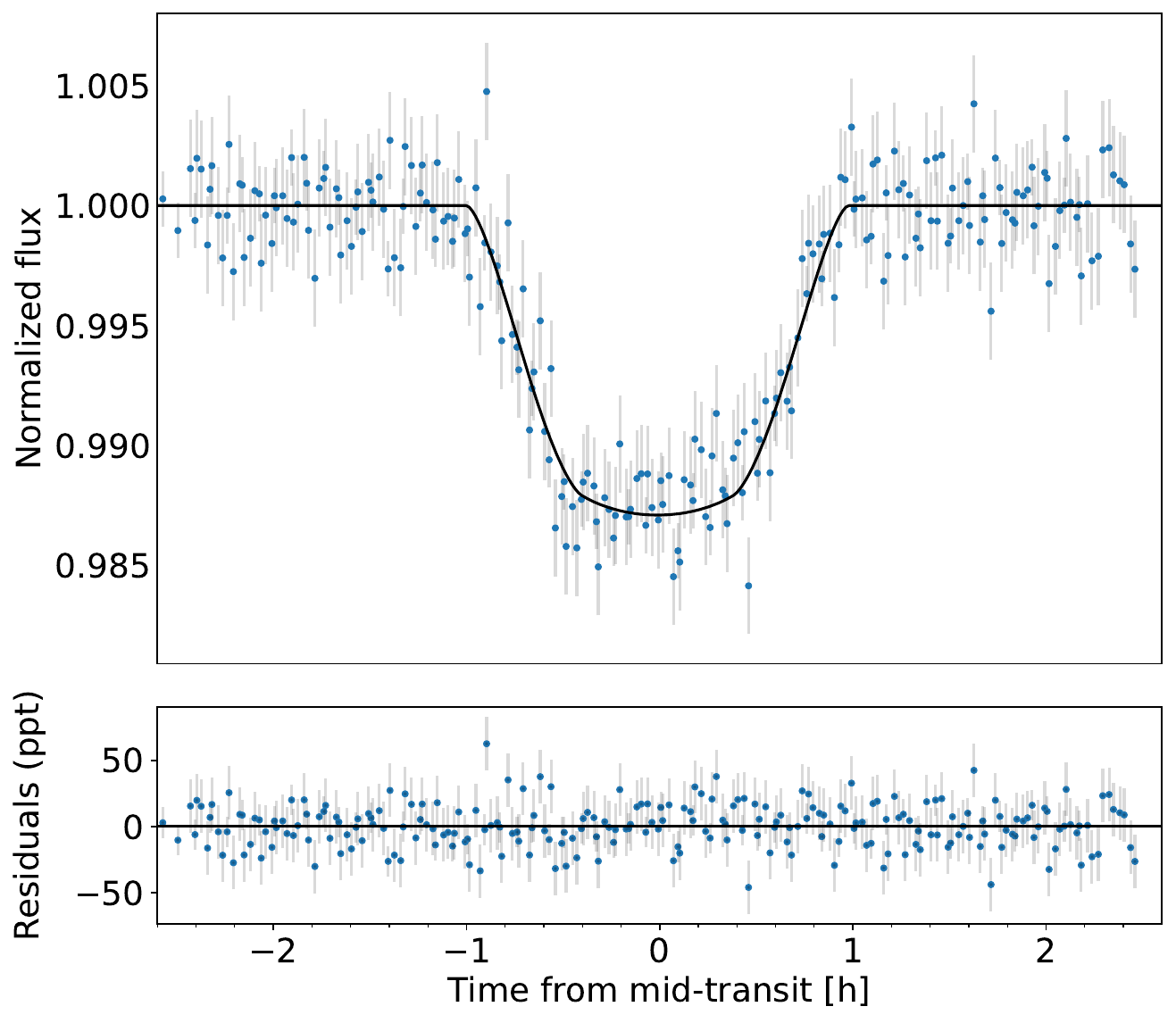}
   \caption{Equal to Fig. \ref{fig:2714lc}, but for TOI-4914 b.}
   \label{fig:4914lc}
\end{figure}

\begin{figure}
   \centering
   \includegraphics[width=\hsize]%
   {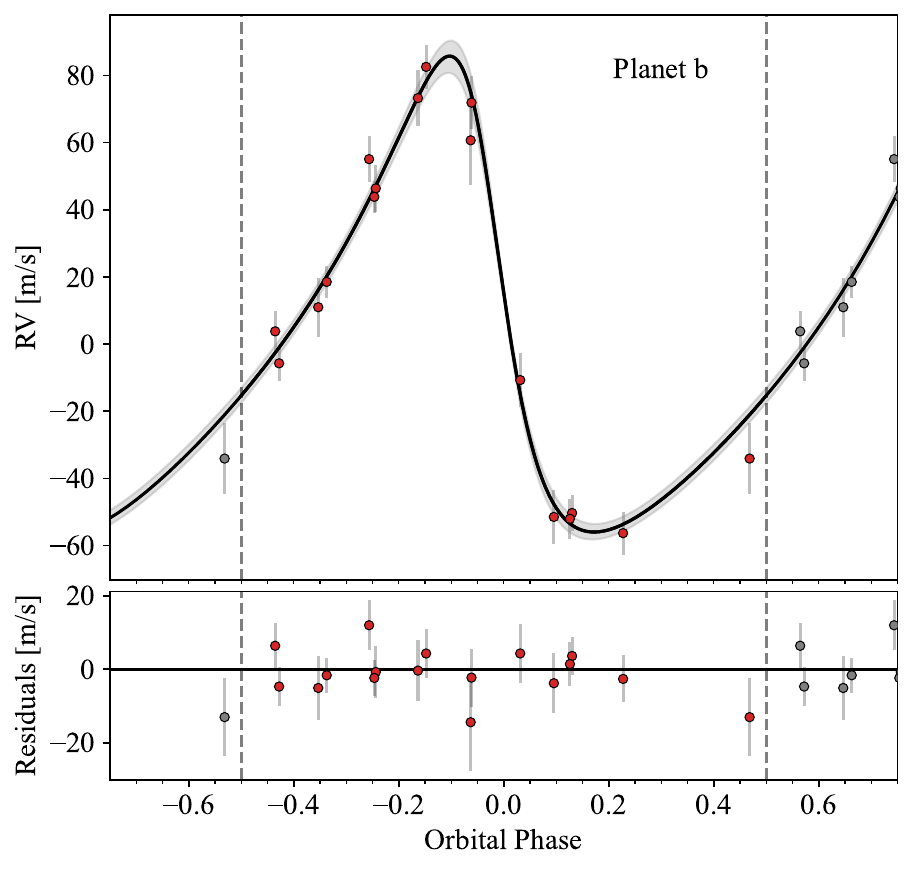}
   \caption{As in Fig. \ref{fig:2981}, but for TOI-4914 b.}
   \label{fig:4914}
\end{figure}

\begin{figure}
   \centering
   \includegraphics[width=\hsize]%
   {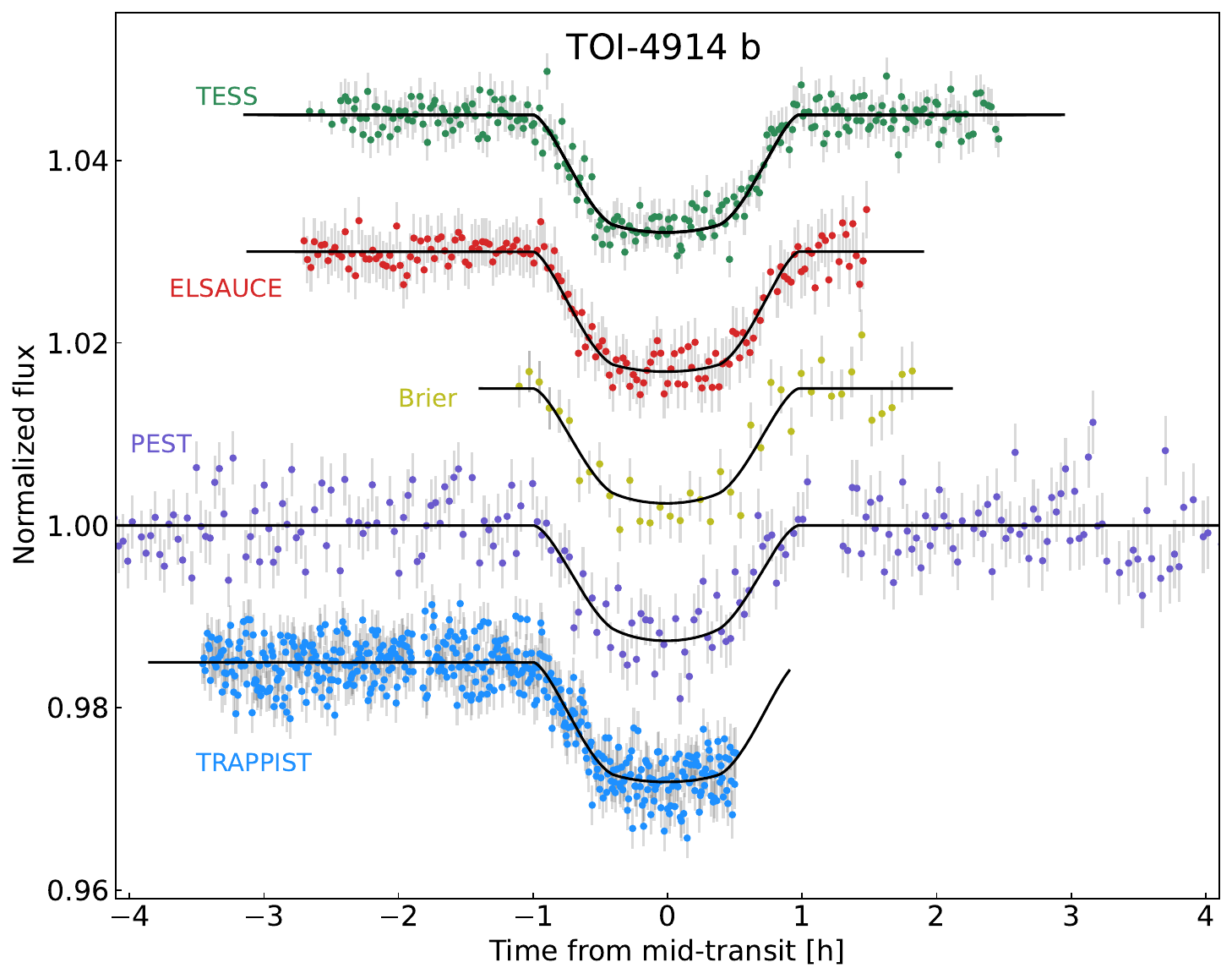}
   \caption{Similar to Fig. \ref{fig:2714lc_tfop}, but for TOI-4914 b.}
   \label{fig:4914lc_tfop}
\end{figure}

We carried out a global optimisation of the parameters by running a differential evolution algorithm (\citealt{1997JGOpt..11..341S}, \texttt{PyDE}\footnote{\url{https://github.com/hpparvi/PyDE}}) and performed a Bayesian analysis. For the latter, we used the affine-invariant ensemble sampler \citep{2010CAMCS...5...65G} for Markov chain Monte Carlo (MCMC), as implemented in the \texttt{emcee} package \citep{2013PASP..125..306F}. We used $4 n_{\rm dim}$ walkers (where $n_{\rm dim}$ represents the dimensionality of the model) for 50 000 generations with \texttt{PyDE}, followed by 100 000 steps with \texttt{emcee} -- where we applied a thinning factor of 200 to mitigate the effect of chain auto-correlation. We discarded the first 25 000 steps (burn-in) after checking the convergence of the chains using the Gelman–Rubin (GR) statistic (\citealt{1992StaSc...7..457G}, with a threshold $\hat{R}$= 1.01). Figures from \ref{fig:2714lc} to \ref{fig:4914lc_tfop} and Table \ref{table:model-lcrv} present the results of the modelling.

\begin{table*}
{\small
\caption{Priors and outcomes of spectroscopic plus photometric modelling.}             
\label{table:model-lcrv}      
\centering          
\begin{tabular}{l c| c c  | c c | c c}     
\hline\hline     

\multicolumn{2}{c|}{Stellar parameters} & \multicolumn{2}{c|}{TOI-2714} & \multicolumn{2}{c|}{TOI-2981} & \multicolumn{2}{c}{TOI-4914} \rule{0pt}{2.0ex} \rule[-1ex]{0pt}{0pt}\\ 
\hline    
\multicolumn{1}{l}{Parameter} & Unit & Prior & Value & 
Prior & Value  & Prior & Value \rule{0pt}{2.0ex} \rule[-1ex]{0pt}{0pt}\\ 
\hline 
   Density ($\rho_{\star}$) & $\rho_{\sun}$ & $\mathcal{N}$(0.548, 0.079) & 0.532$^{+0.075}_{-0.078}$ & 
   $\mathcal{N}$(0.82, 0.09)& 0.89$^{+0.08}_{-0.08}$ &$\mathcal{N}$(1.03, 0.09) &1.10$^{+0.08}_{-0.08}$ \rule{0pt}{2.0ex} \rule[-1ex]{0pt}{0pt}\\
   \textit{TESS} quad. LD coeff. ($u_1$) & &$\mathcal{U}$(0, 1) & 0.21$^{+0.21}_{-0.15}$ & 
   $\mathcal{U}$(0, 1) & 0.25$^{+0.13}_{-0.14}$ & $\mathcal{U}$(0, 1) & 0.50$^{+0.24}_{-0.27}$\rule{0pt}{2.0ex} \rule[-1ex]{0pt}{0pt}\\
   \textit{TESS} quad. LD coeff. ($u_2$) & &$\mathcal{U}$(0, 1) & 0.29$^{+0.32}_{-0.34}$  & 
   $\mathcal{U}$(0, 1) & 0.01$^{+0.22}_{-0.14}$ & $\mathcal{U}$(0, 1) & $-$0.01$^{+0.35}_{-0.26}$ \rule{0pt}{2.0ex} \rule[-1ex]{0pt}{0pt}\\
   \textit{ELSAUCE} LD coeff. ($u_1$) & & -- & -- & 
   $\mathcal{N}$(0.43, 0.12) & 0.41$^{+0.09}_{-0.09}$ & $\mathcal{N}$(0.35, 0.11) & 0.34$^{+0.10}_{-0.10}$ \rule{0pt}{2.0ex} \rule[-1ex]{0pt}{0pt}\\
   \textit{ELSAUCE} LD coeff. ($u_2$) & & -- & -- & 
   $\mathcal{N}$(0.15, 0.15) & 0.15$^{+0.12}_{-0.12}$ & $\mathcal{N}$(0.14, 0.14) & 0.14$^{+0.12}_{-0.12}$ \rule{0pt}{2.0ex} \rule[-1ex]{0pt}{0pt}\\
   \textit{PEST} LD coeff. ($u_1$) & & -- & -- & 
   -- & -- & $\mathcal{N}$(0.45, 0.12) & 0.52$^{+0.10}_{-0.10}$ \rule{0pt}{2.0ex} \rule[-1ex]{0pt}{0pt}\\
   \textit{PEST} LD coeff. ($u_2$) & &-- & -- & 
   -- & -- & $\mathcal{N}$(0.15, 0.15) & 0.22$^{+0.12}_{-0.13}$ \rule{0pt}{2.0ex} \rule[-1ex]{0pt}{0pt}\\
   \textit{Teid} LD coeff. ($u_1$) & &-- & --  & 
   $\mathcal{N}$(0.64, 0.13) & 0.58$^{+0.09}_{-0.09}$ & -- & -- \rule{0pt}{0.0ex} \rule[-1ex]{0pt}{0pt}\\
   \textit{Teid} LD coeff. ($u_2$) & &-- & -- & 
   $\mathcal{N}$(0.14, 0.15) & 0.10$^{+0.13}_{-0.12}$ & -- & -- \rule{0pt}{2.0ex} \rule[-1ex]{0pt}{0pt}\\
   \textit{TRAPPIST} LD coeff. ($u_1, B$) & &-- & --  &
   $\mathcal{N}$(0.68, 0.13) & 0.54$^{+0.10}_{-0.10}$ & -- & -- \rule{0pt}{2.0ex} \rule[-1ex]{0pt}{0pt}\\
   \textit{TRAPPIST} LD coeff. ($u_2, B$) & &-- & -- &
   $\mathcal{N}$(0.12, 0.16) & 0.00$^{+0.13}_{-0.13}$ & -- & -- \rule{0pt}{2.0ex} \rule[-1ex]{0pt}{0pt}\\
   \textit{TRAPPIST} LD coeff. ($u_1, z$) & & -- & --  &
   $\mathcal{N}$(0.33, 0.11) & 0.39$^{+0.09}_{-0.10}$ & $\mathcal{N}$(0.35, 0.11) & 0.35$^{+0.10}_{-0.10}$ \rule{0pt}{2.0ex} \rule[-1ex]{0pt}{0pt}\\
   \textit{TRAPPIST} LD coeff. ($u_2, z$) & & -- & -- & 
   $\mathcal{N}$(0.14, 0.14) & 0.31$^{+0.12}_{-0.13}$ & $\mathcal{N}$(0.14, 0.14) & 0.15$^{+0.12}_{-0.12}$ \rule{0pt}{2.0ex} \rule[-1ex]{0pt}{0pt}\\
   \textit{Brier} LD coeff. ($u_1$) & & -- & --  & 
   -- & -- & $\mathcal{N}$(0.72, 0.13) & 0.62$^{+0.11}_{-0.12}$ \rule{0pt}{1.0ex} \rule[-1ex]{0pt}{0pt}\\
   \textit{Brier} LD coeff. ($u_2$) & & -- & -- &
   -- & -- & $\mathcal{N}$(0.09, 0.16) & 0.02$^{+0.14}_{-0.14}$ \rule{0pt}{2.0ex} \rule[-1ex]{0pt}{0pt}\\
   \textit{McD} LD coeff. ($u_1, i$) & &$\mathcal{N}$(0.43, 0.12) & 0.47$^{+0.11}_{-0.11}$  &
   -- & -- & -- & -- \rule{0pt}{0.0ex} \rule[-1ex]{0pt}{0pt}\\
   \textit{McD} LD coeff. ($u_2, i$) & &$\mathcal{N}$(0.13, 0.15) & 0.19$^{+0.14}_{-0.14}$  &
   -- & -- & -- & -- \rule{0pt}{2.0ex} \rule[-1ex]{0pt}{0pt}\\
   \textit{McD} LD coeff. ($u_1, g$) & &$\mathcal{N}$(0.70, 0.13) & 0.69$^{+0.11}_{-0.12}$ & 
   -- & -- & -- & -- \rule{0pt}{2.0ex} \rule[-1ex]{0pt}{0pt}\\
   \textit{McD} LD coeff. ($u_2, g$) & &$\mathcal{N}$(0.09, 0.16) & 0.08$^{+0.13}_{-0.14}$ & 
   -- & -- & -- & -- \rule{0pt}{2.0ex} \rule[-1ex]{0pt}{0pt}\\
   \textit{TESS} Sect. 5 jitter & ppt 
 & ... & 0.2$^{+0.2}_{-0.1}$ & 
 --&--& -- & --\rule{0pt}{1.8ex} \rule[-1ex]{0pt}{0pt}\\
   \textit{TESS} Sect. 9 jitter & ppt 
 & -- & --  & 
 ... & 0.5$^{+0.3}_{-0.3}$& --&--\rule{0pt}{0.8ex} \rule[-1ex]{0pt}{0pt}\\
   \textit{TESS} Sect. 31 jitter & ppt 
 & ... & 0.3$^{+0.3}_{-0.2}$ & 
 --& --&-- &--\rule{0pt}{0.8ex} \rule[-1ex]{0pt}{0pt}\\
   \textit{TESS} Sect. 36 jitter & ppt 
 & -- & -- &
 ... & 0.5$^{+0.4}_{-0.3}$& --&--\rule{0pt}{0.8ex} \rule[-1ex]{0pt}{0pt}\\
   \textit{TESS} Sect. 37 jitter & ppt 
 &-- & -- & 
 -- & --& ...& 0.2$^{+0.2}_{-0.1}$\rule{0pt}{0.8ex} \rule[-1ex]{0pt}{0pt}\\
   \textit{TESS} Sect. 63 jitter & ppt 
 & -- & --&
 ... & 0.3$^{+0.3}_{-0.2}$& --& --\rule{0pt}{0.8ex} \rule[-1ex]{0pt}{0pt}\\
   \textit{TESS} Sect. 64 jitter & ppt 
 & -- &-- & 
 -- & -- & ...& 0.3$^{+0.3}_{-0.2}$\rule{0pt}{1.8ex} \rule[-1ex]{0pt}{0pt}\\
   \textit{ELSAUCE} jitter & ppt 
 & -- & -- & 
 ... & 0.6$^{+0.4}_{-0.4}$ & ...& 0.2$^{+0.2}_{-0.1}$\rule{0pt}{2.0ex} \rule[-1ex]{0pt}{0pt}\\
   \textit{PEST} jitter & ppt 
 & -- & -- &
 -- & -- & ...& 1.2$^{+0.4}_{-0.5}$\rule{0pt}{2.0ex} \rule[-1ex]{0pt}{0pt}\\
   \textit{Teid (1)} jitter & ppt 
 & -- & -- & 
 ... & 1.0$^{+0.8}_{-0.6}$ & --& --\rule{0pt}{0.0ex} \rule[-1ex]{0pt}{0pt}\\
   \textit{Teid (2)} jitter & ppt 
 & -- & -- &
 ... & 0.7$^{+0.7}_{-0.5}$ & --& --\rule{0pt}{2.0ex} \rule[-1ex]{0pt}{0pt}\\
   \textit{TRAPPIST (1)} jitter & ppt 
 & -- & -- & 
 ... & 1.7$^{+0.5}_{-0.5}$ & --& --\rule{0pt}{2.0ex} \rule[-1ex]{0pt}{0pt}\\
   \textit{TRAPPIST (2)} jitter & ppt 
 & -- & -- &
 ... & 1.6$^{+0.3}_{-0.4}$ & ... & 1.0$^{+0.2}_{-0.2}$ \rule{0pt}{2.0ex} \rule[-1ex]{0pt}{0pt}\\
   \textit{Brier} jitter & ppt 
 & -- & -- & 
 -- & -- & ...& 0.6$^{+0.6}_{-0.4}$\rule{0pt}{1.8ex} \rule[-1ex]{0pt}{0pt}\\
   \textit{McD (i)} jitter & ppt 
 & ... & 0.7$^{+0.7}_{-0.5}$ &
 -- & -- & -- & --\rule{0pt}{0.0ex} \rule[-1ex]{0pt}{0pt}\\
   \textit{McD (g)} jitter & ppt 
 & ... & 2.0$^{+1.0}_{-1.0}$ & 
 -- & -- & --& -- \rule{0pt}{2.0ex} \rule[-1ex]{0pt}{0pt}\\
   Uncorrelated RV jitter & m s$^{-1}$
    & ... & 8.9$^{+11}_{-6.2}$ &
    ... & 11.9$^{+9.3}_{-7.6}$ & ... & 2.1$^{+2.3}_{-1.4}$\rule{0pt}{2.0ex} \rule[-1ex]{0pt}{0pt}\\
   RV offset HARPS-N & m s$^{-1}$ & ... & 45874$\pm$10 & 
   ... & 13110.6$^{+6.4}_{-6.6}$ & ... & 11606$^{+2}_{-2}$  \rule{0pt}{1.0ex}\rule[-1ex]{0pt}{0pt}\\
\hline

 
\multicolumn{2}{c|}{Planet} & \multicolumn{2}{c|}{TOI-2714 b} & \multicolumn{2}{c|}{TOI-2981 b} & \multicolumn{2}{c}{TOI-4914 b} \rule{0pt}{2.0ex} \rule[-1ex]{0pt}{0pt}\\ 
\hline    
   Orbital period ($P$) & days & $\mathcal{U}$(2.3, 2.7) & 2.499387 &
   $\mathcal{U}$(3.5, 3.7)& 3.601501 & $\mathcal{U}$(10.58, 10.62) & 10.60057\rule{0pt}{2.0ex} \rule[-1ex]{0pt}{0pt}\\
   & &   & $\pm$0.000004 & 
    & $\pm$0.000002 &  & $\pm$0.00001\rule{0pt}{0.0ex} \rule[-1ex]{0pt}{0pt}\\
   Central time of transit ($T_{\rm 0}$) & BTJD & $\mathcal{U}$(2168.5, & 2168.9479 &
   $\mathcal{U}$(2302.5, & 2302.7195 & $\mathcal{U}$(2317.1, & 2317.2727 \rule{0pt}{0.0ex} \rule[-1ex]{0pt}{0pt}\\
   & &   2169.4) & $\pm$0.0008 &
   2302.9) & $\pm$0.0004 & 2317.4) & $\pm$0.0006\rule{0pt}{0.0ex} \rule[-1ex]{0pt}{0pt}\\
   Scaled semi-maj. axis ($\frac{a}{R_{\star}}$) & & ... & 6.28$^{+0.28}_{-0.32}$& 
   ... & 9.51$^{+0.27}_{-0.27}$ & ... & 20.98$^{+0.51}_{-0.53}$\rule{0pt}{0.0ex}  \rule[-1ex]{0pt}{0pt}\\
   Orbital semi-maj. axis ($a$) & AU & ... & 0.036$^{+0.002}_{-0.002}$ & 
   ... & 0.048$^{+0.002}_{-0.002}$ & ... & 0.098$^{+0.003}_{-0.003}$\rule{0pt}{2.0ex} \rule[-1ex]{0pt}{0pt}\\
   Orbital inclination ($i$) & deg & ... & 86.3$^{+2.0}_{-1.3}$ &
   ... & 87.98$^{+0.79}_{-0.56}$ & ... & 86.35$^{+0.20}_{-0.22}$\rule{0pt}{2.0ex} \rule[-1ex]{0pt}{0pt}\\
   Orbital eccentricity ($e$) &  & $\mathcal{U}$(0, 0.9) & $\leq$ 0.164\tablefootmark{a} &
   $\mathcal{U}$(0, 0.9) & $\leq$ 0.035\tablefootmark{a} & $\mathcal{U}$(0, 0.9) & 0.408$^{+0.023}_{-0.023}$ \rule{0pt}{2.0ex} \rule[-1ex]{0pt}{0pt}\\
   Impact parameter ($b$) & & $\mathcal{U}$(0, 2) & 0.41$^{+0.18}_{-0.23}$ & 
   $\mathcal{U}$(0, 2) & 0.34$^{+0.09}_{-0.13}$ & $\mathcal{U}$(0, 2) & 0.827$^{+0.012}_{-0.013}$\rule{0pt}{1.0ex} \rule[-1ex]{0pt}{0pt}\\
   Planet/star rad. ratio ($\frac{R_{\rm p}}{R_{\star}}$) & & $\mathcal{U}$(0, 0.5) & 0.101$^{+0.003}_{-0.002}$ &\rule{0pt}{1.5ex}
   $\mathcal{U}$(0, 0.5) & 0.114$^{+0.001}_{-0.001}$ & $\mathcal{U}$(0, 0.5) & 0.118$^{+0.002}_{-0.002}$\rule[-1ex]{0pt}{0pt}\\
   Argument of periastron ($\omega$) & deg & ... & $-$146$^{+55}_{-95}$ & 
   ... & 28$^{+98}_{-134}$  & ... & 58.7$^{+4.2}_{-4.0}$\rule{0pt}{2.0ex} \rule[-1ex]{0pt}{0pt}\\
   Transit duration ($T_{14}$)\tablefootmark{b} & days & ... & 0.131$^{+0.007}_{-0.011}$  &
   ... & 0.128$^{+0.002}_{-0.002}$ & ... & 0.121$^{+0.003}_{-0.003}$\rule{0pt}{2.0ex} \rule[-1ex]{0pt}{0pt}\\
   RV semi-amplitude ($K$) & m s$^{-1}$ & $\mathcal{U}$(0.01, 500) & 103$^{+14}_{-14}$ &
   $\mathcal{U}$(0.01, 500) & 260.6$^{+8.7}_{-8.6}$ & $\mathcal{U}$(0.01, 500)& 70.9$^{+3.0}_{-3.1}$\rule{0pt}{2.0ex} \rule[-1ex]{0pt}{0pt}\\
   Planetary radius ($R_{\rm p}$) & $R_{\oplus}$ & ... & 13.72$^{+0.70}_{-0.67}$  & 
   ... & 13.40$^{+0.42}_{-0.42}$ &... & 12.87$^{+0.36}_{-0.36}$\rule{0pt}{2.0ex}\rule[-1ex]{0pt}{0pt}\\
   Planetary mass ($M_{\rm p}$) & $M_{\oplus}$ & ... & 228$^{+33}_{-32}$   &
   ... & 638$^{+33}_{-33}$ & ...& 227$^{+13}_{-13}$ \rule{0pt}{2.0ex}\rule[-1ex]{0pt}{0pt}\\
   Planetary density ($\rho_{\rm p}$) & g cm$^{-3}$ & ... & 0.49 $\pm$ 0.11 & ... & 1.46 $\pm$0.16 & ... & 0.59 $\pm$0.06 \rule{0pt}{2.0ex}\rule[-1ex]{0pt}{0pt}\\
\hline   
\end{tabular}
 }
 \tablefoot{\tablefoottext{a}{84th percentile.} \tablefoottext{b}{From \cite{2010exop.book...55W}.}}
\end{table*}

\subsubsection{TOI-4914 b has a well-constrained high eccentricity}

Our analysis indicates a high eccentricity for TOI-4914 b ($e = 0.408 \pm 0.023$).
To confirm this eccentricity detection, we repeated the analysis by forcing a circular orbit. We used the Bayesian information criterion \citep[BIC,][]{1978AnSta...6..461S} to compare the two different analyses. Our analysis showed a strong preference for case 1 (eccentric orbit) over case 2 (circular orbit), with a substantial $\Delta \rm BIC_{21}$ value of 96 \citep{doi:10.1080/01621459.1995.10476572}. To further prove the detection, we sampled the semi-amplitude of the RV signal in the base 2 and base 10 logarithmic scales. The resulting eccentricities agree with the value given in Table \ref{table:model-lcrv}. 

\subsubsection{Treatment of potential activity}

Since the \textit{TESS} light curves show no clear, strong modulation, the contribution of stellar activity in the RVs should not exceed $\sim$ 20 m s$^{-1}$ for TOI-2981 b and $\sim$ 10 m s$^{-1}$ for the other two planets (e.g., \citealt{2016A&A...588A.118M}) and can be treated as uncorrelated jitter noise.

\subsection{Search for additional planets in the systems}
\label{sec:additionalplanets}

Finally, as a further test, we searched for the presence of additional planets in our RV datasets by applying broad uniform priors on the period and RV semi-amplitude of the possible companions. No solution with additional planets converged. With the available data, it is not possible to determine whether the extra scatter in the residuals (i.e., the uncorrelated RV jitter) is due to stellar activity, an undetectable planetary companion, or residual instrumental systematics. We emphasise that the solution for each planet in our systems shows little variation, further strengthening the validity of their detections.

To explore the possible presence of dynamical interactions in the three systems described in this paper, we performed a search for transit timing variations \citep[TTVs; e.g.][]{2005MNRAS.359..567A, 2005Sci...307.1288H,2014A&A...571A..38B,2019MNRAS.484.3233B,2021MNRAS.506.3810B} of TOI-2714 b, TOI-2981 b and TOI-4914 b, using a \texttt{PyORBIT} model based on \texttt{BATMAN}. In particular, this model allows us to fit each individual transit time $T_0$ (fixing the orbital periods found in Sect. \ref{sec:modelling}).

\begin{figure*}[!th]
   \centering
   \minipage{0.333\textwidth}
   \includegraphics[width=\hsize]{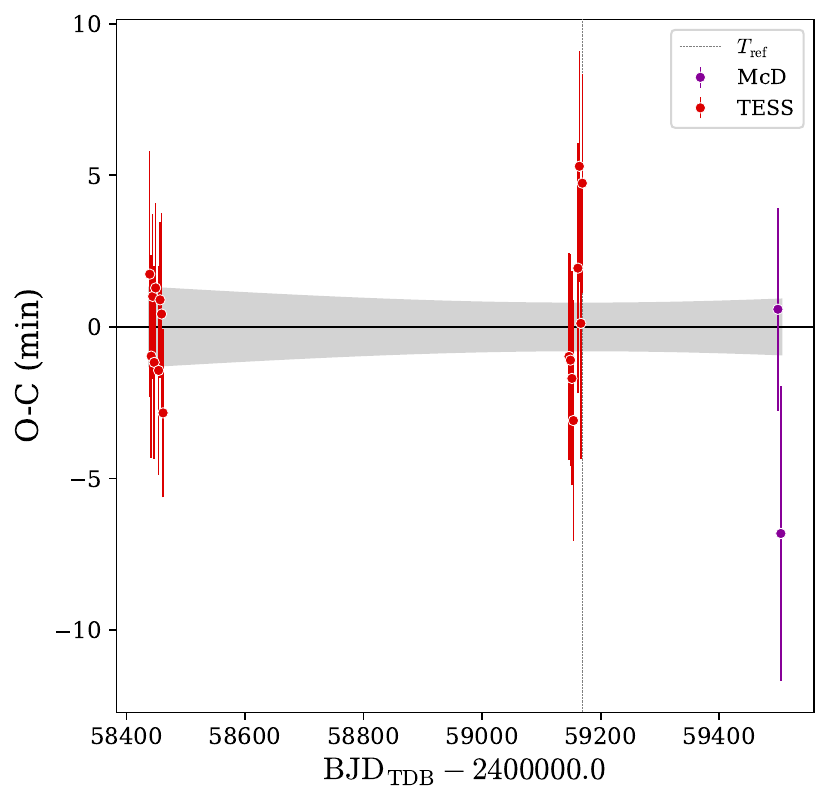}   
   \endminipage\hfill
   \minipage{0.333\textwidth}
   \includegraphics[width=\hsize]{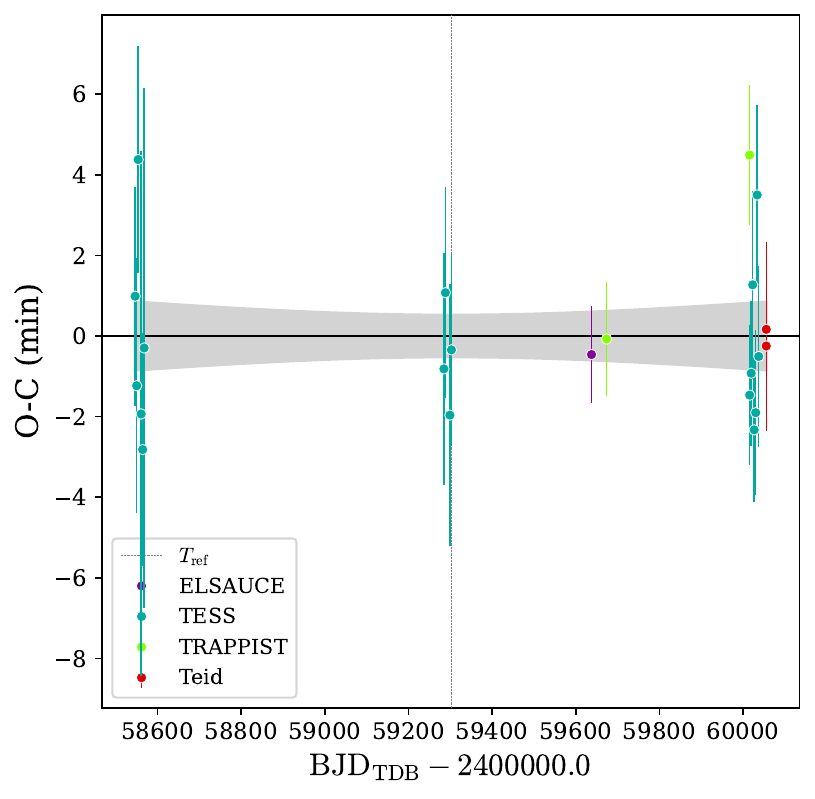} 
   \endminipage\hfill
   \minipage{0.333\textwidth}
   \includegraphics[width=\hsize]{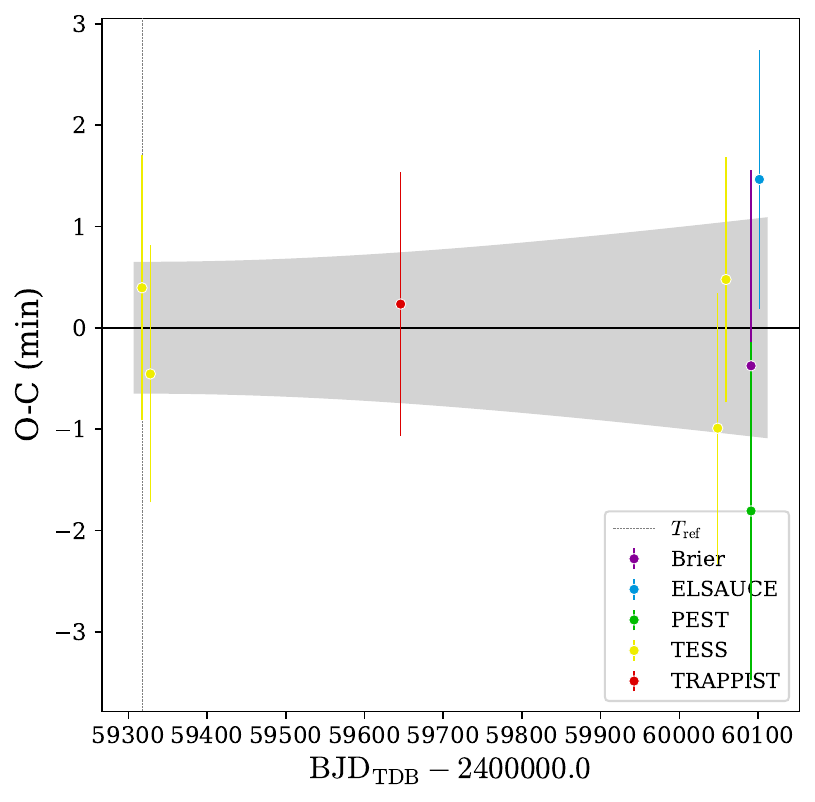}   
   \endminipage\hfill
   \caption{\textit{Left}: O-C plot representing the observed (O) and calculated (C) transit times for the linear ephemeris of TOI-2714 b (see Table~\ref{table:model-lcrv}). Each dataset is shown in a distinct colour. The shaded area represents the formal uncertainty of the linear ephemeris. \textit{Middle}: TOI-2981 b. \textit{Right}: TOI-4914 b.}
   \label{fig:ttv}
\end{figure*}

We computed the observed (O) $-$ calculated (C) diagrams for each planet, removing the linear ephemeris (in Table~\ref{table:model-lcrv}) for each transit time. See the O-C diagrams in Fig.~\ref{fig:ttv} for TOI-2714 b, TOI-2981 b and TOI-4914 b respectively. The possible TTV amplitude ($A_\mathrm{TTV}$), computed as the semi-amplitude of the O-C, is $6.1 \pm 3.4$ minutes for planet TOI-2714 b, $3.7 \pm 4.3$ minutes for planet TOI-2981 b, and $1.6 \pm 1.2$ minutes for planet TOI-4914 b. For each O-C we generated 10\,000 gaussians centred on the O-C with uncertainty, then calculated the semi-amplitude of the O-C for each gaussian ($A_\mathrm{TTV, gauss}$). The error has been computed as the 68.27th percentile of $|A_\mathrm{TTV, gauss} - A_\mathrm{TTV}|$. 

More than 90 per cent of the points in the O-C diagrams fall within the formal uncertainty (weighted least square, shaded areas) of the linear ephemeris at 1$\sigma$. We can therefore conclude that the available data for TOI-2714 b, TOI-2981 b and TOI-4914 b do not show clear TTVs, supporting the hypothesis that there are no other detectable planetary companions in the systems. 

\subsection{Equilibrium temperature, scale height and TSM}

We calculated the equilibrium temperature ($T_{\rm eq}$) of the three planets assuming zero albedo and full day-night heat redistribution:
\begin{equation}
    T_{\rm eq} = T_{\rm eff} \sqrt{\frac{R_\star}{a}} \left ( \frac{1}{4}\right )^{1/4},
\end{equation} where $a$ is the orbital semi-major axis, $R_\star$ the stellar radius and $T_{\rm eff}$ is the host star effective temperature. The equilibrium temperatures are listed in Table \ref{table:tsm}.

Assuming an H$_2$-dominated solar-abundance, cloud-free atmosphere, we can estimate the scale height of the planetary atmosphere $H = \frac{k_{\rm b} T_{\rm eq}}{\mu g}$, where $k_{\rm b}$ is the Boltzmann constant, $\mu = 2.3$ amu the mean molecular weight and $g$ is the surface gravity of the planet. The amplitude of the spectral features in transmission ($\delta_\lambda$) is $\sim 4 R_{\rm p} H /R_\star^2$ \citep{2018haex.bookE.100K}. The calculated $H$ and $\delta_\lambda$ are listed in Table \ref{table:tsm}. 

We then calculated the TSM following \cite{2018PASP..130k4401K}:
\begin{equation}
    {\rm TSM} = S \times \frac{R_{\rm p}^3 T_{\rm eq}}{M_{\rm p} R_\star^2} \times 10^{-m_{\rm J}/5},
\end{equation} where $S$ is a normalisation constant to match the more detailed work of \cite{2018PASP..130d4401L} and $m_{\rm J}$ is the apparent magnitude of the host star in the $J_{\rm 2MASS}$ band. The calculated TSM values are listed in Table \ref{table:tsm}. Compared to the predicted TSM values estimated by the \textit{TESS} ACWG when the planetary masses were unknown, the calculated TSM values are lower and fall below the proposed cut-off for follow-up efforts suggested by \cite{2018PASP..130k4401K}. This can be easily explained by noting that each predicted mass coming from the deterministic mass-radius relation \citep{2017AJ....153...77C} was 2 to 6$\times$ smaller than the measured values. In the giant planet regime, the mass-radius estimates suffer from a large degeneracy, since their radii are almost independent of their masses \citep[e.g.,][and references therein]{2024A&A...686A.296M}. Nevertheless, as we will show in Sect. \ref{sec:discussion}, TOI-4914 b has the second highest TSM value among known warm giant planets orbiting stars more metal-poor than most of the WJ hosting stars, and we will demonstrate the feasibility of its atmospheric characterisation in Sect. \ref{sec:atm_prosp}.

\begin{table}
\caption{Predicted and calculated TSM values.}             
\label{table:tsm}      
\centering                          
\begin{tabular}{l | c c c}        
\hline\hline                 
  & TOI-2714 &  TOI-2981 & TOI-4914\rule{0pt}{2.5ex} \rule[-1ex]{0pt}{0pt} \\
\hline 
Parameter & Value & Value & Value \rule{0pt}{2.5ex} \rule[-1ex]{0pt}{0pt} \\    
\hline                        

     $T_{\rm eq}$ (K)& 1603$\pm$65 & 1358$\pm$39 & 894$\pm$20 \rule{0pt}{2.5ex} \rule[-1ex]{0pt}{0pt} \\
     $g$ (m s$^{-2}$)& 11.6$\pm$2.3 & 37$\pm$4 & 14.2$\pm$1.3 \rule{0pt}{2.5ex} \rule[-1ex]{0pt}{0pt} \\
     $H$ (km)& 501$\pm$108 & 132$\pm$17 & 228$\pm$24 \rule{0pt}{2.5ex} \rule[-1ex]{0pt}{0pt} \\
     $\delta_\lambda$ (ppm)& 230$\pm$51 & 80$\pm$11 & 155$\pm$17 \rule{0pt}{2.5ex} \rule[-1ex]{0pt}{0pt} \\
     predicted~ TSM   & 90 & 104 & 116 \rule{0pt}{2.5ex} \rule[-1ex]{0pt}{0pt} \\
     calculated TSM & 49$\pm$11 & 19$\pm$2 & 60$\pm$7     \rule{0pt}{2.5ex} \rule[-1ex]{0pt}{0pt} \\
\hline                                   
\end{tabular}
\end{table}

\section{Discussion}
\label{sec:discussion}

\subsection{Characteristics of the confirmed planets and bulk density-metallicity correlation}
\label{sec:density-met}

TOI-2714 b is a hot Jupiter planet with a radius of $R_{\rm p} = 1.22 \pm 0.06~R_{\rm J}$ and a mass of $M_{\rm p} = 0.72 \pm 0.10~M_{\rm J}$, which orbits a metal-rich star ([Fe/H] $= 0.30$). On the contrary, TOI-2981 b ($R_{\rm p} = 1.20 \pm 0.04~R_{\rm J}$, $M_{\rm p} = 2.00 \pm 0.10~M_{\rm J}$) and TOI-4914 b ($R_{\rm p} = 1.15 \pm 0.03~R_{\rm J}$, $M_{\rm p} = 0.72 \pm 0.04~M_{\rm J}$) are a hot and a warm Jupiter respectively, both orbiting stars more metal-poor than most of the stars known to host giant planets.

Among these, TOI-4914 b is of particular interest. First, it orbits a star more metal-poor than most of the WJ hosting stars. In this sub-sample, it has the second highest TSM (Table \ref{table:tsm}), making it an interesting target for atmospheric characterisation with JWST. The target with the highest TSM, similar $P_{\rm orb}$ and metallicity is WASP-117 b \citep{2014A&A...568A..81L}. 

\begin{figure*}
   \centering
   \includegraphics[width=\hsize]%
   {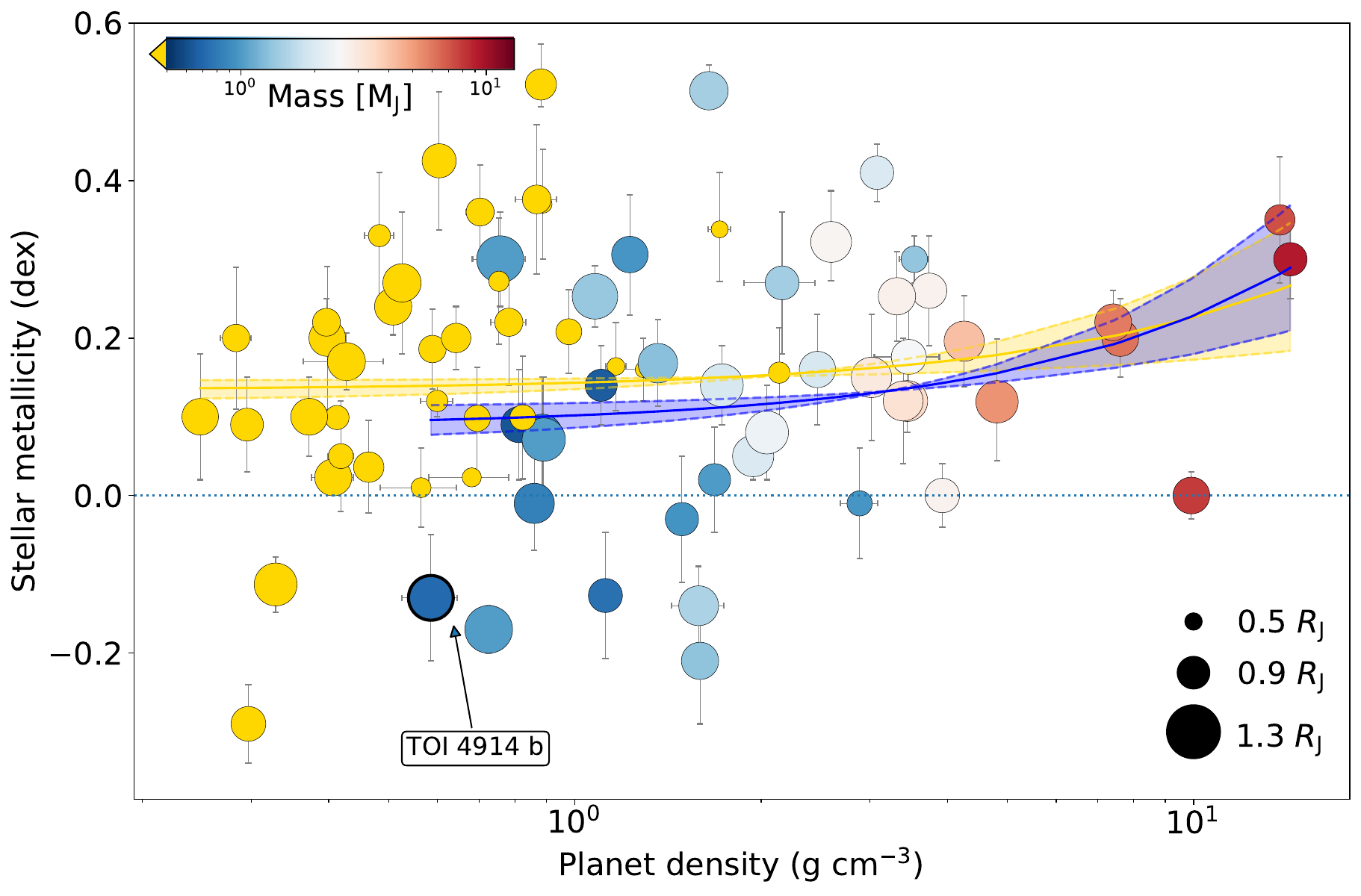}
   \caption{Planet bulk density versus stellar metallicity for lightly irradiated ($F_\star < 2 \times 10^8~{\rm erg~s^{-1}~cm^{-2}}$) giant planets ($M_{\rm p} > 0.1 M_{\rm J}$). Only well-characterised planets are shown, i.e., with a planet density uncertainty of $<$ 25 per cent and a stellar metallicity uncertainty of $<$ 0.1 dex. The planet mass is colour-coded, while the planet density is shown on a logarithmic scale. We colour-coded planets with masses below $0.5 M_{\rm J}$ in yellow to distinguish them from more massive ones. The yellow shaded area corresponds to the correlation distribution when looking at all lightly irradiated giant planets, while the blue one corresponds to the correlation distribution when looking only at planets with masses $> 0.5 M_{\rm J}$. }
   \label{fig:dens-feh}
\end{figure*}

Motivated by the observational trend first noted in \cite{2022MNRAS.511.1043W} -- and reinforced by recent discoveries \citep{2023MNRAS.520.3649H,2023AJ....166....7K,2023A&A...673A..42N,2024MNRAS.tmp.1415H} -- between planet bulk density and stellar metallicity for lightly irradiated ($F_\star < 2 \times 10^8~{\rm erg~s^{-1}~cm^{-2}}$) sub-Neptunes, which postulates that planets orbiting metal-rich stars have metal-rich atmospheres with reduced photo-evaporation \citep{2012MNRAS.425.2931O}, we tested whether a correlation exists for more massive planets. We compared TOI-4914 b with the entire family of well-characterised (uncertainty on the planet density $<$ 25 per cent), lightly irradiated giant planets. We used a Bayesian correlation tool \citep{2016OLEB...46..385F} to measure a possible correlation between bulk density and host star metallicity (Fig. \ref{fig:dens-feh}) in lightly irradiated giant planets. The resulting correlation distribution given by the tool corresponds to a Pearson's coefficient value, and what we found is a median of the correlation posterior distribution of $0.15 \pm 0.11$ ($1.4 \sigma$), with 95 per cent lower and upper bounds of $-$0.05 and 0.37. Most interestingly, as we gradually moved the lower mass limit from 0.1 to $1 M_{\rm J}$ (in steps of $0.1 M_{\rm J}$) and measured the correlation, we found that if we select only giants with $M_{\rm p} > 0.5 M_{\rm J}$, the resulting median correlation becomes $0.275 \pm 0.140$ ($2 \sigma$), with 95 per cent lower and upper bounds of 0.01 and 0.55. We found almost the same peak when selecting only higher-mass planets, but this also increases the coefficient uncertainty. The increasing uncertainty could be due to the fact that there are fewer and fewer planets as we move towards the higher mass cuts. On the basis of the present sample of lightly-irradiated planets, we see no significant dependence of planet density on stellar metallicity for planets less massive than 3 or 4 times the mass of Jupiter. At masses greater than 4 $M_{J}$, the sparseness and scatter of the data preclude any meaningful conclusions.

\begin{figure}
   \centering
   \includegraphics[width=\hsize]%
   {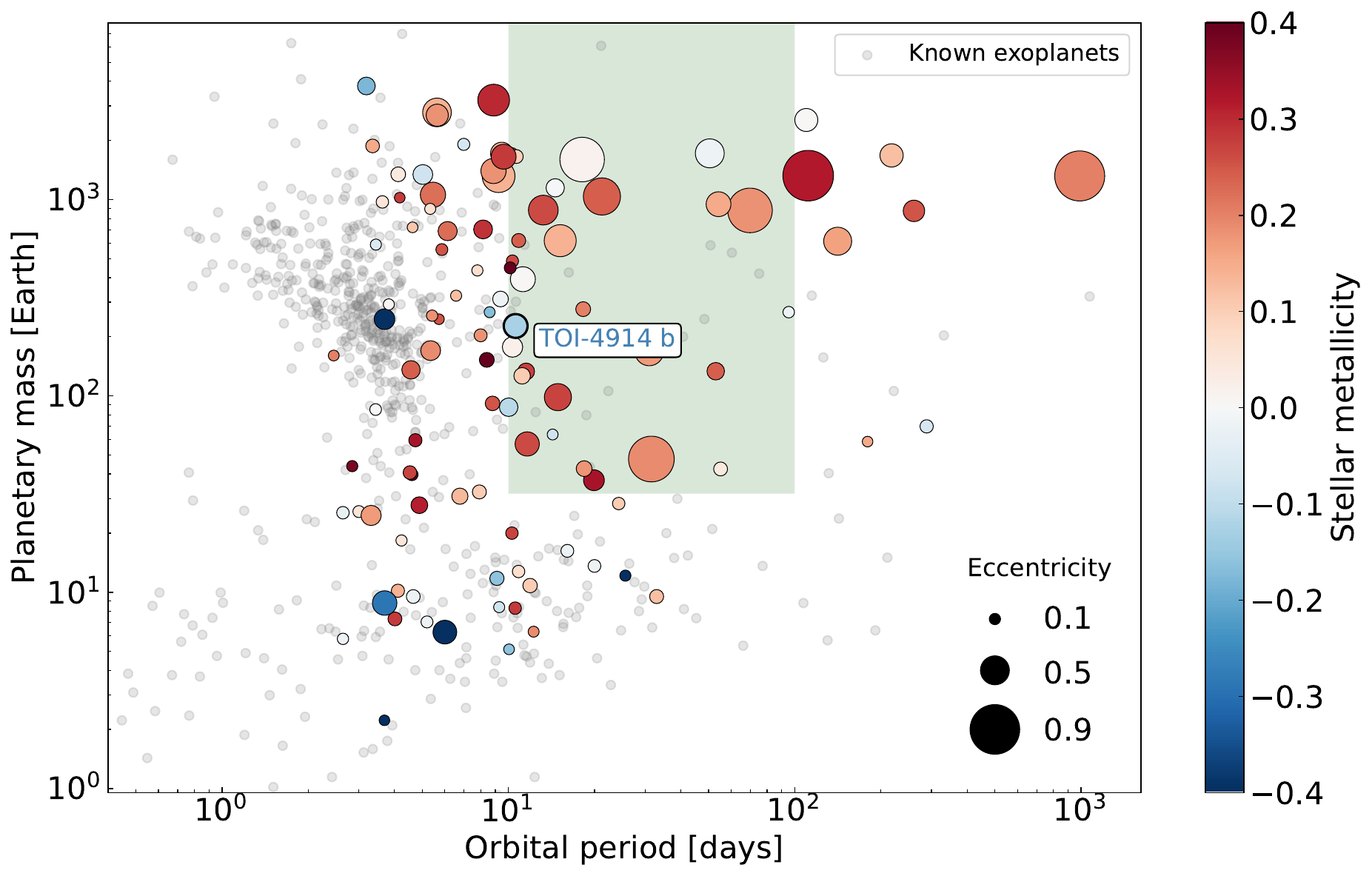}
   \caption{Period--mass distribution of all confirmed transiting planets from the TEPCat catalogue \citep{2011MNRAS.417.2166S}. The dot size tracks the eccentricity, while the host star metallicity is colour-coded (only for planets with $e > 0.1$). The green shaded area represents the location of WJs ($10 < P_{\rm orb} < 100$ d, $M_{\rm p} > 0.1 M_{\rm J}$).}
   \label{fig:ecc_mass_period}
\end{figure}

\begin{figure}
   \centering
   \includegraphics[width=\hsize]%
   {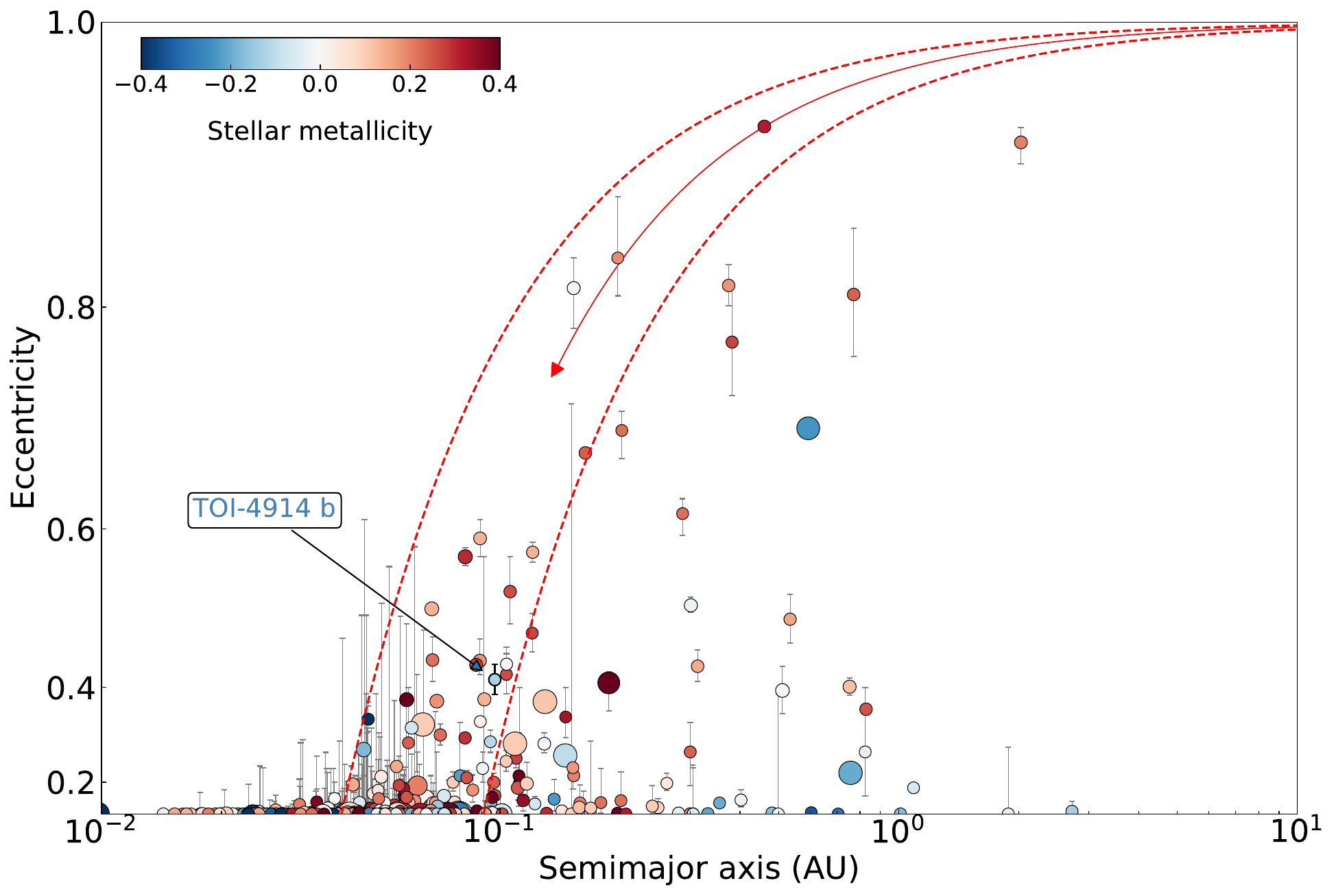}
   \caption{Semimajor axis versus eccentricity of all confirmed transiting planets. The metallicity of the host star is colour coded. The red arrow is an example of a planetary orbit evolution after a high eccentricity migration. TOI-4914 b lies in this region. This diagram is adapted from figure 4 of \citep{2018ARA&A..56..175D}. }
   \label{fig:ecc_semi}
\end{figure}

Second, TOI-4914 b is an eccentric ($e = 0.408 \pm 0.023$) WJ which belongs to a rare sub-sample of high eccentricity WJs orbiting stars more metal-poor than many gas giant hosting stars (Fig. \ref{fig:ecc_mass_period} and \ref{fig:ecc_semi}). This finding seems to contradict previous results \citep[e.g.,][]{2013ApJ...767L..24D,2018ARA&A..56..175D} where giants orbiting metal-poor stars are restricted to low eccentricities, whereas those orbiting metal-rich stars exhibit a range of eccentricities. Figure \ref{fig:ecc_mass_period} shows the period--mass distribution of all the planets, colour-coding those with $e > 0.1$ (and uncertainty on $e \leq 0.1$) by stellar metallicity -- those with smaller eccentricities are left in grey. As recently reviewed in \cite{2024A&A...682A.135C}, the absence of giant planets on eccentric orbits when $P_{\rm orb} < 3$ days is evident. This is most likely a consequence of the effective tidal dissipation they experience during their lifetime. However, as the orbital period increases, we start seeing giant planets on eccentric orbits, with a tendency to be more eccentric at longer orbital periods. This trend can be attributed to the reduction in tidal circularisation rates as orbital distances increase \citep[][Eq. 1]{2008ApJ...678.1396J}. Therefore, in order to gain a deeper understanding of the origin of TOI-4914 b eccentric orbit, we calculated the circularisation timescale ($\tau_{\rm circ}$) using equation 6 from \cite{2008ApJ...686L..29M}. Assuming a circular orbit and a modified tidal quality factor $Q$ of $10^5$ \citep{2014ARA&A..52..171O}, the calculated value for $\tau_{\rm circ}$ is $3.68 \pm 0.97$ Gyr, which is compatible with the age of the system estimated in Sect. \ref{sec:age} ($5.3 \pm 3.4$ Gyr). The large uncertainty in the age of the system prevents us from drawing a conclusion on whether the current eccentric orbit may require an excitation along the lifetime of the system. Figure \ref{fig:ecc_semi} instead is adapted from figure 4 of \cite{2018ARA&A..56..175D} and displays the semi-major axis versus eccentricity of all confirmed transiting planets. In particular, planets that are located in the red dashed region may have formed at high eccentricities (and large semi-major axes) without experiencing tidal disruption \citep{2018ARA&A..56..175D}. The metallicity of the host star is colour-coded. The red arrow is an example of a planetary orbit change after high-eccentricity migration \citep{2003ApJ...589..605W, 2006A&A...453..341M}. 

Encouraged by the observed orbital parameters and the above discussion, we can place a few constraints on the formation history of TOI-4914 b. First, such a high eccentricity cannot be explained by disk migration alone \citep{1996Natur.380..606L}, even though disk migration could have been important in the early stages of evolution of the system. It could have driven the planets close enough to trigger a period of dynamical instability dominated by planet-planet scattering \citep{1996Natur.384..619W, chambers1996, rasioford1996,1997ApJ...477..781L,2002Icar..156..570M,Chatterjee2008,naga08,raymond2009, davies2014,mustill2014,petrovich_rafikov2014,deienno2018}, capable of exciting the eccentricity of TOI-4914 b. Given our discussion in Sect. \ref{sec:kinematics} and \ref{sec:additionalplanets} and the lack of a trend in the RVs, which makes the presence of close stellar companions unlikely, the Kozai-Lidov scenario \citep{2003ApJ...589..605W} is less likely than the planet-planet scattering. The presence of a second planet in the system would support this hypothesis, but with the available data we have no evidence.

\subsection{Challenging radius and interior of TOI-4914 b}

\begin{figure}
   \centering
   \includegraphics[width=\hsize]%
   {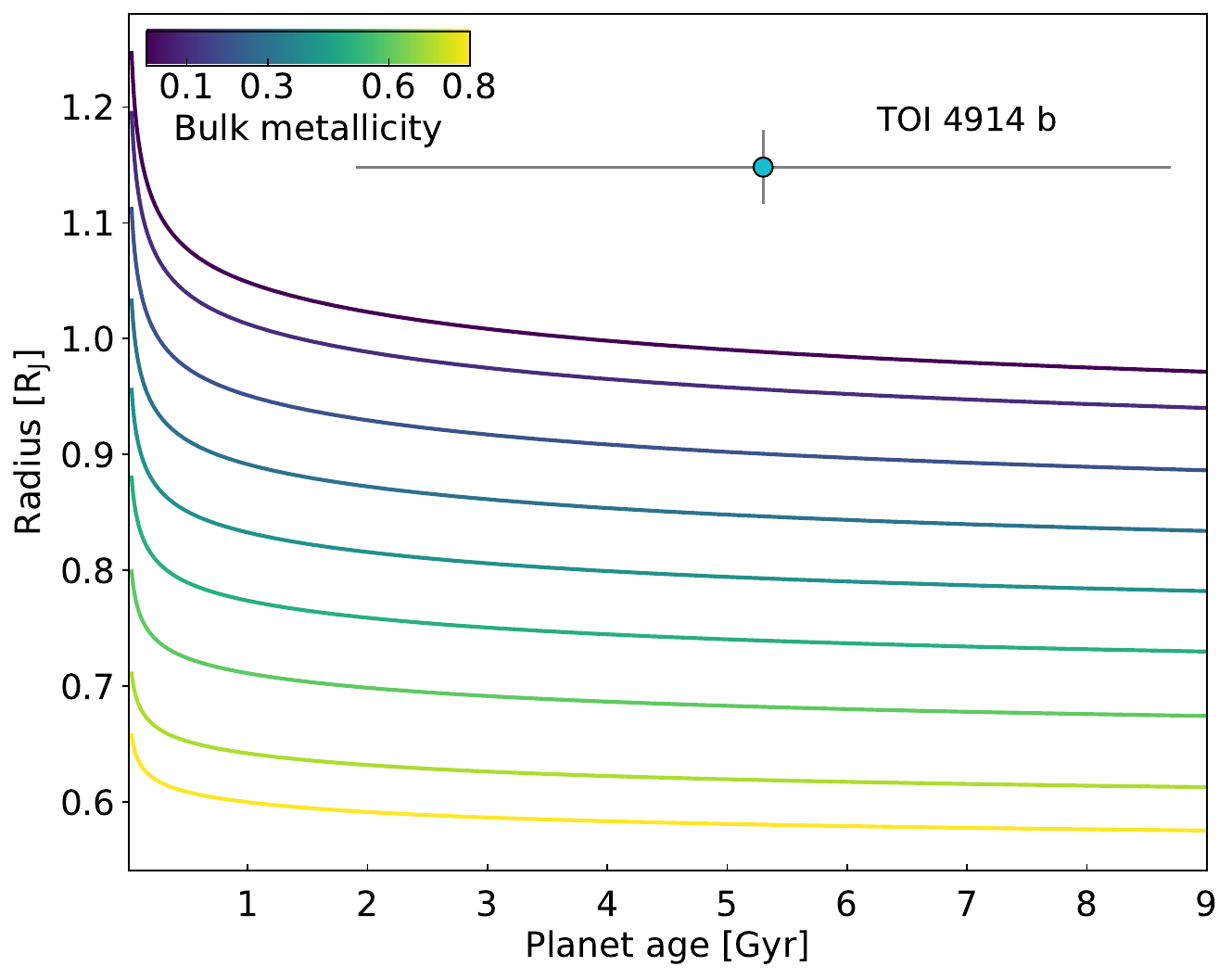}
   \caption{Radius evolution for various bulk metallicities (coloured lines, in units of TOI-4914 b masses). The blue dot represents TOI-4914 b.}
   \label{fig:bulk}
\end{figure}

Using evolutionary models\footnote{\url{https://github.com/tiny-hippo/planetsynth/blob/main}} from \cite{2021MNRAS.507.2094M}, we estimated the planetary bulk heavy element mass fraction (or bulk metallicity) of TOI-4914 b. The data reported in Fig. \ref{fig:bulk} suggest that TOI-4914 b appears to be inflated beyond what is supported by current theoretical models for giant planets (e.g., \citealt{2012A&A...541A..97M}), hindering our ability to effectively use existing planetary interior and evolutionary models to determine a bulk metallicity \citep[see e.g.][]{2024ApJ...962L..22D}. Using equation 3 from \cite{2024arXiv240505307T}, we hence estimated the radius of TOI-4914 b in the absence of anomalous heating, finding $R_{\rm uninflated} = 0.93 \pm 0.11~R_{\rm J}$. This value is much smaller and inconsistent than our estimated radius of $R_{\rm p} = 1.15 \pm 0.03~R_{\rm J}$.

\begin{figure}
   \centering
   \includegraphics[width=\hsize]%
   {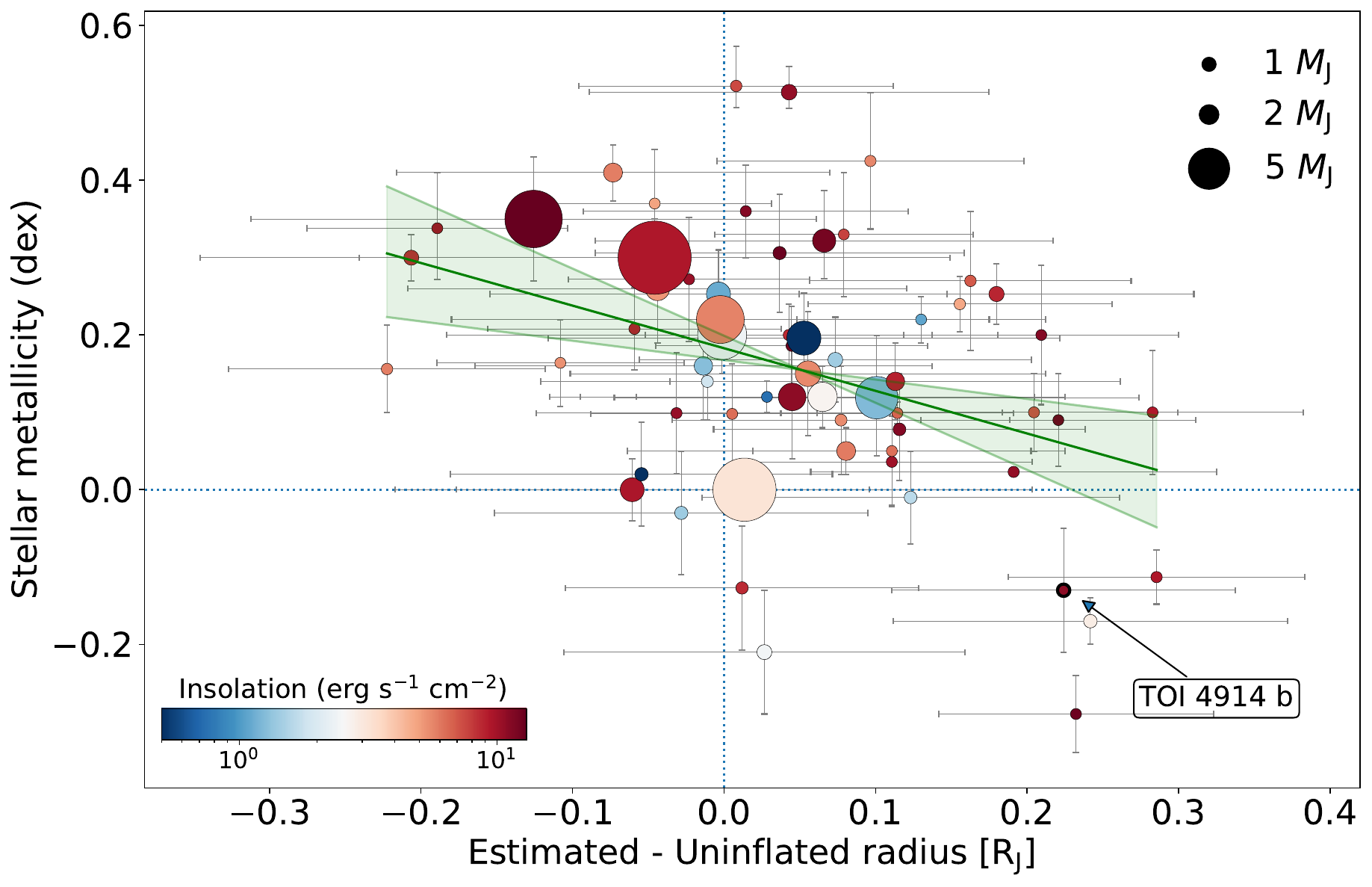}
   \caption{Stellar metallicity as a function of the difference between the observed planetary radius and the one predicted in the absence of anomalous heating.  The dot size tracks the planetary mass, while the incident stellar flux is colour-coded. The green shaded area corresponds to the correlation distribution, with 95 per cent lower and upper bounds.}
   \label{fig:feh_raddiff}
\end{figure}

Following our discussion in Sect. \ref{sec:density-met}, we could link this challenge to the fact that TOI-4914 is more metal-poor than most WJ hosting stars. To test this, we again considered the entire sub-sample of well-characterised, lightly irradiated giant planets, and plotted (in Fig. \ref{fig:feh_raddiff}) the stellar metallicity as a function of the difference between the observed planetary radius and the one predicted in the absence of anomalous heating \citep{2024arXiv240505307T}. Again, we used a Bayesian correlation tool, and what we found is a median correlation of $-0.37 \pm 0.11$ ($3.5 \sigma$), with 95 per cent lower and upper bounds of $-$0.17 and $-$0.59. The strong anti-correlation we see could be evidence that metallicity impacts the atmosphere of WJs. This could confirm that planets around metal-poor stars have enhanced photo-evaporation atmospheres (because they are metal-poor as well) and therefore appear to be more inflated than the models suggest. The reverse is true for metal-rich stars and planets.

\begin{figure*}
   \centering
   \includegraphics[width=\hsize]%
   {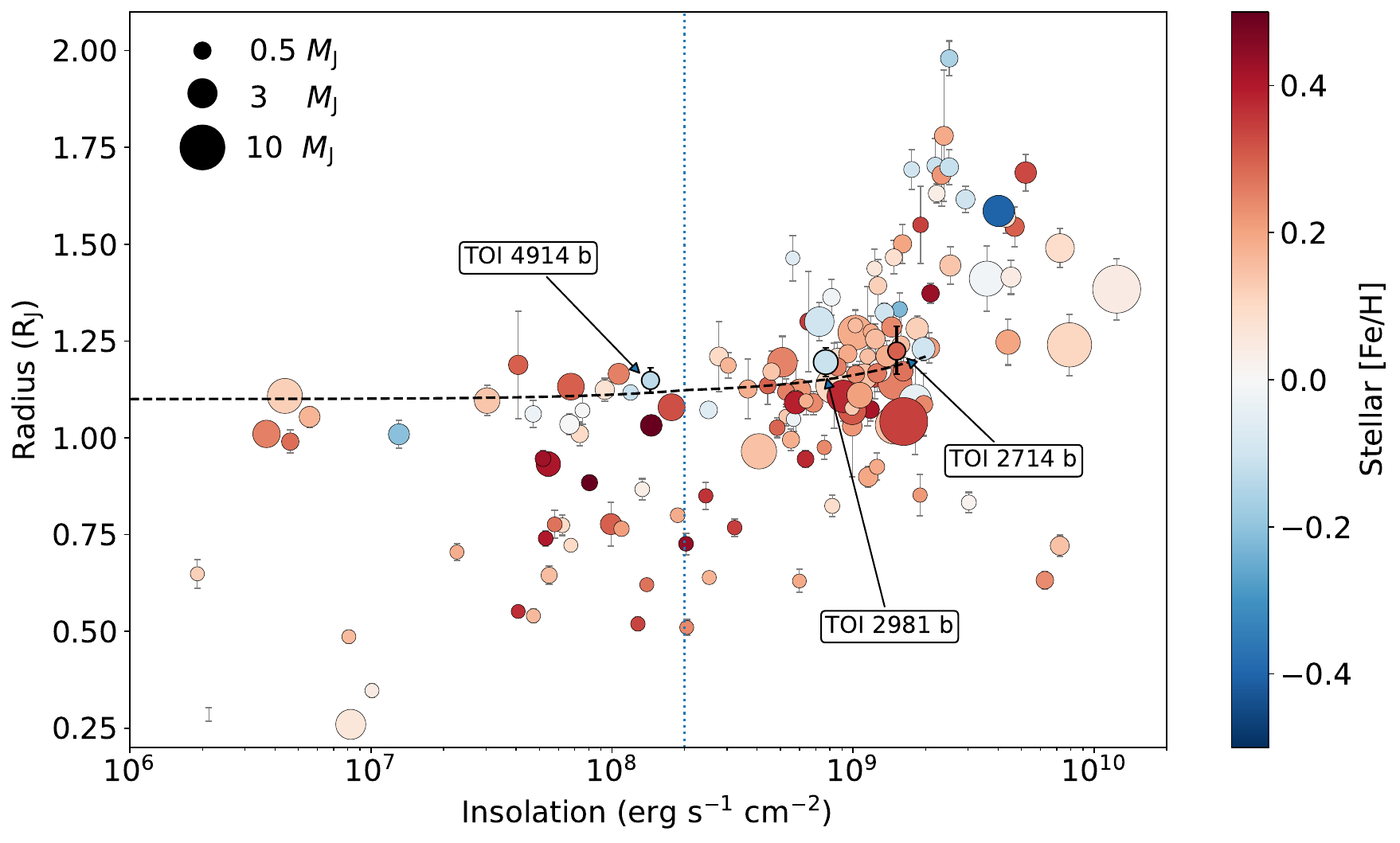}
   \caption{Planet radius of warm (left of the vertical line) and hot giants (right of the line) as a function of incident stellar flux. The stellar metallicity is colour coded, while the size of the dots tracks their planetary masses. The dashed line is the radius of a 4.5 Gyr old, 1 $M_{\rm J}$, pure H/He planet (with no metals), without additional internal heating. Figure adapted from \cite{2024arXiv240505307T}. Only planets with masses above 0.1 $M_{\rm J}$ are shown. }
   \label{fig:insol}
\end{figure*}

Motivated by these findings, we placed TOI-4914 b and the other two planets analysed in this paper in a wider context by plotting the radius of hot and warm giants as a function of the incident flux (Fig. \ref{fig:insol}, adapted from \citealt{2011ApJ...736L..29M, 2016ApJ...831...64T,2024arXiv240505307T} and references therein). We calculated the TOI-4914 b incident flux as $F_\star = (R_\star/R_\odot)^2 (T_{\rm eff}/5777)^4 / (a/{\rm AU})^2$ finding $F_\star = 106.2~ S_\oplus$ where $S_\oplus$ is the incident flux received by the Earth. It appears that although TOI-4914 b has an incident flux below the $F_\star = 2 \times 10^8~{\rm erg~s^{-1}~cm^{-2}}$ threshold for planets not affected by the anomalous radius inflation mechanism (e. g., \citealt{2016ApJ...831...64T}), it seems to be inflated, with a planetary radius above that of a $1 M_{\rm J}$ pure H/He object without additional internal heating.

\begin{figure*}[!th]
   \centering
   \minipage{0.5\textwidth}
   \includegraphics[width=\hsize]{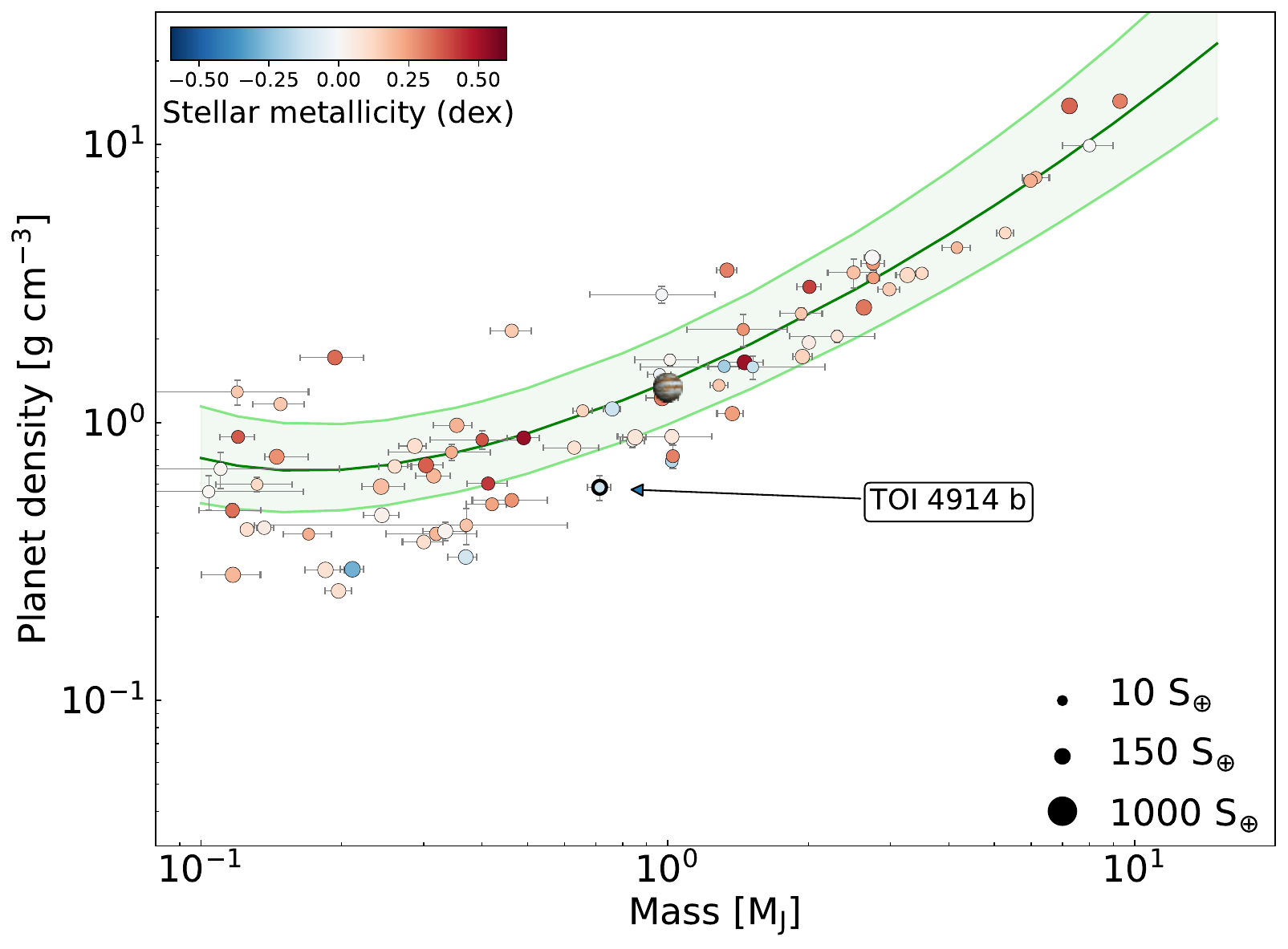}   
   \endminipage\hfill
   \minipage{0.5\textwidth}
   \includegraphics[width=\hsize]{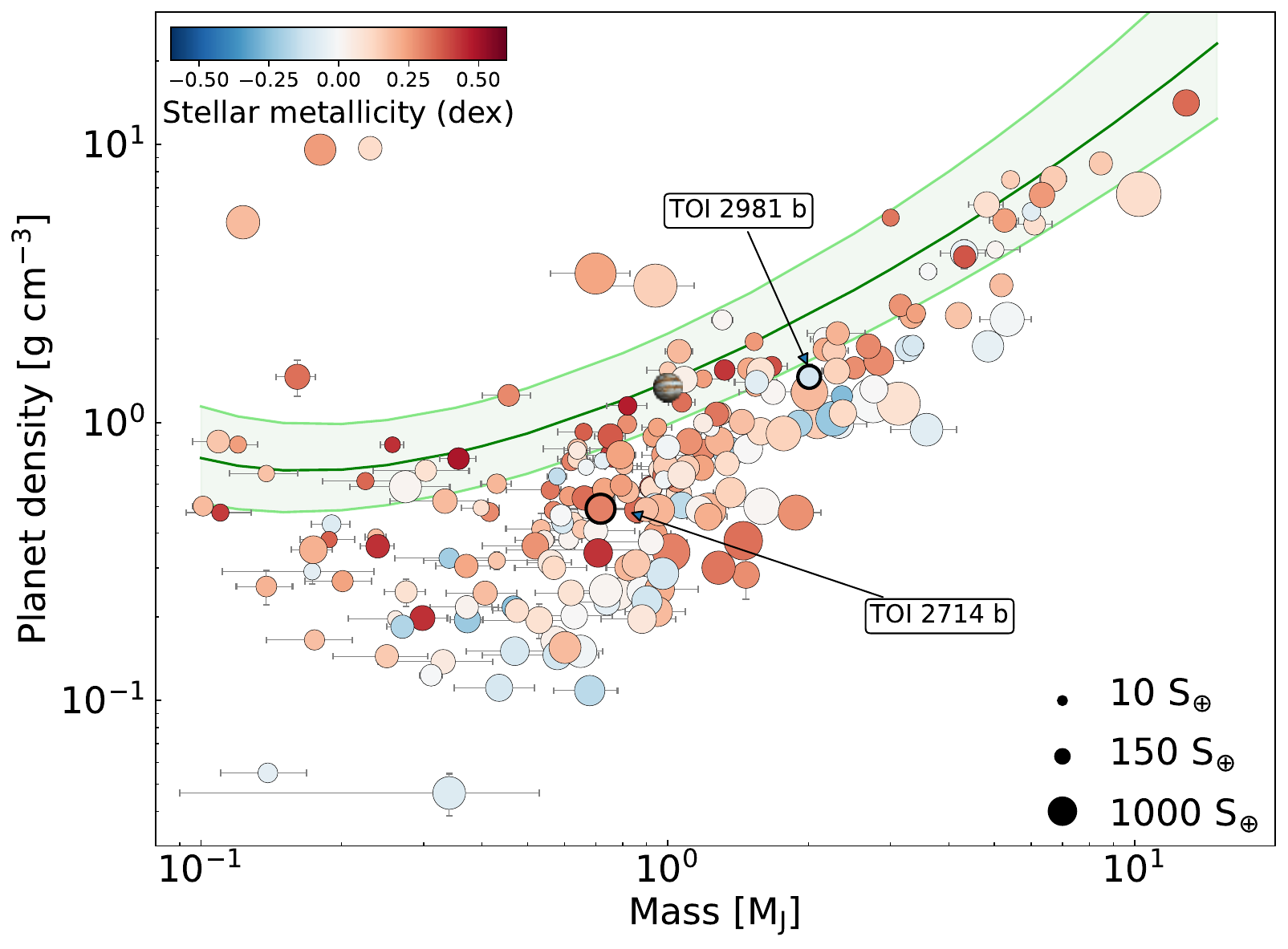} 
   \endminipage\hfill
   
       \caption{\textit{Left}: Planet bulk density as a function of planet mass for lightly irradiated ($F_\star < 2 \times 10^8~{\rm erg~s^{-1}~cm^{-2}}$) giant planets ($M_{\rm p} > 0.1 M_{\rm J}$). Only well-characterised planets are shown, as in Fig. \ref{fig:dens-feh}. The stellar metallicity is colour-coded, while the symbol size scales with insolation. The green shaded area corresponds to the mass-radius relation for giant planets in the absence of anomalous heating, i.e. assuming no inflation \citep[from][]{2024arXiv240505307T}. \textit{Right}: Same plot, but for irradiated ($F_\star > 2 \times 10^8~{\rm erg~s^{-1}~cm^{-2}}$) giants.}
   \label{fig:wj_hj}
\end{figure*}

For the same reasons, we included the three confirmed planets in mass versus bulk density plots (Fig. \ref{fig:wj_hj}). We compared TOI-4914 b with the entire family of well-characterised (planet density uncertainty < 25 per cent), lightly irradiated giant planets, while we compared TOI-2714 b and TOI-2981 b with each well-characterised, irradiated giant. In both comparisons, we included the mass-radius relation from \cite{2024arXiv240505307T} for giant planets in the absence of anomalous heating, i.e. assuming no inflation (green shaded area). TOI 4914 b appears to be inflated compared to other lightly irradiated giants (left panel of Fig. \ref{fig:wj_hj}), but not as extreme as the majority of irradiated HJs (i.e. those in the right panel of Fig. \ref{fig:wj_hj}). Looking at the lightly irradiated giants, there appear to be very few dense planets among the metal-poor systems. Moreover, the majority of planets orbiting metal-poor stars tend to be below the green shaded area. For the irradiated giants, it seems that very metal-rich ones are generally not inflated, with some exceptions in the case of low-mass planets. Conversely, planets around metal-poor stars often appear to be inflated.

\subsection{Ephemeris improvement}
An important step of our analysis is the derivation of new and updated mean ephemeris for TOI-4914 b. Our best-fit relation for the warm giant is:

\begin{equation}
\begin{split}
    T_{0}\,(\mathrm{BJD_{TDB}}) &= 2459317.2727 \pm 0.0006  \\
    &\quad +N \times (10.600567 \pm 0.000010),
\end{split}
\end{equation} 
where the variable $N$ is an integer number commonly referred to as the `epoch' and set to zero at our reference transit time $T_{\rm ref}$. We emphasise that if we propagate the new ephemeris at 2025 January 1, the level of uncertainty is significantly reduced to $\sim$ 2 minutes compared to the previous $\sim$ 251 minutes for TOI-4914 b when only \textit{TESS} photometry was available. This means that when we extend the baseline with ground-based photometry, the error bar for TOI-4914 b is 99\% smaller than when using \textit{TESS} data alone. Accurately identifying the transit windows is crucial for upcoming space-based observations, given the significant investment in observing time and the time-critical nature of such observations. It is crucial to note that no further observations of TOI-4914 are planned in the current \textit{TESS} Extended Mission\footnote{As it results from the Web \textit{TESS} Viewing Tool \url{https://heasarc.gsfc.nasa.gov/wsgi-scripts/TESS/TESS-point_Web_Tool/TESS-point_Web_Tool/wtv_v2.0.py/}}. 

\subsection{Rossiter-McLaughlin measurement prospects}
The measurement of exoplanet sky-projected obliquity, which refers to the angle between the orbital axis of a planet and the spin axis of its host star, provides crucial insights into how planets form and migrate \citep{2011Natur.473..187N}. This can be detected with in-transit RVs by the Rossiter-McLaughlin (RM) effect (e.g., \citealt{2000A&A...359L..13Q,2005ApJ...622.1118O}). Observations of tidally young transiting planets (e.g., with $a/R_\star > 11$, \citealt{2023AJ....166..217W}), which have not yet experienced significant tidal alterations, provide a unique opportunity to study their initial obliquity configuration (e.g., \citealt{2022PASP..134h2001A,2024A&A...684L..17M}). Moreover, they could serve as a primordial distribution of stellar obliquities if they share a common origin with hot giant planets -- affected by the final step in which the star can realign, making the hot giant formation mechanism inaccessible. We estimated the obliquity damping timescales under the influence of tides using the approach of \cite{2012MNRAS.423..486L} and the convective tidal realignment timescale following \cite{2012ApJ...757...18A}, and found a decay time exceeding the Hubble time for TOI-4914 b. This implies that the obliquity should still be unaltered by tidal effects, thus providing a direct diagnostic for the formation path of the planetary system. 

The ``long'' orbital period ($\sim10.6$ days) and high eccentricity ($\sim 0.4$, Table \ref{table:model-lcrv}) of TOI-4914 b, make it an intriguing target for measuring the obliquity between the system's orbital plane and the equatorial plane of the star. The discovery of an aligned system would support planet-planet scattering, as scattering is less efficient in exciting mutual inclination compared to the secular, Kozai-Lidov process \citep{2001Icar..152..205C}.

The measurement and interpretation of the stellar obliquities of warm giants is a remarkable problem that requires a larger sample size of observations \citep{2018ARA&A..56..175D}. We thus determined the expected amplitude of the RV anomaly produced by the RM effect when TOI-4914 b transits $\Delta V_{\rm RM} = 18~{\rm m~s}^{-1}$ using eq. 40 from \cite{2010exop.book...55W}. The predicted amplitude is about three times larger than the individual RV errors for TOI-4914 (see Table \ref{table:obsHARPS}, even for shorter exposure times, better suited for RM observations) and the activity jitter, which is expected to be much smaller on the individual transit timescale compared to that of the stellar rotation period ($\gtrsim$ 25.8 days, Sect. \ref{sec:rot_per}) and is likely to appear as a slope. HARPS can provide the required RV precision to measure the expected RV anomaly.

\subsection{Prospects for atmospheric characterisation}
\label{sec:atm_prosp}
We quantified the atmospheric characterisation using JWST of TOI-4914\,b, the planet with the highest TSM in this work. 

We investigated three different atmospheric scenarios considering equilibrium chemistry as a function of temperature and pressure using FastChem \citep{stock2018} with three different C/O ratios: 0.25 (sub-solar), 0.55 (solar), and 1.0 (super-solar). We did this to constrain models of planet formation. We used FastChem within TauREx3 with the \texttt{taurex-fastchem}\footnote{\url{https://pypi.org/project/taurex-fastchem}} plugin. TauREx \citep{alrefaie2021} is a retrieval code that uses a Bayesian approach to infer atmospheric properties from observed data, utilising a forward model to generate synthetic spectra by solving the radiative transfer equation throughout the atmosphere. We used all the possible gases contributions within FastChem and cross-sections from the ExoMol catalogue\footnote{\url{https://www.exomol.com/data/molecules/}} \citep{2013AIPC.1545..186T,2020JQSRT.25507228T}.

\begin{figure*}
    \centering
    \includegraphics[width=\hsize]{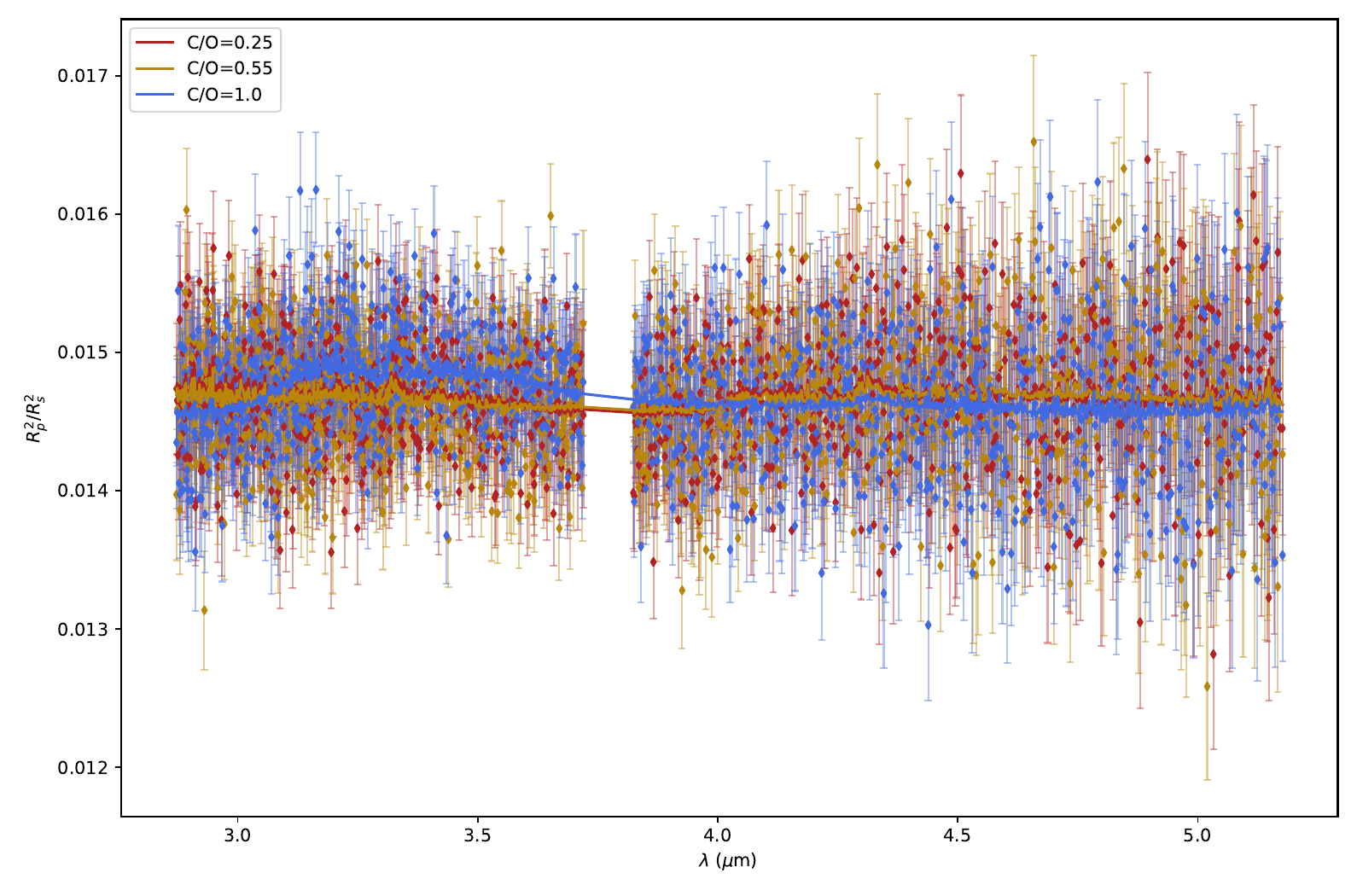}
    \caption{NIRSpec observation simulation using the g395m disperser with f290lp filter (scatter points) and best-fit models from TauREx (lines). The three colours represent three scenarios: C/O = 0.25 in red, C/O = 0.55 in yellow and C/O = 1.0 in blue.}
    \label{fig:retrieval-spectra}
\end{figure*}

After generating the transmission spectra, we simulated the JWST observations using Pandexo \citep{batalha2017}, a tool specifically developed for the JWST mission. We simulated NIRSpec observations in BOTS mode, using the s1600a1 aperture with g395m disperser, sub2048 subarray, nrsrapid read mode, and f290lp filter. We simulated a single transit and an observation 1.75 times $T_{14}$ long to ensure a robust baseline coverage. We fixed this instrumental configuration for all three scenarios. In Fig. \ref{fig:retrieval-spectra}, we show the resulting spectra for the different C/O ratios and their best-fit models.

\begin{figure}
    \centering
    \includegraphics[width=\hsize]{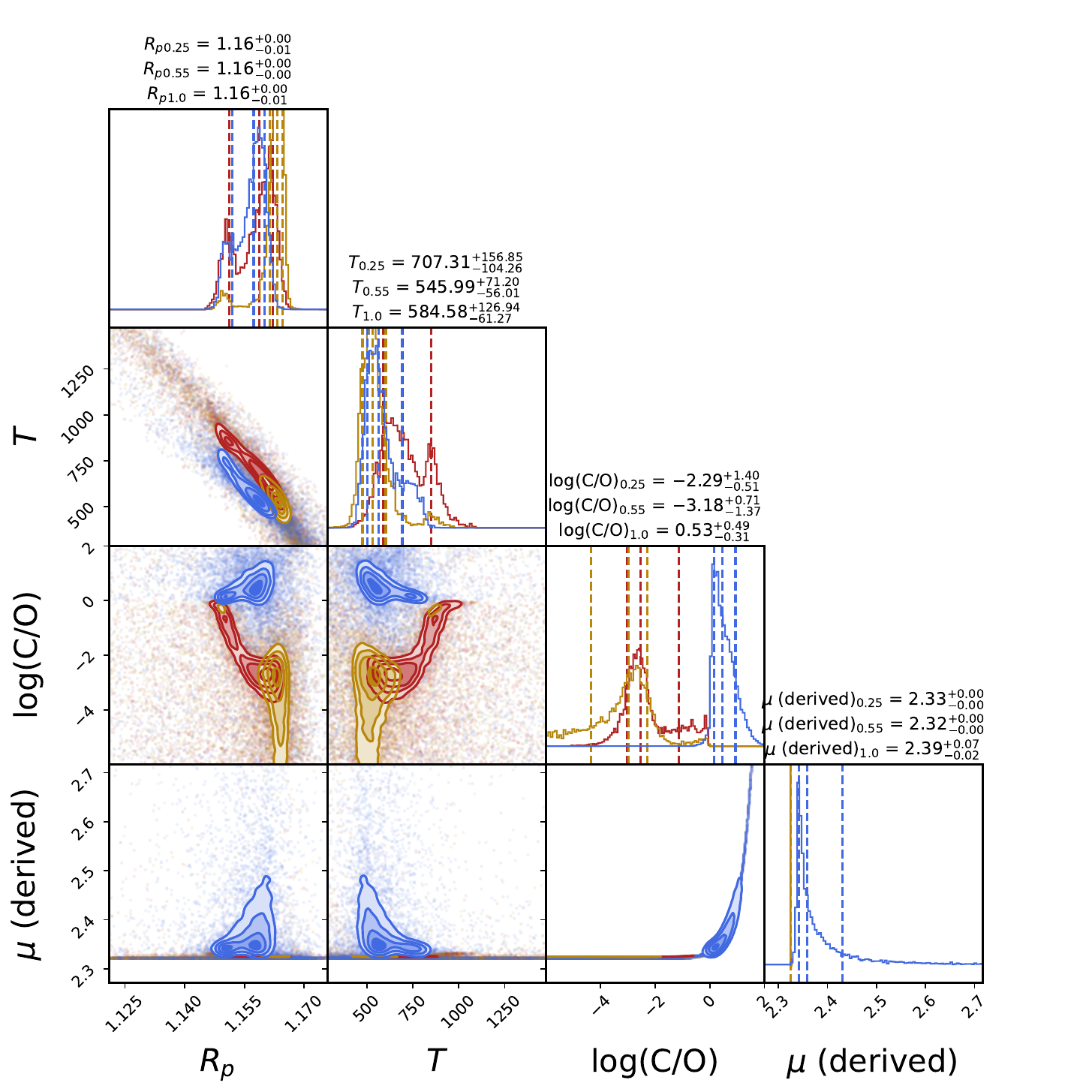}
    \caption{Posterior distributions for the three different scenarios. We show C/O = 0.25 in red, C/O = 0.55 in yellow and C/O = 1.0 in blue.}
    \label{fig:retrieval-posteriors}
\end{figure}

We proceeded with the atmospheric retrievals on the NIRSpec/JWST simulations using a Nested Sampling algorithm with the \texttt{MULTINEST} \citep{feroz2009} library with 1000 live points. We fitted the radius of the planet $R_p$, the equilibrium temperature of the atmosphere $T_{\rm eq}$, and the C/O ratios. Figure \ref{fig:retrieval-posteriors} shows the results of the atmospheric retrieval. 

Using NIRSpec with the g395m disperser wavelength range (2.87\,$\mu$m -- 5.17\,$\mu$m), we can assess the C/O ratio under three different assumptions (see Table \ref{tab:retrieval-results}). In particular, when assuming C/O ratios lower than 1, our sensitivity to precise C/O ratio estimates decreases, and all solutions lead to a C/O ratio compatible with 0. A precise atmospheric configuration, in this case, would be possible only for high atmospheric C/O ratios. It is important to note that the three scenarios are distinguishable between themselves as shown by the different Bayesian evidence in Table \ref{tab:retrieval-results}. Furthermore, if TOI-4914 b is found to have high atmospheric C/O ratios, we could retrieve these values and possibly rule out the disk instability \citep{1997Sci...276.1836B, 2007prpl.conf..607D} scenario \citep[see e.g.,][for a discussion]{2022MNRAS.516.1032H, 2024ApJ...969L..21B}.

Future observations of TOI-4914\,b would be feasible, but accurate constraints on formation and evolution models will depend critically on the particular chemical composition of its atmosphere.

\begin{table}[h]
\caption{Retrieval results for the three different scenarios.}
\centering
\begin{tabular}{p{1.53cm}| p{1.8cm}| p{1.8cm} |p{1.92cm}}
\hline \hline
Parameter & C/O = 0.25 & C/O = 0.55 & C/O = 1.0 \rule{0pt}{2.3ex} \rule[-1.5ex]{0pt}{0pt} \\
\hline 
$R_p$ $(R_J)$ &  $1.158_{-0.008}^{+0.003}$ & $1.163_{-0.002}^{+0.001}$ & $1.156_{-0.005}^{+0.003}$ \rule{0pt}{2.5ex} \rule[-1.5ex]{0pt}{0pt} \\
$T_{\rm eq}$ (K) & $700_{-100}^{+160}$ & $550_{-60}^{+70}$ & $580_{-60}^{+130}$ \rule{0pt}{2.3ex} \rule[-2.2ex]{0pt}{0pt} \\
C/O & $0.005_{-0.006}^{+0.016}$ & $0.001_{-0.002}^{+0.001}$ &  $\geq 1$\tablefootmark{a} \rule{0pt}{2.3ex} \rule[-2.0ex]{0pt}{0pt} \\
$\mu$ (derived) & $2.3251\pm0.0006$ &  $2.32471\pm0.00004$ & $2.39\pm0.07$ \rule{0pt}{2.3ex} \rule[-1.5ex]{0pt}{0pt} \\
$E$\tablefootmark{b} & 6914 & 6834 & 6882 \rule{0pt}{2.3ex} \rule[-1.5ex]{0pt}{0pt} \\
\hline
\end{tabular}
\tablefoot{\tablefoottext{a}{16th percentile.} \tablefoottext{b}{Bayesian evidence.}}
\label{tab:retrieval-results}
\end{table}

\section{Conclusions}
\label{sec:conclusions}

In this work, we present the discovery of two hot Jupiters on nearly circular orbits (TOI-2714 b and TOI-2981 b) and one warm Jupiter with a significant eccentricity (TOI-4914 b). Thanks to HARPS radial velocity time series, we measure the masses of each of the candidates identified in the \textit{TESS} light curves and further detected with ground-based photometry, confirming their planetary nature. In addition, we accurately estimate the parameters of the three host stars. 

Placing these new planets into a wider context, we can see that TOI-4914 b orbits a star that is more metal-poor than most gas giant planets hosting stars. It joins a small group of eccentric, warm and lightly irradiated giant planets. 

The well-constrained high eccentricity of TOI-4914 b provides insights into its formation history. In particular, we find that planet-planet scattering is the most likely scenario capable of producing such an excitation. Future measurements of the Rossiter-McLaughlin effect would provide crucial information in this regard. If planet-planet scattering is confirmed, this finding could support the idea that even metal-poor stars can form systems with multiple gas giant planets (see e.g., \citealt{2023AJ....165..171W} for metal-rich stars). 

Motivated by previous observational trends between planetary bulk density and stellar metallicity for lightly irradiated sub-Neptunes, we test whether a correlation also exists for lightly irradiated gas giant planets. We find no significant evidence of a correlation between stellar metallicity and planet density for lightly-irradiated planets less massive than 4 $M_{J}$. The available data are too sparse to draw conclusions about planets with higher masses. The explanation for TOI-4914 b's location below the main density-mass relation in Fig. \ref{fig:wj_hj}, in a region that it shares with planets of similar mass orbiting more metal-rich stars, must lie elsewhere. 

We estimate the radius of TOI-4914 b in the absence of anomalous heating \citep{2024arXiv240505307T} and find a value ($R_{\rm uninflated}$) much smaller and inconsistent with our measured radius ($R_{\rm p}$). To investigate this discrepancy, we test the possible influence of the metallicity of TOI-4914. Taking into account the entire sub-sample of lightly irradiated gas giants, we find a strong anti-correlation (Pearson's coefficient = $-$0.37 with a 2-sided p-value of 0.001) between the observed minus predicted radius difference ($R_{\rm p} - R_{\rm uninflated}$) and the stellar metallicity. Our finding may suggest that metallicity affects the atmospheres of warm Jupiter planets. This may confirm that planets orbiting metal-poor stars experience enhanced photo-evaporation, making them appear more inflated than the models suggest. We could further test these hypotheses by measuring the atmospheric chemistry with JWST and obtaining information on the metal content of TOI-4914 b and other planets orbiting metal-poor stars.

\begin{acknowledgements}
GMa, LBo, TZi, VNa, and GPi acknowledge support from CHEOPS ASI-INAF agreement n. 2019-29-HH.0. TGW would like to acknowledge the University of Warwick and UKSA for their support. The postdoctoral fellowship of KB is funded by F.R.S.-FNRS grant T.0109.20 and by the Francqui Foundation. This publication benefits from the support of the French Community of Belgium in the context of the FRIA Doctoral Grant awarded to MTi. Author F.J.P acknowledges financial support from the Severo Ochoa grant CEX2021-001131-S funded by MCIN/AEI/10.13039/501100011033 and Ministerio de Ciencia e Innovación through the project PID2022-137241NB-C43. ACC acknowledges support from STFC consolidated grant number ST/V000861/1, and UKSA grant number ST/X002217/1.

This paper made use of data collected by the \textit{TESS} mission and are publicly available from the Mikulski Archive for Space Telescopes (MAST) operated by the Space Telescope Science Institute (STScI). Funding for the \textit{TESS} mission is provided by NASA's Science Mission Directorate. We acknowledge the use of public \textit{TESS} data from pipelines at the \textit{TESS} Science Office and at the \textit{TESS} Science Processing Operations Center. Resources supporting this work were provided by the NASA High-End Computing (HEC) Program through the NASA Advanced Supercomputing (NAS) Division at Ames Research Center for the production of the SPOC data products.

Based in part on observations obtained at the Southern Astrophysical Research (SOAR) telescope, which is a joint project of the Minist\^erio da Ci\^encia, Tecnologia e Inova\c{c}\~oes do Brasil (MCTI/LNA), the US National Science Foundation's NOIRLab, the University of North Carolina at Chapel Hill (UNC), and Michigan State University (MSU).

TRAPPIST-South is funded by the Belgian National Fund for Scientific Research (FNRS) under the grant PDR T.0120.21. 
\end{acknowledgements}

\bibliographystyle{aa} 
\section*{Data Availability}
The RV spectroscopic time series will be available in electronic format as supplementary material to the paper at the CDS.
\bibliography{references} 
%

\end{document}